\documentclass[11pt]{article}
\usepackage{amssymb}
\usepackage{graphicx}
\usepackage{amsmath}
\usepackage{makeidx}
\usepackage{indentfirst}
\usepackage[T1]{fontenc}
\usepackage[utf8]{inputenc}
\usepackage{authblk}
\usepackage{rotating}

\newcounter{resultnum}[section]
\newcounter{conclusionnum}[section]
\newcounter{conditionnum}[section]
\newcounter{conjecturenum}[section]
\newcounter{examplenum}[section]
\newcounter{exercisenum}[section]
\newcounter{lemmanum}[section]
\newcounter{notationnum}[section]
\newcounter{theoremnum}[section]
\newcounter{definitionnum}[section]
\newcounter{corollarynum}[section]
\newcounter{remarknum}[section]
\newcounter{propositionnum}[section]
\newcounter{acknowledgementnum}[section]
\newcounter{algorithmnum}[section]
\newcounter{axiomnum}[section]
\newcounter{casenum}[section]
\newcounter{claimnum}[section]
\newcounter{summarynum}[section]
\newcounter{problemnum}[section]
\begin{document}

\title{Off--Diagonal Deformations of Kerr Metrics and\\
Black Ellipsoids in Heterotic Supergravity}
\date{December 16, 2016}
\author{ 
Sergiu I. Vacaru \footnote{{\it Address for post correspondence:\ } Flat 4, Brefney house, Fleet street, Ashton-under-Lyne, Lancashire, OL6 7PG, the UK }\\
{\small \textit{Quantum Gravity Research; 101 S. Topanga Canyon Blvd \#
1159. Topanga, CA 90290, USA}} \\
{\small and } {\small \textit{Project IDEI , University "Al. I. Cuza" Ia\c si, Romania}} \\
{\small \textit{email: sergiu.vacaru@gmail.com}} \\
${}$ \\
Klee Irwin \\
{\small \textit{Quantum Gravity Research; 101 S. Topanga Canyon Blvd \#
1159. Topanga, CA 90290, USA}} \\
{\small \textit{email: klee@quantumgravityresearch.org }} }
\maketitle

\begin{abstract}
Geometric methods for constructing exact solutions of motion equations with first order $\alpha ^{\prime }$ corrections to the heterotic supergravity action implying a non-trivial Yang-Mills sector and six  dimensional, 6-d, almost-K\"{a}hler internal spaces are studied. In 10-d spacetimes, general parametrizations for generic off--diagonal metrics, nonlinear and linear connections and matter sources, when the equations of motion decouple in very general forms are considered. This allows us to construct a variety of exact solutions when the coefficients of fundamental geometric/physical objects depend on all higher dimensional spacetime coordinates via corresponding classes of generating and integration functions, generalized effective sources and integration constants. Such generalized solutions are determined by generic off--diagonal metrics and nonlinear and/or linear connections. In particular, as configurations which are warped/compactified to lower dimensions and for Levi--Civita connections. The corresponding metrics can have (non) Killing and/or Lie algebra symmetries and/or describing (1+2)-d and/or (1+3)-d domain wall configurations, with possible warping nearly almost-K\"{a}hler manifolds, with gravitational and gauge instantons for nonlinear vacuum configurations and effective polarizations of cosmological and interaction constants encoding string gravity effects. A series of examples of exact solutions describing generic off-diagonal supergravity modifications to black hole/ ellipsoid and solitonic configurations are provided and analyzed. We prove that it is possible to reproduce the Kerr and other type black solutions in general relativity (with certain types of string corrections) in 4-d and to generalize the solutions to non-vacuum configurations in (super) gravity/ string theories. %

\vskip0.1cm


\vskip3pt MSC2010:\ 8C15, 8D99, 83E99;\   PACS2008:\ 04.20.Jb, 04.50.-h, 04.20.Cv
\end{abstract}

\newpage

\tableofcontents


\section{Introduction}

The problem of constructing exact solutions (in particular, with parametric
dependence on some deformation parameters) of equations of motion in
ten--dimensional, 10-d, (super) string and gravity is of great importance.
Recent approaches for solutions of generalized gravitational and matter
field equations in modified gravity theories, (MGTs, with
bi-metric/-connection structure, possible nontrivial mass terms for
graviton, locally anisotropic effects etc.) including generic off--diagonal
solutions in general relativity, GR, have been elaborated upon. This
involves phenomenological applications in high energy physics, various
approaches to quantum gravity and attempts to explain observational data in
modern accelerating cosmology. For review of such subjects, we cite
respectively \cite{lecht1,harl,gran,wecht, doug,lust, samt}, on
superstrings, flux compactifications, D-branes, instantons etc; \cite%
{vepl,vggr1,vjgp,vwitten}, on geometric methods in quantum gravity; and \cite%
{ferrara1,kounnas3,kehagias,stavr,saridak,mavromat,basstavr,mavr,
odints1,capoz,drg1,hr1,vgrg,vsingl2,vcosmsol1,vcosmsol2,vcosmsol3,vcosmsol4,vcosmsol5}%
, on MGTs and applications, see also references therein.

The supergravity/superstring and MGT gravitational field equations are
formulated as sophisticated systems of nonlinear partial differential
equations, PDEs. Various types of advanced analytic and numeric methods for
constructing exact and approximate solutions of such equations have been
explored. For GR, a number of examples of exact and physically important
solutions are summarized in monographs \cite{kramer,griff}. In generalized
(super) gravity theories, most of the solutions with a variety of different
vacua, such as in string and brane theories, comes from corresponding
choices of the internal manifold. Toroidal compactifications, special
geometric cases as Calabi-Yau and, more generally but with less
supersymmetry, with $SU(3)$ structure manifolds were considered. The bulk of
solutions in various super/noncommutative/extra dimension/modified gravity
theories are described by metrics, frames and connections with coefficients
of fundamental geometric/physical objects depending on one and/or two
coordinates in 4-d -- 10-d spacetimes. For well--known classes of solutions,
diagonalizations of metrics are possible via coordinate transforms and the
linear connection structures are mostly of Levi Civita, LC, and K\"{a}hler
type. Additional distortions by torsion structures are also considered.
Various generalizations of well known and physically important exact
solutions for the Schwarzschild, Kerr, Friedman-Lema\^{\i}%
tre-Robertson-Worker (FLRW), wormhole spacetimes etc were constructed. These
classes of diagonalizable metrics (the off--diagonal terms in the Kerr
solutions are determined by rotations and respective frame/coordinate
systems) are generated by a certain ansatz where motion equations are
transformed into certain systems of nonlinear second order ordinary
equations (ODE), 2-d solitonic equations etc. These systems of ODEs have
Killing vector symmetries which result in additional parametric symmetries
\cite{ger1,ger2,vpars} and depend on integration constants.

A number of physically important solutions with black hole, wormhole,
cosmological, monopole and instanton configurations etc. were constructed in
different (super) string/ gravity and MGTs for a diagonalizable ansatz
depending generically on one spacetime coordinate (in certain cases, with
dependencies on two coordinates and with Killing or other type symmetries).
For the majority of such solutions, the corresponding motion equations
transform with the corresponding diagonal ansatz for metrics and frame
transforms from nonlinear systems of PDE in nonlinear systems of ODEs. The
integrals of such ODEs depend on certain integration constants which are
defined from some boundary/asymptotic/initial and prescribed symmetry
conditions. For many years, physicists and mathematicians concentrated their
efforts on constructing further generalizations and applications of
"diagonalizable" solutions in string and (super) gravity theories because it
was more "easy" to find analytic and numeric solutions of resulting systems
of ODEs. In these approaches, the integration constants can be related to
certain physical constants considering the Cauchy problem, or by using
various assumptions on asymptotic/boundary/symmetry conditions.

All versions of (supersymmetric) modified Einstein equations consist of very
sophisticated off--diagonal nonlinear systems of PDEs. In general form, the
main properties of nonlinear/nonholonomic/parametric interactions of such
(super) gravitational and matter field systems are described by PDEs and not
by approximations to ODEs. To find general classes of solutions in
analytical form, understand their geometric properties and search for
possible physical implications is of great importance in modern mathematics,
physics and cosmology. Imposing only a "simple" diagonalizable ansatz of
higher symmetry (as it is usually considered for generating new classes of
exact solutions in various gravity theories), we "cut" the possibility to
find infinite numbers of classes of generic off-diagonal solutions
determined by generating and integrations functions depending on 3,4,....
spacetime coordinates, with various commutative and noncommutative
parameters etc from the very beginning. A typical example is that of
solitonic wave solutions depending on three variables, which can not be
generated for a very "simple" diagonal ansatz and factorized dependent on
certain generated functions. It is possible that various problems in modern
acceleration cosmology (with structure formation, dark energy and dark mater
etc.) could be solved by certain (super) string/gravity solutions related to
generic off--diagonal nonlinear configurations in GR and it may not be
necessary to radically modify the standard gravity and particle physics
theories.

The problem of constructing generic off--diagonal exact solutions\footnote{%
the metric fields corresponding to such solutions can not be diagonalized in
a finite or infinite spacetime region via coordinate transforms} with
coefficients of metrics and connections and other physically important
geometric objects depending on three, four and extra dimensional spacetime
coordinates is much more difficult. For instance, there are a maximum of six
independent components of a metric tensor from ten components in a 4-d
(pseudo) Riemannian spacetime\footnote{%
four components from a maximum of ten can be fixed to be zero using
coordinate transforms, which is related to the Bianchi identities}. Any such
ansatz with metrics depending on 3-4 spacetime coordinates transforms the
Einstein equations into systems of nonlinear PDEs, which can not be
decoupled and integrated in a general analytic form if the constructions are
performed with respect to local coordinate frames and for the
LC--connection. The condition of zero torsion imposes various types of
contractions between the coefficients of the linear connection, reference
frames and various tensor fields which do not allow any general decoupling
of the corresponding systems of PDEs. To generate solutions with generic
off--diagonal metrics and generalized connections for higher dimensional
configurations (for instance, in 5d -10-d string gravity and MGTs) is a
technically more difficult task than in 3d- 4d theories.

In a series of publications \cite%
{vpars,sv2001,vapexsol,vex1,vex2,vex3,veym,svvvey,tgovsv,vtamsuper}, the
so-called anholonomic frame deformation method, AFDM, of constructing exact
solutions in commutative and noncommutative (super) gravity and geometric
flow theories has been explored. By straightforward analytic computations,
it was proven that it is possible to decouple the gravitational field
equations and generate general classes of solutions in various theories of
gravity with metric and nonlinear, N-, and linear connections structures.
The geometric formalism is based on spacetime fibrations determined by
nonholonomic distributions with splitting of dimensions, 2 (or 3) + 2 + 2 +
.... In explicit form, certain classes of N--elongated frames of reference,
considered formal extensions/embeddings of 4-d spacetimes into higher
dimensional spacetimes are introduced and necessary types of adapted linear
connections are defined. These connections are called distinguished,
d--connections, defined in forms which preserve the N--connection splitting.
In Einstein gravity, a d--connection is considered auxiliary, which in
certain canonical forms can be uniquely defined by the metric structure
following the conditions of metric compatibility and some other geometric
conditions (for instance, that certain zero values for "pure" horizontal and
vertical components contain nonholonomically induced torsion fields).
Surprisingly, such a canonical d--connection allows us to decouple the
motion equations in general forms and generate various classes of exact
solutions in generalized/modified string and gravity theories. Having
constructed a class of generalized solutions in explicit form (depending on
generating and integration functions, generalized effective sources and
integration constants), we can constrain the induced torsion fields to zero
and "extract" solutions for LC--configurations and/or Einstein gravity. It
should be emphasized that it is important to impose the zero--torsion
conditions at the end, i.e. after we have found a class of generalized
solutions. We can not decouple and solve the corresponding systems of PDEs
in general forms if we use the LC--connection from the very beginning. Here
it should be noted that it is important to work with nontrivial torsion
configurations in order to find exact solutions in string gravity and gauge
gravity models.

In this paper, we apply methods in the geometry of nonhlonomic and almost-K%
\"{a}hler manifolds in order to study heterotic supergravity derived in the
low--energy limit of heterotic string theory \cite{grsch,gross1,gross2}.
This publication is associated with another paper\cite{partner1}, where an
approach to heterotic string gravity is formulated in the language of
nonholonomic and almost-K\"{a}hler geometry. We cite also section 4.4 \cite%
{lecht1} for a summary of previous results and similar conventions on warped
configurations and modified gravitational equations.\footnote{%
Nevertheless, we shall elaborate a different system of notations with
N--connections and auxiliary d--connections which allows us to define
geometric objects on higher order shells of nonholonomically splitted 10-d
spacetimes.} The main goal of this work is to develop a geometric method for
integrating in generic off--diagonal forms, and for generalized connections,
the equations of motion of heterotic supergravity, up to and including terms
of order $\alpha ^{\prime }.$ As a secondary goal, we shall construct
explicit examples of exact solutions describing nonholonomic deformations of
the Kerr metric. In general, it is possible to formulate conditions for
effective sources and generic off-diagonal when string gravity may
encode/mimic equivalent solutions in massive gravity and/or modified $f(R,T)$
gravity. For reviews and original results related to massive and other types
MGTs, we cite references \cite{odints1,capoz,drg1,hr1,nieu,koyam}.

In this work, a series of exact and/or small parameter depending solutions
which for small deformations mimic rotoid Kerr-de Sitter like black
holes/ellipsoids, self--consistently embedded into generic off--diagonal
backgrounds of 10 dimensional spacetimes are constructed. Such backgrounds
can be of solitonic/vertex/instanton type. We study string gravity
modifications with respect to nonholonomic frames and via re--definition of
generating and integrations functions and coefficients of sources. These
modifications can be analyzed in the framework of Einstein gravity but
modelled by effective polarized cosmological constants and off--diagonal
terms. Using solutions for heterotic string gravity, it is possible to mimic
physically important effects in modified gravity. In a series of associated
papers, we studied the acceleration of the universe, certain dark energy and
dark matter locally anisotropic interactions, effective renormalization of
quantum gravity models (\cite{vgrg,vbranef,vepl}) via nonlinear generic
off--diagonal interactions on effective Einstein spaces. These constructions
can developed for models elaborated in the framework of string theory.

The solutions of heterotic supergravity which are constructed in this and
the associated \cite{partner1} work describe (1+3)--dimensional walls
endowed with generic, off--diagonal metrics, warped to an almost-K\"{a}hler
6-d internal space in the presence of nonholonomically deformed
gravitational and gauge instantons. The generalized instanton contributions
are adapted to a nontrivial nonlinear connection structure determined by
generic off--diagonal interactions which allows us to solve the Yang--Mills,
YM, sector and the corresponding Bianchi identity at order $\alpha ^{\prime }
$ (related to the gravitational constant in 10-d). Such 10-d solutions
preserve two real supercharges, which correspond to the $\mathcal{N}=1/2$
supersymmetry. The almost-K\"{a}hler internal 6-d structure can be defined
for various classes of solutions in 10-d gravity if we prescribe an
effective Lagrange type generating function. In this approach we can work
both with real nonholonomic gravitational and YM instanton configurations to
consider deformed $SU(3)$ structures.

The paper is organized as follows:\ We begin section \ref{s2} with a summary
on formulation of the heterotic supergravity theory in nonholonomic
variables which was performed in \cite{partner1}. This will allow us to
derive a general decoupling property of motion equations in further
sections. The geometric formalism on nonlinear and distinguished connections
and adapted metrics to nonholonomic 2+2+2+... splitting of higher dimension
spacetimes is outlined. We develop the AFDM as a geometric method for
constructing exact solutions of motion equations in heterotic string theory
and related 4d - 10d modified Einstein gravitational equations. The
contributions of gauge like NS 3-forms, curvature of interior almost-K\"{a}%
hler configurations, effective scalar and gauge fields etc. are encoded into
certain effective sources and generic off--diagonal terms of metrics, with
effective N--connection structure, defining a nontrivial vacuum structure.
Such generalized/modified gravitational equations are formulated in adapted
variables and with a very general off--diagonal ansatz for the metrics,
(non) linear connections and effective matter fields, which allows a
decoupling of corresponding nonlinear systems of PDEs in very general forms.
We show that using this nonholonomic decoupling property, it is possible to
construct classes of exact solutions depending on various sets of generating
and integration functions and integrations constants, on all 10-d spacetime
coordinates. The existence of a very important nonlinear symmetry is proven.
This allows a re-definition of the generating functions and effective
sources to other equivalent data with effective cosmological constants
induced by off--diagonal, or warped, and effective sources interactions. It
is shown how the geometric constructions can be performed for the "simplest"
case of one Killing symmetry in 4-d and then generalized to non-Killing
configurations and higher dimensions. The conditions of generating solutions
with zero torsion are analyzed. A self-consistent formalism of constructing
small N--adapted nonholonomic stationary deformations is also explored.

Section \ref{s3} is devoted to a rigorous geometric study of nonholonomic
generic off--diagonal deformations of exact solutions in heterotic string
gravity containing the Kerr solution as a "primary" configuration. We show
by using the AFDM it is possible to generate the Kerr solution as a
particular nonholonomically constrained case and small parametric
deformations with a well-defined physical interpretation. Then we construct
solutions with general off--diagonal deformations of the Kerr metrics in
4d--10d effective gravity with heterotic string modifications. We provide
examples of (non--Einstein) metrics with nonholonomically induced torsions
and study small off-diagonal modifications of the Kerr metrics determined by
warped and general almost-K\"{a}hler internal space structures. Separate
subsections are devoted to ellipsoidal 4--d, 6-d and 10-d deformations of
the Kerr metric resulting in target vacuum rotoid, or Kerr--de Sitter,
configurations, all self-consistently defining exact solutions of motion
equations in heterotic string theory.

Finally (in section \ref{s4}), we summarize the paper, provide conclusions
and speculate on physical meanings of solutions with generic off-diagonal
metrics and generalized connections constructed using the AFDM for the
heterotic string theory.


\section{Heterotic Supergravity in Nonholonomic Variables}

\label{s2}In this section, we outline a nonholonomic geometric approach to
the heterotic supergravity modelled in the low--energy limit of heterotic
string theory as a $\mathcal{N}=1$ and 10-d supergravity coupled to super
Yang--Mills theory, see details in \cite{partner1}. The nonholonomic
variables will be parameterized in forms which allow a general decoupling of
motion equations and generating exact solutions depending, in principle, on
all spacetime coordinates. Such a higher dimension spacetime is modelled as
a 10-d manifold $\mathcal{M}$, equipped with a Lorentzian metric $\check{g}$
of signature $(++-+++++++)$, with a time like third coordinate. The
heterotic supergravity theory is defined by a couple $(\mathcal{M},\check{g}%
),$ an NS 3-form $\check{H},$ a dilaton field $\check{\phi}$ and a gauge
connection $^{A}\check{\nabla},$ with gauge group $SO(32)$ or $E_{8}\times
E_{8}.$ In our approach, we elaborate a system of notations which is adapted
correspondingly for applications of geometric methods of constructing exact
solutions in \cite{vex3,veym,vtamsuper,svvvey,tgovsv}. The notations in \cite%
{partner1} and this work are different from the "standard" system of
notations in string theory (see, for instance, \cite{lecht1}). A so--called
nonholonomic "shell by shell", or 2+2+..., splitting should be elaborated
with corresponding left shell lables and shell indices. This minimizes the
procedure of separating the motion equations in certain general forms and
for constructing exact solutions. In "shell" diadic variables, certain
important (non) linear symmetries are explicitly shown and the type of
generating and integration functions which can be considered are emphasized.
Using only standard 4-d, 6-d and/or 10-d indices, like in former superstring
and supergravity works, it is not possible to understand how the AFDM can be
applied for generating off-diagonal solutions in (super) string/gravity
theories.

\subsection{Geometric conventions on nonholonomic 2+2+.... splitting}

For geometric spacetime models on a 10-d pseudo-Riemannian spacetime $\
\mathcal{M}$ with a time like coordinate $u^{3}=t$ and other coordinates
being space-like\footnote{%
this parameterization of coordinates is convenient for constructing various
classes of stationary solutions with warping on coordinate $u^{4}$ in a
"minimal" form}, we consider conventional splitting of dimensions $\dim
\mathcal{M}=4+2s=10;s=0,1,2,3.$ The AFDM, allows us to construct exact
solutions with arbitrary signatures of metrics $\check{g},$ but our goal is
to consider extra dimensional space generalizations of the Einstein theory
to heterotic supergravity models. In most general forms, this is possible if
we use the formalism of nonlinear connection splitting for higher
dimensional (super) spaces and strings, which was originally considered in
(super) Lagrange-Finsler theory \cite{vnpfins,vapfins}. We shall not
consider Finsler type gravity in this work, but follow an approach
re--defined for nonholonomic distributions on (super) manifolds \cite%
{tgovsv,vtamsuper,vex3}. The same geometric technique can be applied for
tangent (super) bundles or for (super) manifolds enabled with certain
classes of nonholonomic distributions like nonholonomic frames, nonlinear
connections etc. For vector/tangent (super) bundles, certain $x$%
--coordinates are used on the base (super) manifold, but $y$--coordinates
are considered for the typical fiber (super) vector space. In the case of
fibrations, the $(x,y)$--coordinates are used for the definition of a
fibered structure.

We consider (abstract, or coordinate) indices and coordinates $u^{\alpha
_{s}}=(x^{i_{s}},y^{a_{s}})$ for an oriented number of two dimensional, 2-d,
"shells" added to a 4--d spacetime of signature $(++-+)$. For $s=0,$ we
write $u^{\alpha }=(x^{i},y^{a})$ and then extend "shell by shell" to a
local system of 10-d coordinates,
\begin{eqnarray}
s &=&0:u^{\alpha _{0}}=(x^{i_{0}},y^{a_{0}})=(x^{i},y^{a});\ s=1:u^{\alpha
_{1}}=(x^{\alpha }=u^{\alpha },y^{a_{1}})=(x^{i},y^{a},y^{a_{1}});
\label{coordconv} \\
s &=&2:u^{\alpha _{2}}=(x^{\alpha _{1}}=u^{\alpha
_{1}},y^{a_{2}})=(x^{i},y^{a},y^{a_{1}},y^{a_{2}});\ s=3:u^{\alpha
_{3}}=(x^{\alpha _{2}}=u^{\alpha
_{2}},y^{a_{3}})=(x^{i},y^{a},y^{a_{1}},y^{a_{2}},y^{a_{3}}),  \notag
\end{eqnarray}%
The corresponding subsets of indices are labeled in the form: $%
i=i_{0},j=j_{0},...=1,2;a=a_{0},b=b_{0},...=3,4,$ when $u^{3}=y^{3}=t$; $%
a_{1},b_{1}...=5,6;a_{2},b_{2}...=7,8;$ $a_{3},b_{3}...=9,10;$ and, for
instance, $i_{1},j_{1},...=1,2,3,4;i_{2},j_{2},...=1,2,3,4,5,6;$\ $%
i_{3},j_{3},...=1,2,3,4,5,6,7,8,$ or we shall write only $i_{s}.$ For brief
denotations, we shall write $\ ^{0}u=(\ ^{0}x,\ ^{0}y);\ ^{1}u=(\ ^{0}u,\
^{1}y)=(\ ^{0}x,\ ^{0}y,\ ^{1}y),\ ^{2}u=(\ ^{1}u,\ ^{2}y)=(\ ^{0}x,\
^{0}y,\ ^{1}y,\ ^{2}y)$ and $\ ^{3}u=(\ ^{2}u,\ ^{3}y)=(\ ^{0}x,\ ^{0}y,\
^{1}y,\ ^{2}y,\ ^{3}y).$ Here we note in modern gravity, the so--called ADM
(Arnowit--Deser--Misner) formalism is largely used with 3+1 splitting, or
any $n+1$ splitting, see details in \cite{misner}. For the purposes of this
work, such splittings are not convenient because it is not possible to
elaborate a technique of general decoupling of the gravitational field
equations and generating off--diagonal solutions. In these cases, the
conventional one dimensional "fibers" result in certain degenerate systems
of equations. To construct exact solutions in 4-10 dimensional theories, it
is more convenient to work with a corresondingly defined non--integrable
2+2+... splitting, see details in \cite{tgovsv,vex2,vex3}. For the heterotic
supergravity, such geometric constructions are explored in greater detail in
\cite{partner1}. In order to connect "shell by shell" indices and
coordinates to standard ones for supergravity theories (see \cite%
{lecht1,harl}), we can consider small Greek \ indices without sub indices,
and respective coordinates $x^{\mu },$ when $\alpha ,\mu ,...=0,1,...,9$.
The identification of shell coordinates with the standard ones follows such
a rule: $x^{0}$ $=u^{3}=t$ (timelike coordinate) and (for spacelike
coordinates): $%
x^{1}=u^{1},x^{2}=u^{2},x^{3}=u^{4},x^{4}=u^{5},x^{5}=u^{6},x^{6}=u^{7},x^{7}=u^{8},x^{8}=u^{9},x^{9}=u^{10}
$.

On $\mathcal{M},$ we can consider local frames/ bases, $e_{\alpha _{s}}=e_{\
\alpha _{s}}^{\underline{\alpha }_{s}}(\ ^{s}u)\partial /\partial u^{%
\underline{\alpha }_{s}},$ where the partial derivatives $\partial _{\beta
_{s}}:=\partial /\partial u^{\beta _{s}}$ define local coordinate bases. We
shall underline indices if it is necessary to emphasize that such values are
defined with respect to a coordinate frame. The (co) frames, $e^{\alpha
_{s}}=e_{\ \underline{\alpha }_{s}}^{\ \alpha _{s}}(\ ^{s}u)du^{\underline{%
\alpha }_{s}},$ can be defined as dual to respective $e_{\alpha _{s}}.$

For our purposes, it is convenient to work with nonholonomic
(non--integrable) distributions defining a $2+2+...$ spacetime splitting.
Such a distribution can be introduced to define a nonlinear connection,
N--connection, structure via a Whitney sum
\begin{equation}
\ ^{s}\mathbf{N}:T\mathcal{M}=\ ^{0}h\mathcal{M}\oplus \ ^{0}v\mathcal{M}%
\oplus \ ^{1}v\mathcal{M}\oplus \ ^{2}v\mathcal{M}\oplus \ ^{3}v\mathcal{M}.
\label{whitney}
\end{equation}%
This formula states a conventional horizontal (h) and vertical (v) "shell by
shell" splitting. We shall use boldface letters in order to define spaces
and geometric objects enabled/adapted to a N--connection structure. In local
form, an N--connection is defined by its coefficients $N_{i_{s}}^{a_{s}}$
when $\ ^{s}\mathbf{N}=N_{i_{s}}^{a_{s}}(\ ^{s}u)dx^{i_{s}}\otimes \partial
/\partial y^{a_{s}}.$

A nonholonomic manifold is enabled with a nonholonomic distribution of type (%
\ref{whitney}), when (for instance, for the "zero" shell) $\mathbf{V\simeq }h%
\mathcal{M}\oplus v\mathcal{M}$. The term N--anholonomic manifold is also
used. This definition comes form the fact that the N--connection
coefficients determine a system of N--adapted local bases, with N-elongated
partial derivatives, $\mathbf{e}_{\nu _{s}}=(\mathbf{e}_{i_{s}},e_{a_{s}}),$
and cobases with N--adapted differentials, $\mathbf{e}^{\mu _{s}}=(e^{i_{s}},%
\mathbf{e}^{a_{s}}),$ For $s=0,$
\begin{eqnarray}
&&\mathbf{e}_{i_{0}}=\mathbf{e}_{i}=\frac{\partial }{\partial x^{i_{0}}}-\
N_{i_{0}}^{a_{0}}\frac{\partial }{\partial y^{a_{0}}},\ e_{a_{0}}=\frac{%
\partial }{\partial y^{a_{0}}},\ e^{i_{0}}=dx^{i},\mathbf{e}^{a}=\mathbf{e}%
^{a_{0}}=dy^{a}+\ N_{i}^{a}dx^{i};\mbox{ or/ and }  \label{nadaptb} \\
&&\mathbf{e}_{i_{s}}=\frac{\partial }{\partial x^{i_{s}}}-\ N_{i_{s}}^{a_{s}}%
\frac{\partial }{\partial y^{a_{s}}},\ e_{a_{s}}=\frac{\partial }{\partial
y^{a_{s}}},e^{i_{s}}=dx^{i_{s}},\mathbf{e}^{a_{s}}=dy^{a_{s}}+\
N_{i_{s}}^{a_{s}}dx^{i_{s}}\mbox{ for }s=1,2,3.  \notag
\end{eqnarray}%
The anholonomy relations
\begin{equation}
\lbrack \mathbf{e}_{\alpha _{s}},\mathbf{e}_{\beta _{s}}]=\mathbf{e}_{\alpha
_{s}}\mathbf{e}_{\beta _{s}}-\mathbf{e}_{\beta _{s}}\mathbf{e}_{\alpha
_{s}}=W_{\alpha _{s}\beta _{s}}^{\gamma _{s}}\mathbf{e}_{\gamma _{s}}
\label{anhrel}
\end{equation}%
are computed $W_{i_{s}a_{s}}^{b_{s}}=\partial _{a_{s}}N_{i_{s}}^{b_{s}}$ and
$W_{j_{s}i_{s}}^{a_{s}}=\Omega _{i_{s}j_{s}}^{a_{s}},$ where the curvature
of N--connection is defined as the Neijenhuis tensor, $\Omega
_{i_{s}j_{s}}^{a_{s}}:=\mathbf{e}_{j_{s}}\left( N_{i_{s}}^{a_{s}}\right) -%
\mathbf{e}_{i_{s}}\left( N_{j_{s}}^{a_{s}}\right) .$

Any metric structure on $\mathcal{M}$ can be parametrized via nonholonomic
frame transforms as a distinguished metric (d--metric, in boldface form), $\
^{s}\mathbf{g}=\{\mathbf{g}_{\alpha _{s}\beta _{s}}\},$
\begin{eqnarray}
\ \ ^{s}\mathbf{g} &=&\ g_{i_{s}j_{s}}(\ ^{s}u)\ e^{i_{s}}\otimes
e^{j_{s}}+\ g_{a_{s}b_{s}}(\ ^{s}u)\mathbf{e}^{a_{s}}\otimes \mathbf{e}%
^{b_{s}}  \label{dm} \\
&=&g_{ij}(x)\ e^{i}\otimes e^{j}+g_{ab}(u)\ \mathbf{e}^{a}\otimes \mathbf{e}%
^{b}+g_{a_{1}b_{1}}(\ ^{1}u)\ \mathbf{e}^{a_{1}}\otimes \mathbf{e}%
^{b_{1}}+....+\ g_{a_{s}b_{s}}(\ ^{s}u)\mathbf{e}^{a_{s}}\otimes \mathbf{e}%
^{b_{s}}.  \notag
\end{eqnarray}%
Introducing dual coframes $\mathbf{e}^{\mu _{s}}=(e^{i_{s}},\mathbf{e}%
^{a_{s}})$ decomposed with respect to N--elongated differentials $du^{a_{s}}$
as in (\ref{nadaptb}), we obtain a generic off--diagonal form $\ \ ^{s}%
\mathbf{g}$ with respect to a dual coordinate frame. A metric $\ ^{s}\mathbf{%
g}$ can be diagonalized by coordinate transforms in a finite region if and
only if the anholonomy coefficients in (\ref{anhrel}) vanish, i.e. $%
W_{\alpha _{s}\beta _{s}}^{\gamma _{s}}=0$

A linear connection $\ ^{s}D$ can be introduced on $\mathcal{M}$ in standard
form. A distinguished connection, d--connection, $\ ^{s}\mathbf{D}$ is a
liniear connection preserving the N--connection spitting under parallelism.
One defines the curvature $\mathcal{R}_{~\beta _{s}}^{\alpha _{s}}$ and
torsion $\mathcal{T}^{\alpha _{s}}$ in standard forms, see details in \cite%
{partner1}. In this paper, we shall work with two very important linear
connection structures determined by the same metric structure. These linear
connections are uniquely defined following the geometric conditions: {\small
\begin{equation}
\ ^{s}\mathbf{g}\rightarrow \left\{
\begin{array}{ccccc}
\ ^{s}\nabla : &  & \ ^{s}\nabla \ (\ ^{s}\mathbf{g})\ =0;\ \ _{\nabla }^{s}%
\mathcal{T}=0, &  & \mbox{ the Levi--Civita connection;} \\
\ ^{s}\widehat{\mathbf{D}}: &  & \ ^{s}\widehat{\mathbf{D}}\ (\ ^{s}\mathbf{%
g)}=0;\ h\widehat{\mathcal{T}}=0,\ ^{1}v\widehat{\mathcal{T}}=0,\ \ ^{2}v%
\widehat{\mathcal{T}}=0,\ ^{3}v\widehat{\mathcal{T}}=0. &  &
\mbox{ the
canonical d--connection.}%
\end{array}%
\right.  \label{lcconcdcon}
\end{equation}%
} Here we note that the LC--connection $\ ^{s}\nabla =\{\ _{\shortmid
}\Gamma _{\beta _{s}\gamma _{s}}^{\alpha _{s}}\}$ can be introduced without
any N--connection structure. It can always be canonically distorted to a
necessary type of d--connection $\ ^{s}\mathbf{D}$ completely defined by $\
^{s}\mathbf{g}$ following certain geometric principles. The canonical
d--connection $\ ^{s}\widehat{\mathbf{D}}$ is characterized by a
nonholonomically induced torsion\ d--tensor which is completely defined by $%
\ ^{s}\mathbf{g}$ for any chosen $\ ^{s}\mathbf{N=\{}N_{i_{s}}^{a_{s}}\}.$
The N--adapted coefficients are parameterized "shell by shell" by formulas
{\small
\begin{equation}
\ ^{s}\widehat{\mathcal{T}}=\{\widehat{\mathbf{T}}_{\ \alpha _{s}\beta
_{s}}^{\gamma _{s}}:\widehat{T}_{\ j_{s}k_{s}}^{i_{s}}=\widehat{L}%
_{j_{s}k_{s}}^{i_{s}}-\widehat{L}_{k_{s}j_{s}}^{i_{s}},\widehat{T}_{\
j_{s}a_{s}}^{i_{s}}=\widehat{C}_{j_{s}b_{s}}^{i_{s}},\widehat{T}_{\
j_{s}i_{s}}^{a_{s}}=-\Omega _{\ j_{s}i_{s}}^{a_{s}},\ \widehat{T}%
_{a_{s}j_{s}}^{c_{s}}=\widehat{L}%
_{a_{s}j_{s}}^{c_{s}}-e_{a_{s}}(N_{j_{s}}^{c_{s}}),\widehat{T}_{\
b_{s}c_{s}}^{a_{s}}=\ \widehat{C}_{b_{s}c_{s}}^{a_{s}}-\ \widehat{C}%
_{c_{s}b_{s}}^{a_{s}}\}.  \label{dtorss}
\end{equation}
} It should be noted that this torsion is different form torsions in
Einstein-Cartan gauge type and string gravity theories with absolutely
antisymmetric torsion. Additional sources are not necessary because a
d--torsion (\ref{dtorss}) is determined by the nonholonomic structures. In
generalized theories, the torsion fields which are independent from the
metric and vielbein fields may posses proper sources. Considering additional
assumptions, we can relate the values, (\ref{dtorss}) for instance, to a
subclass of nontrivial coefficients of an absolute antisymmetric torsion in
string gravity. We can always extract LC--configurations with zero torsion
if we additionally impose for (\ref{dtorss}) the conditions
\begin{equation}
\widehat{\mathbf{T}}_{\ \alpha _{s}\beta _{s}}^{\gamma _{s}}=0,\mbox{ i.e. }
\ ^{s}\widehat{\mathbf{D}}_{\mid \widehat{\mathcal{T}}=0}\rightarrow \
^{s}\nabla .  \label{zerotors}
\end{equation}
Such additional nonholonomic constraints may be stated in non--explicit
forms and without certain limits with small parameters and smooth functions.
It should be noted that, in general, $W_{\alpha _{s}\beta _{s}}^{\gamma
_{s}} $ (\ref{anhrel}) may not be zero even if the conditions (\ref{zerotors}%
) are satisfied.

Any (pseudo) Riemannian geometry can be equivalently formulated in
nonholonomic variables $(\ ^{s}\mathbf{g,}\ ^{s}\mathbf{N,}\ ^{s}\widehat{%
\mathbf{D}})$ or using the standard data $(\ ^{s}\mathbf{g,}\ ^{s}\nabla ).$
Because both linear connections $\ ^{s}\nabla $ and $\ ^{s}\widehat{\mathbf{D%
}}$ are defined by the same metric structure, there is a canonical
distortion relation
\begin{equation}
\ ^{s}\widehat{\mathbf{D}}=\ ^{s}\nabla +\ ^{s}\widehat{\mathbf{Z}}.
\label{distorsrel}
\end{equation}

The N--adapted coefficients of a curvature d--tensor $\mathcal{R}_{~\beta
_{s}}^{\alpha _{s}}=\{\mathbf{\mathbf{R}}_{\ \ \beta _{s}\gamma _{s}\delta
_{s}}^{\alpha _{s}}\}$ of $\ $the canonical d--connection $\ ^{s}\widehat{%
\mathbf{D}}$ and $\ ^{s}\mathbf{g}$ can be computed respectively for all
shells $s=0,1,2,3,$ see details in \cite{partner1}. \ The Ricci d--tensor $%
\widehat{R}ic=\{\widehat{\mathbf{R}}_{\alpha _{s}\beta _{s}}:=\widehat{%
\mathbf{R}}_{\ \alpha _{s}\beta _{s}\tau _{s}}^{\tau _{s}}\}$ is a
respective contracting of the coefficients of curvature tensor,
\begin{equation}
\widehat{\mathbf{R}}_{\alpha _{s}\beta _{s}}=\{\widehat{R}_{h_{s}j_{s}}:=%
\widehat{R}_{\ h_{s}j_{s}i_{s}}^{i_{s}},\ \ \widehat{R}_{j_{s}a_{s}}:=-%
\widehat{P}_{\ j_{s}i_{s}a_{s}}^{i_{s}},\ \widehat{R}_{b_{s}k_{s}}:=\widehat{%
P}_{\ b_{s}k_{s}a_{s}}^{a_{s}},\widehat{R}_{\ b_{s}c_{s}}=\widehat{S}_{\
b_{s}c_{s}a_{s}}^{a_{s}}\}.  \label{dricci}
\end{equation}%
Consindering the inverse d--metric to$\ ^{s}\mathbf{g,}$ we define and
compute the scalar curvature of $\ ^{s}\widehat{\mathbf{D}}\mathbf{,}$
\begin{eqnarray}
\ ^{s}\widehat{R}:= &&\mathbf{g}^{\alpha _{s}\beta _{s}}\widehat{\mathbf{R}}%
_{\alpha _{s}\beta _{s}}=g^{i_{s}j_{s}}\widehat{R}%
_{i_{s}j_{s}}+g^{a_{s}b_{s}}\widehat{R}_{a_{s}b_{s}}  \notag \\
&=&\widehat{R}+\widehat{S}+\ ^{1}\widehat{S}+\ ^{2}\widehat{S}+\ ^{3}%
\widehat{S},  \label{rdsc}
\end{eqnarray}%
with respective h-- and v--components of scalar curvature, $\widehat{R}%
=g^{ij}\widehat{R}_{ij},$ $S=g^{ab}S_{ab},$ $\
^{1}S=g^{a_{1}b_{1}}S_{a_{1}b_{1}},\ ^{2}S=g^{a_{2}b_{2}}S_{a_{2}b_{2}},\
^{3}S=g^{a_{3}b_{3}}R_{a_{3}b_{3}}.$

\subsection{The AFDM for heterotic supergravity}

We develop the "anholonomic frame deformation method", AFDM,\ and apply
these geometric techniques for applications to 10--d gravity and heterotic
supergravity formulated in noholonomic variables.

The heterotic supergravity field equations were formulated in N--adapted
form in \cite{partner1}. They include terms of order $\alpha ^{\prime },$
equivalent to the equations of motion of heterotic nonholonomic supergravity
considered in \cite{lecht1}. In explicit form,%
\begin{eqnarray}
\widehat{\mathbf{R}}_{\mu _{s}\nu _{s}}+2(\ ^{s}\widehat{\mathbf{D}}\widehat{%
\mathbf{d}}\widehat{\phi })_{\mu _{s}\nu _{s}}-\frac{1}{4}\widehat{\mathbf{H}%
}_{\alpha _{s}\beta _{s}\mu _{s}}\widehat{\mathbf{H}}_{\nu _{s}}^{\quad
\alpha _{s}\beta _{s}}+\frac{\alpha ^{\prime }}{4}\left[ \widetilde{\mathbf{R%
}}_{\mu _{s}\alpha _{s}\beta _{s}\gamma _{s}}\widetilde{\mathbf{R}}_{\nu
_{s}}^{\quad \alpha _{s}\beta _{s}\gamma _{s}}-tr\left( \widehat{\mathbf{F}}%
_{\mu _{s}\alpha _{s}}\widehat{\mathbf{F}}_{\nu _{s}}^{\quad \alpha
_{s}}\right) \right] &=&0,  \label{hs1} \\
\ ^{s}\widehat{R}+4\widehat{\square }\widehat{\phi }-4|\widehat{\mathbf{d}}%
\widehat{\phi }|^{2}-\frac{1}{2}|\widehat{\mathbf{H}}|^{2}+\frac{\alpha
^{\prime }}{4}tr\left[ |\widetilde{\mathbf{R}}|^{2}-|\widehat{\mathbf{F}}|%
\right] &=&0,  \label{hs2} \\
e^{2\widehat{\phi }}\widehat{\mathbf{d}}\widehat{\ast }(e^{-2\widehat{\phi }}%
\widehat{\mathbf{F}})+\widehat{\mathbf{A}}\wedge \widehat{\ast }\widehat{%
\mathbf{F}}-\widehat{\ast }\widehat{\mathbf{F}}\wedge \widehat{\mathbf{A}}+%
\widehat{\ast }\widehat{\mathbf{H}}\wedge \widehat{\mathbf{F}} &=&0,
\label{hs3} \\
\widehat{\mathbf{d}}\widehat{\ast }(e^{-2\widehat{\phi }}\widehat{\mathbf{H}}%
) &=&0,  \label{hs4}
\end{eqnarray}%
where the Hodge operator $\widehat{\ast },$ $\ ^{s}\widehat{\mathbf{D}}=\{%
\widehat{\mathbf{D}}_{\mu _{s}}\},$ the canonical nonholonomic d 'Alambert
wave operator $\widehat{\square }:=\widehat{\mathbf{g}}^{\mu _{s}\nu _{s}}%
\widehat{\mathbf{D}}_{\mu _{s}}\widehat{\mathbf{D}}_{\nu _{s}},$ the Ricci
d--tensor $\widehat{\mathbf{R}}_{\mu _{s}\nu _{s}}$ and scalar$\ ^{s}%
\widehat{R}$ are determined by a d--metric $\widehat{\mathbf{g}}$ (\ref{dm}%
). \ The curvature d--tensor $\widetilde{\mathbf{R}}_{\mu _{s}\alpha
_{s}\beta _{s}\gamma _{s}}$ is taken for an almost-K\"{a}hler structure on
shells $s=1,2,3$ as we described above. The gauge field $\widehat{\mathbf{A}}
$ corresponds to the N--adapted operator $\ _{A}^{s}\widehat{\mathbf{D}}\ $%
and curvature $\widehat{\mathbf{F}}=\mathcal{F}(\ ^{1}\psi )$ via a map
constructed in \cite{lecht1,partner1}). \

The equations (\ref{hs1}) \ can be written as effective Einstein equations
for the canonical d--connection $\ ^{s}\widehat{\mathbf{D}},$%
\begin{eqnarray}
&&\ ^{s}\widehat{\mathbf{R}}_{\ \beta _{s}\delta _{s}}=\mathbf{\Upsilon }%
_{\beta _{s}\delta _{s}},  \label{cdeinst} \\
&&\widehat{L}_{a_{s}j_{s}}^{c_{s}}=e_{a_{s}}(N_{j_{s}}^{c_{s}}),\ \widehat{C}%
_{j_{s}b_{s}}^{i_{s}}=0,\ \Omega _{\ j_{s}i_{s}}^{a_{s}}=0.  \label{lcconstr}
\end{eqnarray}%
The sources $\mathbf{\Upsilon }_{\beta _{s}\delta _{s}}$ can be formally
defined as in GR but for extra dimensions and \ in N-adapted form, when
\begin{equation*}
\mathbf{\Upsilon }_{\beta _{s}\delta _{s}}\rightarrow \varkappa (T_{\beta
_{s}\delta _{s}}\frac{1}{2}\mathbf{g}_{\beta _{s}\delta _{s}}\ ^{s}T)%
\mbox{
for }\ ^{s}\widehat{\mathbf{D}}\rightarrow \ ^{s}\nabla .
\end{equation*}%
The system (\ref{cdeinst}) can be derived and formulated in variational
N-adapted form by considering a nonholonomic gravitational Lagrange density
of type $\ ^{g}\widehat{L}=\widehat{\mathbf{R}}$ and an effective Lagrange
density for matter$~\ ^{m}\widehat{L}.$ For simplicity, we shall consider
Lagrangians depending only on the coefficients of metric field and matter
field but not on their derivatives (for such configurations, it will be
possible to construct exact solutions in explicit form). The
energy--momentum d--tensor is computed by definition
\begin{equation}
\ ^{m}\widehat{\mathbf{T}}_{\alpha \beta }:=-\frac{2}{\sqrt{|\ ^{s}\mathbf{g}%
|}}\frac{\delta (\sqrt{|\ ^{s}\mathbf{g}|}\ \ ^{m}\widehat{L})}{\delta \ ^{s}%
\mathbf{g}^{\alpha \beta }}=\ ^{m}\widehat{L}\ ^{s}\mathbf{g}^{\alpha \beta
}+2\frac{\delta (\ ^{m}\widehat{L})}{\delta \ ^{s}\mathbf{g}_{\alpha \beta }}%
,  \label{ematter}
\end{equation}%
for $|\ ^{s}\mathbf{g}|=\det |\ ^{s}\mathbf{g}_{\mu \nu }|.$ \ We conclude
that by following an N--adapted variational calculus with action
\begin{equation*}
\ ^{g}\mathcal{S+\ }^{m}\mathcal{S}=\int d^{4}u\sqrt{|\ ^{s}\mathbf{g}|}(\
^{g}\widehat{L}+~^{m}\widehat{L}),
\end{equation*}%
we can elaborate a 10-d nonholonomic gravity theory with gravitational field
equations (\ref{cdeinst}). A changing of geometric data $(\ ^{s}\mathbf{g},\
^{s}\widehat{\mathbf{D}})\rightarrow (\ ^{s}\mathbf{g,\ ^{s}\nabla )}$ is
possible via canonical distorting relations (\ref{distorsrel}), or imposing
the zero torsion condition at the end $\widehat{\mathcal{T}}=0$ (\ref%
{zerotors}) for extracting LC--configurations $\ ^{s}\widehat{\mathbf{D}}%
_{\mid \widehat{\mathcal{T}}=0}=\ ^{s}\nabla $ stated by (\ref{lcconstr}).
The matter Lagrange density $\ ^{m}\widehat{L}$ can be chosen in such a form
that, via corresponding frame transforms, the source $\mathbf{\Upsilon }%
_{\beta _{s}\delta _{s}}$ will encode the terms with effective matter fields
$\widehat{\phi },\widehat{\mathbf{H}}_{\alpha _{s}\beta _{s}\mu _{s}},$
interior space fields $\widetilde{\mathbf{R}}_{\mu _{s}\alpha _{s}\beta
_{s}\gamma _{s}}$ and gauge fields $\widehat{\mathbf{F}}_{\mu _{s}\alpha
_{s}}.$

The nonholonomic motion equations in heterotic string gravity (\ref{hs1})
can be written in the form (\ref{cdeinst}) with an effective source (related
via nonlinear transforms of generating functions to certain effective
cosmological constants)
\begin{eqnarray}
\mathbf{\Upsilon }_{\mu _{s}\nu _{s}} &=&\ ^{\phi }\mathbf{\Upsilon }_{\mu
_{s}\nu _{s}}+\ ^{H}\mathbf{\Upsilon }_{\mu _{s}\nu _{s}}+\ ^{F}\mathbf{%
\Upsilon }_{\mu _{s}\nu _{s}}+\ ^{R}\mathbf{\Upsilon }_{\mu _{s}\nu _{s}},%
\mbox{ where }  \label{es} \\
\ ^{\phi }\mathbf{\Upsilon }_{\mu _{s}\nu _{s}} &=&-2(\ ^{s}\widehat{\mathbf{%
D}}\widehat{\mathbf{d}}\widehat{\phi })_{\mu _{s}\nu _{s}}%
\mbox{ with
effective constant }\ ^{\phi }\Lambda ;  \label{es1} \\
\ ^{H}\mathbf{\Upsilon }_{\mu _{s}\nu _{s}} &=&\frac{1}{4}\widehat{\mathbf{H}%
}_{\alpha _{s}\beta _{s}\mu _{s}}\widehat{\mathbf{H}}_{\nu _{s}}^{\quad
\alpha _{s}\beta _{s}}\mbox{ with effective constant }\ ^{H}\Lambda ;
\label{es2} \\
\ ^{F}\mathbf{\Upsilon }_{\mu _{s}\nu _{s}} &=&\frac{\alpha ^{\prime }}{4}%
tr\left( \widehat{\mathbf{F}}_{\mu _{s}\alpha _{s}}\widehat{\mathbf{F}}_{\nu
_{s}}^{\quad \alpha _{s}}\right) \mbox{ with effective constant }\
^{F}\Lambda ;  \label{es4} \\
\ ^{R}\mathbf{\Upsilon }_{\mu _{s}\nu _{s}} &=&-\frac{\alpha ^{\prime }}{4}%
\widetilde{\mathbf{R}}_{\mu _{s}\alpha _{s}\beta _{s}\gamma _{s}}\widetilde{%
\mathbf{R}}_{\nu _{s}}^{\quad \alpha _{s}\beta _{s}\gamma _{s}}%
\mbox{ with
effective constant }\ ^{\widetilde{R}}\Lambda .  \label{es5}
\end{eqnarray}%
The traces of above sources, and respective effective cosmological
constants, are related via condition (\ref{hs2}).

\subsection{Decoupling \& integration of nonholonomic motion equations}

We show how the heterotic string gravitational field equations (\ref{hs1})
written in the form (\ref{cdeinst}) with sources (\ref{es}) and possible
constraints (\ref{lcconstr}), can be formally integrated in very general
forms for generic off--diagonal metrics with coefficients depending on all
spacetime coordinates. Additional conditions for extracting
LC--confiugrations with $\ ^{s}\nabla $ will be analysed at the end after
certain classes of general solutions have been constructed.

\subsubsection{ Ansatz for metrics, N--connections, and gravitational
polarizations}

In the simplest form, the decoupling property for any shell $s=0,1,2,3$ can
be proven for certain ansatz with at least one Killing symmetry on the
corresponding shell. Using an additional conformal transform on necessary
shells, we can extend the constructions for non--Killing configurations.
Such a procedure is described for arbitrary finite shell $s$ in Refs. \cite%
{tgovsv,vex3,veym,svvvey}. In this work, we state the formulas for 10-d
gravity and heterotic string theory in explicit form when the prime
solutions are stationary ones, i.e. do not depend on the time-like
coordinate $y^{3}=t.$

Let us consider metrics of type (\ref{dm}), which via frame transforms $%
\mathbf{g}_{\alpha _{s}\beta _{s}}=e_{\ \alpha _{s}}^{\alpha _{s}^{\prime
}}e_{\ \beta _{s}}^{\beta _{s}^{\prime }}\mathbf{g}_{\alpha _{s}^{\prime
}\beta _{s}^{\prime }}$ can be parametrized\footnote{%
we shall put a left lable K in order to emphasize that this is a d--metric
with Killing symmetry}
\begin{eqnarray}
\ \ _{K}^{s}\mathbf{g} &=&\ g_{i}(x^{k})dx^{i}\otimes
dx^{i}+h_{a}(x^{k},y^{4})\mathbf{e}^{a}\otimes \mathbf{e}^{b}+  \label{ansk}
\\
&&h_{a_{1}}(u^{\alpha },y^{6})\ \mathbf{e}^{a_{1}}\otimes \mathbf{e}%
^{a_{1}}+h_{a_{2}}(u^{\alpha _{1}},y^{8})\ \mathbf{e}^{a_{2}}\otimes \mathbf{%
e}^{b_{2}}+h_{a_{3}}(\ u^{\alpha _{2}},y^{10})\mathbf{e}^{a_{3}}\otimes
\mathbf{e}^{a_{3}},  \notag
\end{eqnarray}
\begin{eqnarray*}
\mbox{ where }\mathbf{e}^{a} &=&dy^{a}+N_{i}^{a}dx^{i},\mbox{\ for \ }%
N_{i}^{3}=n_{i}(x^{k},y^{4}),N_{i}^{4}=w_{i}(x^{k},y^{4}); \\
\mathbf{e}^{a_{1}} &=&dy^{a_{1}}+N_{\alpha }^{a_{1}}du^{\alpha },%
\mbox{\ for
\ }N_{\alpha }^{5}=\ ^{1}n_{\alpha }(u^{\beta },y^{6}),N_{\alpha }^{6}=\
^{1}w_{\alpha }(u^{\beta },y^{6}); \\
\mathbf{e}^{a_{2}} &=&dy^{a_{2}}+N_{\alpha _{1}}^{a_{2}}du^{\alpha _{1}},%
\mbox{\ for \ }N_{\alpha _{1}}^{7}=\ ^{2}n_{\alpha _{1}}(u^{\beta
_{1}},y^{8}),N_{\alpha _{1}}^{8}=\ ^{2}w_{\alpha _{1}}(u^{\beta _{1}},y^{8});
\\
\mathbf{e}^{a_{3}} &=&dy^{a_{3}}+N_{\alpha _{2}}^{a_{3}}du^{\alpha _{2}},%
\mbox{\ for \ }N_{\alpha _{2}}^{9}=\ ^{3}n_{\alpha _{2}}(u^{\beta
_{2}},y^{10}),N_{\alpha _{2}}^{10}=\ ^{3}w_{\alpha _{3}}(u^{\beta
_{2}},y^{10}).
\end{eqnarray*}%
Such an ansatz has a Killing vector $\partial /\partial y^{9}$ because the
coordinate $y^{9}$ is not contained in the coefficients of such metrics. We
propose that via coordinate transforms we can eliminate dependence on $%
y^{3}=t$ and can parametrize $h_{4}=1$ and $h_{a_{s}}=\ _{\shortmid }h(y^{4})%
\check{h}_{a_{s}}(x^{i},y^{b_{s}})$ for shells $s=1,2,3$, as it was
considered in \cite{partner1}, if the configurations are with warping on $%
y^{4}.$ The ansatz (\ref{ansk}) can be considered as a target d--metric of a
prime d--metric with flat domain wall considered in that associated paper.
In this work, we restrict our considerations to stationary solutions in
heterotic string gravity (which do not depend on $t$). Inhomogeneous and
locally anisotropic cosmological configurations in such string models, with
generic dependence on $t$ (see examples in \cite%
{vcosmsol1,vcosmsol2,vcosmsol3,vcosmsol4,vcosmsol5}) will be studied in our
future publications.

\subsubsection{ Ricci d--tensors and N--adapted sources}

We suppose that via frame transforms it is always possible to introduce
frame and coordinate parametrizations for ansatz (\ref{ansk}) with\footnote{%
we can construct special classes of solutions if such conditions are not
satisfied} $\partial _{4}h_{3}=h_{3}^{\ast }\neq 0,\partial
_{6}h_{5}=h_{5}^{\ast _{1}}\neq 0,\partial _{8}h_{7}=h_{7}^{\ast _{2}}\neq
0,\partial _{10}h_{9}=h_{9}^{\ast _{3}}\neq 0.$ In brief, the partial
derivatives are denoted, for instance, $\partial _{1}h=\partial h/\partial
x^{1}=h^{\cdot },$ $\partial _{2}h=\partial h/\partial x^{2}=h^{\prime
},\partial _{3}h=0,$ and $\partial _{44}h=\partial ^{2}h/\partial
y^{4}\partial y^{4}=h^{\ast \ast },\partial _{66}h=\partial ^{2}h/\partial
y^{6}\partial y^{6}=h^{\ast _{1}\ast _{1}},$ etc. We shall write explicitly $%
\partial _{5}h=\partial h/\partial y^{5},$ $\partial _{6}h=\partial
h/\partial x^{6},...$ without introducing "dot " and "prime" symbols for
partial derivatives on shells $s=1,2,3$ but working with "star" partial
derivatives on these shells, considering respectively $\ast _{1},\ast
_{2},\ast _{3}$ if necessarily written as $\ast _{s}.$ A tedious computation
of the coefficients of the canonical d--connection $\widehat{\mathbf{D}}=\{$
$\widehat{\mathbf{\Gamma }}_{\ \alpha _{s}\beta _{s}}^{\gamma _{s}}\}$ for
the ansatz (\ref{ansk}) and then of corresponding non-trivial coefficients
of the Ricci d--tensor $\mathbf{\hat{R}}_{\alpha _{s}\beta _{s}}$ (\ref%
{dricci}), see similar details in \cite%
{tgovsv,vex3,veym,svvvey,vex1,vpars,vex2}, results in nontrivial N-adapted
coefficients:
\begin{eqnarray}
\widehat{R}_{1}^{1} &=&\widehat{R}_{2}^{2}=-\frac{1}{2g_{1}g_{2}}%
[g_{2}^{\cdot \cdot }-\frac{g_{1}^{\cdot }g_{2}^{\cdot }}{2g_{1}}-\frac{%
(g_{2}^{\cdot })^{2}}{2g_{2}}+g_{1}^{\prime \prime }-\frac{g_{1}^{\prime
}g_{2}^{\prime }}{2g_{2}}-\frac{\left( g_{1}^{\prime }\right) ^{2}}{2g_{1}}],
\label{equ1} \\
\widehat{R}_{3}^{3} &=&\widehat{R}_{4}^{4}=-\frac{1}{2h_{3}h_{4}}%
[h_{3}^{\ast \ast }-\frac{\left( h_{3}^{\ast }\right) ^{2}}{2h_{3}}-\frac{%
h_{3}^{\ast }h_{4}^{\ast }}{2h_{4}}],  \label{equ2} \\
\widehat{R}_{3k} &=&\frac{h_{3}}{2h_{4}}n_{k}^{\ast \ast }+\left( \frac{h_{3}%
}{h_{4}}h_{4}^{\ast }-\frac{3}{2}h_{3}^{\ast }\right) \frac{n_{k}^{\ast }}{%
2h_{4}},  \label{equ3} \\
\widehat{R}_{4k} &=&\frac{w_{k}}{2h_{3}}[h_{3}^{\ast }-\frac{\left(
h_{3}^{\ast }\right) ^{2}}{2h_{3}}-\frac{(h_{3}^{\ast })(h_{4}^{\ast })}{%
2h_{4}}]+\frac{h_{3}^{\ast }}{4h_{3}}(\frac{\partial _{k}h_{3}}{h_{3}}+\frac{%
\partial _{k}h_{4}}{h_{4}})-\frac{\partial _{k}(h_{3}^{\ast })}{2h_{3}};
\label{equ4}
\end{eqnarray}%
on shell $s=1$ with $\tau =1,2,3,4,$
\begin{eqnarray}
\widehat{R}_{5}^{5} &=&\widehat{R}_{6}^{6}=-\frac{1}{2h_{5}h_{6}}%
[h_{5}^{\ast _{1}\ast _{1}}-\frac{(h_{5}^{\ast _{1}})^{2}}{2h_{5}}-\frac{%
h_{5}^{\ast _{1}}h_{6}^{\ast _{1}})}{2h_{6}}],  \label{equ5} \\
\widehat{R}_{5\tau } &=&\frac{h_{5}}{2h_{6}}\ ^{1}n_{\tau }^{\ast
_{1}}+\left( \frac{h_{5}}{h_{6}}h_{6}^{\ast _{1}}-\frac{3}{2}h_{5}^{\ast
_{1}}\right) \frac{^{1}n_{\tau }^{\ast _{1}}}{2h_{6}},  \label{equ6} \\
\widehat{R}_{6\tau } &=&\frac{\ ^{1}w_{\tau }}{2h_{5}}[h_{5}^{\ast _{1}\ast
_{1}}-\frac{\left( h_{5}^{\ast _{1}}\right) ^{2}}{2h_{5}}-\frac{h_{5}^{\ast
_{1}}h_{6}^{\ast _{1}}}{2h_{6}}]+\frac{h_{5}^{\ast _{1}}}{4h_{5}}(\frac{%
\partial _{\tau }h_{5}}{h_{5}}+\frac{\partial _{\tau }h_{6}}{h_{6}})-\frac{%
\partial _{\tau }(h_{5}^{\ast _{1}})}{2h_{5}},  \label{equ7}
\end{eqnarray}%
on shell $s=2$ with $\tau _{1}=1,2,3,4,5,6:$
\begin{eqnarray}
\widehat{R}_{7}^{7} &=&\widehat{R}_{8}^{8}=-\frac{1}{2h_{7}h_{8}}%
[h_{7}^{\ast _{2}\ast _{2}}-\frac{\left( h_{7}^{\ast _{2}}\right) ^{2}}{%
2h_{7}}-\frac{h_{7}^{\ast _{2}}h_{8}^{\ast _{2}}}{2h_{8}}],  \notag \\
\widehat{R}_{7\tau _{1}} &=&\frac{h_{7}}{2h_{8}}\ ^{2}n_{\tau _{1}}^{\ast
_{2}\ast _{2}}+\left( \frac{h_{7}}{h_{8}}h_{8}^{\ast _{2}}-\frac{3}{2}%
h_{7}^{\ast _{2}}\right) \frac{\ ^{2}n_{\tau _{1}}^{\ast _{2}}}{2h_{7}},
\notag \\
\widehat{R}_{8\tau _{1}} &=&\frac{\ ^{2}w_{\tau _{1}}}{2h_{7}}[h_{7}^{\ast
_{2}\ast _{2}}-\frac{\left( h_{7}^{\ast _{2}}\right) ^{2}}{2h_{7}}-\frac{%
h_{7}^{\ast _{2}}h_{8}^{\ast _{2}}}{2h_{8}}]+\frac{h_{7}^{\ast _{2}}}{4h_{7}}%
(\frac{\partial _{\tau _{1}}h_{7}}{h_{7}}+\frac{\partial _{\tau _{1}}h_{8}}{%
h_{8}})-\frac{\partial _{\tau _{1}}(h_{7}^{\ast _{2}})}{2h_{7}},
\label{equ4d}
\end{eqnarray}%
on shell $s=3$ with $\tau _{2}=1,2,3,4,5,6,7,8:$
\begin{eqnarray}
\widehat{R}_{9}^{9} &=&\widehat{R}_{10}^{10}=-\frac{1}{2h_{9}h_{10}}%
[h_{9}^{\ast _{3}\ast _{3}}-\frac{\left( h_{9}^{\ast _{3}}\right) ^{2}}{%
2h_{9}}-\frac{h_{9}^{\ast _{3}}h_{10}^{\ast _{3}})}{2h_{10}}],  \notag \\
\widehat{R}_{9\tau _{2}} &=&\frac{h_{9}}{2h_{10}}\ ^{2}n_{\tau _{2}}^{\ast
_{3}\ast _{3}}+\left( \frac{h_{9}}{h_{10}}h_{10}^{\ast _{3}}-\frac{3}{2}%
h_{9}^{\ast _{3}}\right) \frac{^{2}n_{\tau _{2}}^{\ast _{3}}}{2h_{9}},
\notag \\
\widehat{R}_{10\tau _{2}} &=&\frac{\ ^{2}w_{\tau _{1}}}{2h_{9}}[h_{9}^{\ast
_{3}\ast _{3}}-\frac{\left( h_{9}^{\ast _{3}}\right) ^{2}}{2h_{9}}-\frac{%
h_{9}^{\ast _{3}}h_{10}^{\ast _{3}}}{2h_{10}}]+\frac{h_{9}^{\ast _{3}}}{%
4h_{9}}(\frac{\partial _{\tau _{2}}h_{9}}{h_{9}}+\frac{\partial _{\tau
_{2}}h_{10}}{h_{10}})-\frac{\partial _{\tau _{2}}(h_{9}^{\ast _{3}})}{2h_{9}}%
.  \label{equ5d}
\end{eqnarray}

Using the above formulas, we can compute the Ricci scalar (\ref{rdsc}) for $%
\ ^{s}\widehat{\mathbf{D}}$ when for the ansatz (\ref{ansk}) and $s=0,1,2,3.$
\begin{equation*}
\ ^{0}\widehat{R} = 2(\widehat{R}_{1}^{1}+\widehat{R}_{3}^{3}),\ ^{1}%
\widehat{R}=2(\widehat{R}_{1}^{1}+\widehat{R}_{3}^{3}+\widehat{R}_{5}^{5}),\
^{2}\widehat{R} =2(\widehat{R}_{1}^{1}+\widehat{R}_{3}^{3}+\widehat{R}%
_{5}^{5}+\widehat{R}_{7}^{7}),\ \ ^{3}\widehat{R}=2(\widehat{R}_{1}^{1}+%
\widehat{R}_{3}^{3}+\widehat{R}_{5}^{5}+\widehat{R}_{7}^{7}+\widehat{R}%
_{9}^{9}).
\end{equation*}%
This imposes certain N-adapted symmetries on the Einstein d--tensor for the
ansatz (\ref{ansk}), see details in \cite{tgovsv,vex3}.

\subsubsection{N--adapted sources and nonholonomically modified Einstein
equations}

We will be able to integrate nonholonomic motion equations in heterotic
string gravity in explicit form for very general assumptions if the source $%
\mathbf{\Upsilon }_{\beta _{s}\delta _{s}}$ (\ref{es}) is parametrized in
N-adapted form. This can be determined by 5 independent effective sources,
respectively, on $h$- and $\ ^{s}v$--subspaces, which will be related to
certain effective cosmological constants corresponding to formulas

\begin{eqnarray}
\widehat{\mathbf{\Upsilon }}_{1}^{1} &=&\widehat{\mathbf{\Upsilon }}%
_{2}^{2}=\ _{h}\Upsilon (x^{k})\rightarrow \ _{h}\Lambda =\ _{h}^{\phi
}\Lambda +\ _{h}^{H}\Lambda +\ _{h}^{F}\Lambda +\ _{h}^{R}\Lambda ,
\label{effects} \\
\widehat{\mathbf{\Upsilon }}_{3}^{3} &=&\widehat{\mathbf{\Upsilon }}%
_{4}^{4}=\ \Upsilon (x^{k},y^{4})\rightarrow \ \Lambda =\ ^{\phi }\Lambda +\
^{H}\Lambda +\ ^{F}\Lambda +\ ^{R}\Lambda ,  \notag \\
\widehat{\mathbf{\Upsilon }}_{5}^{5} &=&\widehat{\mathbf{\Upsilon }}%
_{6}^{6}=\ _{1}\Upsilon (x^{k},y^{a},y^{6})\rightarrow \ _{1}\Lambda =\
_{1}^{\phi }\Lambda +\ _{1}^{H}\Lambda +\ _{1}^{F}\Lambda +\ _{1}^{R}\Lambda
,  \notag \\
\widehat{\mathbf{\Upsilon }}_{7}^{7} &=&\widehat{\mathbf{\Upsilon }}%
_{8}^{8}=\ _{2}\Upsilon (x^{k},y^{a},y^{a_{1}},y^{8})\rightarrow \
_{2}\Lambda =\ _{2}^{\phi }\Lambda +\ _{2}^{H}\Lambda +\ _{2}^{F}\Lambda +\
_{2}^{R}\Lambda ,  \notag \\
\widehat{\mathbf{\Upsilon }}_{9}^{9} &=&\widehat{\mathbf{\Upsilon }}%
_{10}^{10}=\ _{3}\Upsilon
(x^{k},y^{a},y^{a_{1}},y^{a_{2}},y^{10})\rightarrow \ _{3}\Lambda =\
_{3}^{\phi }\Lambda +\ _{3}^{H}\Lambda +\ _{3}^{F}\Lambda +\ _{3}^{R}\Lambda
.  \notag
\end{eqnarray}%
For certain general configurations, it will be possible to fix generating
and effective sources of type $\ _{h}\Lambda =\ _{s}\Lambda =\Lambda $ for
all $s,$ when a value for the corresponding contribution of fields can be
zero, or non-zero, for instance, $\ _{h}^{\phi }\Lambda =0$ but $\
_{2}^{H}\Lambda \neq 0.$ This depends on the type of vacuum, non--vacuum, or
effective vacuum model we study. It is possible to compensate contributions
into an effective source of a field with contributions of another field, for
instance, to get $\Lambda =\ ^{\phi }\Lambda +\ ^{H}\Lambda +\ ^{F}\Lambda
+\ ^{R}\Lambda =0$ even when not all the values of such effective
cosmological constants are zero. We note that by prescribing certain values
of sources (\ref{effects}) we can relate via nonholonomic frame transforms
(in coordinate and/or N-adapted form) $\mathbf{\Upsilon }_{\alpha _{s}\beta
_{s}}=e_{\ \alpha _{s}}^{\alpha _{s}^{\prime }}e_{\ \beta _{s}}^{\beta
_{s}^{\prime }}\widehat{\mathbf{\Upsilon }}_{\alpha _{s}^{\prime }\beta
_{s}^{\prime }},$ where $\mathbf{\Upsilon }_{\alpha _{s}\beta _{s}}$ is any
(effective) source (\ref{es}), in heterotic string gravity, or (\ref{ematter}%
), in a 10-d nonholonomic generalization of GR. For any effective $\ ^{m}%
\widehat{L}$ we can solve a system of quadratic algebraic equations for $%
e_{\ \alpha _{s}}^{\alpha _{s}^{\prime }}$ in a form compatible with
transforms to N--adapted frames of the metric/d--metric components,

In this section, we shall construct general solutions of the generalized
nonholonomic Einstein equations (\ref{cdeinst}) with Ricci d--tensors (\ref%
{equ1})--(\ref{equ5d}) and effective sources (encoding contributions from
heterotic supergravity) (\ref{es}) parameterized in the form (\ref{effects})
written in N--adapted form as
\begin{eqnarray}
\widehat{R}_{1}^{1} &=&\widehat{R}_{2}^{2}=-\ _{h}\Upsilon (x^{k}),\
\widehat{R}_{3}^{3}=\widehat{R}_{4}^{4}=-\ \Upsilon (x^{k},y^{4}),\ \widehat{%
R}_{5}^{5}=\widehat{R}_{6}^{6}=-\ \ _{1}\Upsilon (x^{k},y^{a},y^{6}),
\label{sourc1} \\
\ \widehat{R}_{7}^{7} &=&\widehat{R}_{8}^{8}=-\ \ _{2}\Upsilon
(x^{k},y^{a},y^{a_{1}},y^{8}),\ \widehat{R}_{9}^{9}=\widehat{R}_{10}^{10}=-\
\ _{3}\Upsilon (x^{k},y^{a},y^{a_{1}},y^{a_{2}},y^{10}).  \notag
\end{eqnarray}%
Similar equations can be written recurrently for arbitrary finite extra
dimensions.

\subsubsection{The ansatz for effective matter fields in heterotic string
gravity}

\paragraph{Assumptions on nontrivial $\protect\widehat{\protect\phi }$%
-configurations:}

We shall find d--metrics of type (\ref{dm}) for which the contributions of
the field $\widehat{\phi }$ can be included by nonholonomic deformations $%
\mathbf{\mathring{g}}\rightarrow \mathbf{\ }^{s}\mathbf{g}$ into certain
generic off--diagonal terms, i.e. into N--connection coefficients. For
instance if under a prime configuration $\ ^{0}\widehat{\phi }%
(x^{i},y^{a},y^{a_{s}})\rightarrow \ ^{\eta }\widehat{\phi }=\widehat{\phi }%
(x^{i},y^{4})$ for which $\widehat{\mathbf{d}}\widehat{\phi }=0.$ . This
equation is equivalent to a linear system of equations
\begin{equation}
\partial _{i}\widehat{\phi }-w_{i}(x^{k},y^{4})\widehat{\phi }^{\ast }=0,
\label{scfeq}
\end{equation}%
which can be solved in explicit form if the N--connection coefficients $%
w_{i} $ are defined (see next sections how such values can be found in
explicit form). For such configurations, we state that $\ ^{\phi }\mathbf{%
\Upsilon }_{\mu _{s}\nu _{s}}$ and$\ ^{\phi }\Lambda $ in (\ref{es}) and (%
\ref{es1}) are zero and the fields $\widehat{\phi }$ with such
configurations contribute to possible heterotic supergravity effects only
via possible nontrivial off--diagonal interactions and not via effective
sources. The terms with $\widehat{\phi }$ also vanish in all nonholonomic
motion equations (\ref{hs1})-(\ref{hs4}) written with respect to N-adapted
frames if the conditions (\ref{scfeq}) are satisfied. Nontrivial coefficents
with $\widehat{\phi }$ and its partial derivatives will appear if certain
physical equations are written, for instance, in coordinate frames. We have
chosen special type configurations for $\widehat{\phi }$ $\ $in order to
simplify the procedure of finding generic off--diagonal solutions in
heterotic string gravity in explicit form.

\paragraph{Nonholonomic gauge configurations:}

In general form, we can consider any $\ ^{H}\mathbf{\Upsilon }_{\mu _{s}\nu
_{s}},\ ^{F}\mathbf{\Upsilon }_{\mu _{s}\nu _{s}}$ and $\ ^{R}\mathbf{%
\Upsilon }_{\mu _{s}\nu _{s}}$ for which $\mathbf{\Upsilon }_{\mu _{s}\nu
_{s}}=\ ^{H}\mathbf{\Upsilon }_{\mu _{s}\nu _{s}}+\ ^{F}\mathbf{\Upsilon }%
_{\mu _{s}\nu _{s}}+\ ^{R}\mathbf{\Upsilon }_{\mu _{s}\nu _{s}},$ can be
nonholonomicaly transformed into an N--adapted diagonal form
\begin{eqnarray*}
\mathbf{\Upsilon }_{\ \nu _{s}}^{\mu _{s}} &=&diag[\ _{h}\Upsilon (x^{k}),\
_{h}\Upsilon (x^{k}),\ \Upsilon (x^{k},y^{4}),\ \Upsilon (x^{k},y^{4}),\
_{1}\Upsilon (x^{k},y^{a},y^{6}),\ _{1}\Upsilon (x^{k},y^{a},y^{6}), \\
&&\ _{2}\Upsilon (x^{k},y^{a},y^{a_{1}},y^{8}),\ _{2}\Upsilon
(x^{k},y^{a},y^{a_{1}},y^{8}),\ _{3}\Upsilon
(x^{k},y^{a},y^{a_{1}},y^{a_{2}},y^{10}),\ _{3}\Upsilon
(x^{k},y^{a},y^{a_{1}},y^{a_{2}},y^{10})].
\end{eqnarray*}%
Nonholonomic deformations of string gauge fields can be parametrized in the
form
\begin{equation*}
\widehat{\mathbf{H}}_{\alpha _{s}\beta _{s}\mu _{s}}=\ ^{0}\widehat{\mathbf{H%
}}_{\alpha _{s}\beta _{s}\mu _{s}}+\ ^{\eta }\widehat{\mathbf{H}}_{\alpha
_{s}\beta _{s}\mu _{s}},\ \widehat{\mathbf{F}}_{\mu _{s}\alpha _{s}}=\ ^{0}%
\widehat{\mathbf{F}}_{\mu _{s}\alpha _{s}}+\ ^{\eta }\widehat{\mathbf{F}}%
_{\mu _{s}\alpha _{s}},\ \widetilde{\mathbf{R}}_{\mu _{s}\alpha _{s}\beta
_{s}\gamma _{s}}=\ ^{0}\widetilde{\mathbf{R}}_{\mu _{s}\alpha _{s}\beta
_{s}\gamma _{s}}+\ ^{\eta }\widetilde{\mathbf{R}}_{\mu _{s}\alpha _{s}\beta
_{s}\gamma _{s}}.
\end{equation*}%
when, respectively,
\begin{eqnarray}
\ ^{H}\mathbf{\Upsilon }_{\ \nu _{s}}^{\mu _{s}} &=&diag[\ _{h}^{H}\Upsilon
(x^{k}),\ _{h}^{H}\Upsilon (x^{k}),\ ^{H}\Upsilon (x^{k},y^{4}),\
^{H}\Upsilon (x^{k},y^{4}),\ _{1}^{H}\Upsilon (x^{k},y^{a},y^{6}),\
_{1}^{H}\Upsilon (x^{k},y^{a},y^{6}),  \label{effsourcp1} \\
&&\ _{2}^{H}\Upsilon (x^{k},y^{a},y^{a_{1}},y^{8}),\ _{2}^{H}\Upsilon
(x^{k},y^{a},y^{a_{1}},y^{8}),\ _{3}^{H}\Upsilon
(x^{k},y^{a},y^{a_{1}},y^{a_{2}},y^{10}),\ _{3}^{H}\Upsilon
(x^{k},y^{a},y^{a_{1}},y^{a_{2}},y^{10})];  \notag \\
\ ^{F}\mathbf{\Upsilon }_{\ \nu _{s}}^{\mu _{s}} &=&diag[\ _{h}^{F}\Upsilon
(x^{k}),\ _{h}^{F}\Upsilon (x^{k}),\ ^{F}\Upsilon (x^{k},y^{4}),\
^{F}\Upsilon (x^{k},y^{4}),\ _{1}^{F}\Upsilon (x^{k},y^{a},y^{6}),\
_{1}^{F}\Upsilon (x^{k},y^{a},y^{6}),  \notag \\
&&\ _{2}^{F}\Upsilon (x^{k},y^{a},y^{a_{1}},y^{8}),\ _{2}^{F}\Upsilon
(x^{k},y^{a},y^{a_{1}},y^{8}),\ _{3}^{F}\Upsilon
(x^{k},y^{a},y^{a_{1}},y^{a_{2}},y^{10}),\ _{3}^{F}\Upsilon
(x^{k},y^{a},y^{a_{1}},y^{a_{2}},y^{10})];  \notag \\
\ ^{R}\mathbf{\Upsilon }_{\ \nu _{s}}^{\mu _{s}} &=&diag[\ _{h}^{R}\Upsilon
(x^{k}),\ _{h}^{R}\Upsilon (x^{k}),\ ^{R}\Upsilon (x^{k},y^{4}),\
^{R}\Upsilon (x^{k},y^{4}),\ _{1}^{R}\Upsilon (x^{k},y^{a},y^{6}),\
_{1}^{R}\Upsilon (x^{k},y^{a},y^{6}),  \notag \\
&&\ _{2}^{R}\Upsilon (x^{k},y^{a},y^{a_{1}},y^{8}),\ _{2}^{R}\Upsilon
(x^{k},y^{a},y^{a_{1}},y^{8}),\ _{3}^{R}\Upsilon
(x^{k},y^{a},y^{a_{1}},y^{a_{2}},y^{10}),\ _{3}^{R}\Upsilon
(x^{k},y^{a},y^{a_{1}},y^{a_{2}},y^{10})].  \notag
\end{eqnarray}%
For instance, kink-like almost-K\"{a}hler configurations can be encoded as a
prime configuration $\ ^{0}\widehat{\mathbf{H}}_{\alpha _{s}\beta _{s}\mu
_{s}}$ modifying for the target configuration $\widehat{\mathbf{H}}_{\alpha
_{s}\beta _{s}\mu _{s}}$ the solutions for off-diagonal metrics via
effective source.

Such parametrizations are possible by considering the ansatz with effective
cosmological constants%
\begin{equation*}
\widehat{\mathbf{H}}_{\alpha _{s}\beta _{s}\mu _{s}}=\ ^{H}s\sqrt{|\mathbf{g}%
_{\beta _{s}\mu _{s}}|}\epsilon _{\alpha _{s}\beta _{s}\mu _{s}},\ \widehat{%
\mathbf{F}}_{\mu _{s}\alpha _{s}}=\ ^{F}s\sqrt{|\mathbf{g}_{\beta _{s}\mu
_{s}}|}\epsilon _{\mu _{s}\alpha _{s}},\ \widetilde{\mathbf{R}}_{\mu
_{s}\alpha _{s}\beta _{s}\gamma _{s}}=\ ^{R}s\sqrt{|\mathbf{g}_{\beta
_{s}\mu _{s}}|}\epsilon _{\mu _{s}\alpha _{s}\beta _{s}\gamma _{s}},
\end{equation*}%
with absolute anti-symmetric $\epsilon $-tensors (we refer readers to
similar details for 4-d to \cite{svvvey} and references therein). For such
an ansatz, we obtain effective energy--momentum tensors%
\begin{eqnarray}
\ ^{H}\mathbf{\Upsilon }_{\mu _{s}\nu _{s}} &=&-\frac{10}{2}(\ ^{H}s)^{2}%
\mathbf{g}_{\beta _{s}\mu _{s}},\mbox{ for }\ ^{H}\Lambda =-5(\ ^{H}s)^{2},
\label{ansatzsourc} \\
\ ^{F}\mathbf{\Upsilon }_{\mu _{s}\nu _{s}} &=&-\frac{10\alpha ^{\prime }}{2}%
n_{F}(\ ^{F}s)^{2}\mathbf{g}_{\beta _{s}\mu _{s}},\mbox{ for
}\ ^{F}\Lambda =-5n_{F}(\ ^{F}s)^{2},  \notag \\
\ ^{R}\mathbf{\Upsilon }_{\mu _{s}\nu _{s}} &=&\frac{10\alpha ^{\prime }}{2}%
n_{R}(\ ^{R}s)^{2}\mathbf{g}_{\beta _{s}\mu _{s}},\mbox{ for }\ ^{R}\Lambda
=-5trn_{R}(\ ^{R}s)^{2},  \notag
\end{eqnarray}%
were we considered formulas (\ref{effects}) and (\ref{es}) -(\ref{es5}) and,
for instance the numbers $n_{F}=tr[internal$ $F]$ and $n_{R}=tr[internal$ $%
\widetilde{R}]$ depends on the representation of the Lie algebra for $F$ and
on the representation of Lie groups on the internal space.

\subsubsection{Decoupling of nonholonomic motion equations and effective
gravitational field equations}

Considering the ansatz (\ref{ansk}) for $\ g_{i}(x^{k})=\epsilon
_{i}e^{q(x^{k})},\epsilon _{i}=1,$ in (\ref{sourc1}) with respective
sources, we obtain this nonlinear system of PDEs:
\begin{equation}
q^{\cdot \cdot }+q^{\prime \prime }=2\ _{h}\Upsilon ,  \label{e1}
\end{equation}%
\begin{eqnarray}
\varpi ^{\ast }h_{3}^{\ast } &=&2h_{3}h_{4}\ \Upsilon ,\   \label{e2} \\
n_{i}^{\ast \ast }+\gamma n_{i}^{\ast } &=&0,  \label{e3} \\
\beta w_{i}-\alpha _{i} &=&0,\   \label{e4}
\end{eqnarray}%
\begin{eqnarray}
\ ^{1}\varpi ^{\ast _{1}}h_{5}^{^{\ast _{1}}} &=&2h_{5}h_{6}\ _{1}\Upsilon ,
\label{e2aa} \\
\ ^{1}n_{i_{1}}^{\ast _{1}\ast _{1}}+\ ^{1}\gamma \ ^{1}n_{i_{1}}^{\ast
_{1}} &=&0,  \label{e3aa} \\
\ ^{1}\beta \ ^{1}w_{i_{1}}-\ ^{1}\alpha _{1} &=&0,\   \label{e4aa}
\end{eqnarray}%
\begin{eqnarray}
\ \ ^{2}\varpi ^{\ast _{2}}h_{7}^{\ast _{2}} &=&2h_{7}h_{8}\ _{2}\Upsilon ,
\notag \\
\ ^{2}n_{i_{2}}^{\ast _{2}\ast _{2}}+\ ^{2}\gamma \ ^{2}n_{i_{2}}^{\ast
_{2}} &=&0,  \notag \\
\ ^{2}\beta \ ^{2}w_{i_{2}}-\ ^{2}\alpha _{i_{2}} &=&0,\   \label{e4dd}
\end{eqnarray}%
\begin{eqnarray}
\ \ ^{3}\varpi ^{\ast _{3}}h_{7}^{\ast _{3}} &=&2h_{9}h_{10}\ _{3}\Upsilon ,
\notag \\
\ ^{3}n_{i_{3}}^{\ast _{3}\ast _{3}}+\ ^{3}\gamma \ ^{3}n_{i_{3}}^{\ast
_{3}} &=&0,  \notag \\
\ ^{3}\beta \ ^{3}w_{i_{3}}-\ ^{3}\alpha _{i_{3}} &=&0,\   \label{e5dd}
\end{eqnarray}%
In equivalent form, the same equations are obtained recurrently if we write,
for instance, $\ ^{s}w_{i_{s}}\ $instead of $\ ^{s}w_{\tau _{s-1}},\
^{s}n_{i_{s}}\ $instead of $\ ^{s}n_{\tau _{s-1}}$ etc. (some equations and
solutions can be parametrized in more simple forms if we follow the first
convention, other equations will be more "compact" if we follow the second
convention). In these equations, the generating functions
\begin{equation*}
\varpi =\ln |\Phi |,\ ^{1}\varpi =\ln |\ ^{1}\Phi |,\ ^{2}\varpi =\ln |\
^{2}\Phi |,\ ^{3}\varpi =\ln |\ ^{3}\Phi |,
\end{equation*}%
(we shall use this in the next sections, on convenience formulas with $%
\varpi $- and/or $\Phi $-vaules) and $\alpha $- , $\beta $- , $\gamma $%
-coefficients \ on corresponding shells are defined respectively:
\begin{eqnarray}
\varpi &=&\ln |\frac{h_{3}^{\ast }}{\sqrt{|h_{3}h_{4}|}}|,\ \gamma :=(\ln
\frac{|h_{3}|^{3/2}}{|h_{4}|})^{\ast },\ \ \alpha _{i}=\frac{h_{3}^{\ast }}{%
2h_{3}}\partial _{i}\varpi ,\ \beta =\frac{h_{3}^{\ast }}{2h_{3}}\varpi
^{\ast },  \label{c1} \\
\ ^{1}\varpi &=&\ln |\frac{h_{5}^{\ast _{1}}}{\sqrt{|h_{5}h_{6}|}}|,\
^{1}\gamma :=(\ln \frac{|h_{5}|^{3/2}}{|h_{6}|})^{\ast _{1}},\ ^{1}\alpha
_{\tau }=\frac{h_{5}^{\ast _{1}}}{2h_{5}}\partial _{\tau }\ ^{1}\varpi ,\
^{1}\beta =\frac{h_{5}^{\ast _{1}}}{2h_{5}}\partial _{\tau }\ ^{1}\varpi ,
\notag \\
\ ^{2}\varpi &=&\ln |\frac{h_{7}^{\ast _{2}}}{\sqrt{|h_{7}h_{8}|}}|,\
^{2}\gamma :=(\ln \frac{|h_{7}|^{3/2}}{|h_{8}|})^{\ast _{2}},\ \ ^{2}\alpha
_{\tau _{1}}=\frac{h_{7}^{\ast _{2}}}{2h_{7}}\partial _{\tau _{1}}\
^{2}\varpi ,\ \ ^{2}\beta =\frac{h_{7}^{\ast _{2}}}{2h_{7}}\partial _{\tau
_{1}}\ ^{2}\varpi ,  \notag \\
\ ^{3}\varpi &=&\ln |\frac{h_{9}^{\ast _{3}}}{\sqrt{|h_{7}h_{8}|}}|,\
^{3}\gamma :=(\ln \frac{|h_{9}|^{3/2}}{|h_{10}|})^{\ast _{3}},\ \ ^{3}\alpha
_{\tau _{2}}=\frac{h_{9}^{\ast _{3}}}{2h_{9}}\partial _{\tau _{2}}\
^{3}\varpi ,\ \ ^{3}\beta =\frac{h_{9}^{\ast _{3}}}{2h_{9}}\partial _{\tau
_{2}}\ ^{3}\varpi ,  \notag
\end{eqnarray}%
when the frame/coordinate systems are chosen in such a way that nonzero
conditions for the partial derivatives are satisfied: $\varpi ^{\ast
},h_{a}^{\ast },\ ^{1}\varpi ^{\ast _{1}},h_{a_{1}}^{\ast _{1}},\ ^{2}\varpi
^{\ast _{2}},h_{a_{2}}^{\ast _{2}},\ ^{3}\varpi ^{\ast _{3}},h_{a_{3}}^{\ast
_{3}}\neq 0$ (this allows us to avoid singular nonholonomic deformations).

The equations (\ref{e1})-- (\ref{e5dd}) prove a very important decoupling
property of the heterotic string equations (\ref{hs1}) and 4d-10d
(generalized) Einstein equations (\ref{cdeinst}) with respect to
correspondingly N--adapted frames. In explicit and in certain simple forms,
such formulas can be obtained for metrics with at least one Killing
symmetry. Nevertheless, the constructions can be generalized for
non--Killing configurations in any finite extra dimension gravity, see a
corresponding technique in \cite{tgovsv,vex3}. In brief the decoupling
property of the AFDM can be explained for 4--d configurations:

\begin{enumerate}
\item The equation (\ref{e1}) is just a 2-d Laplacian, which can be solved
for any $h$-source $\ _{h}\Upsilon (x^{k}).$

\item The equation (\ref{e2}) contains only the partial derivative $\partial
_{4},$ equivalently $\ast $-derivative, and constraints by a system of two
equations, together with the formula for the value $\varpi $ in (\ref{c1}),
four values $h_{3}(x^{i},y^{4}),$ $h_{4}(x^{i},y^{4})$ and $\varpi
(x^{i},y^{4})$ and source $\ \Upsilon (x^{k},y^{4}).$ Prescribing any two
such functions, we can define (integrating on $y^{4})$ another two such
functions. We note that $h_{a}$ are coefficients of a d-metric, $\varpi $ is
a generating function and $\Upsilon $ is related to extra dimensional string
contributions

\item Using $h_{3}$ and $\varpi ,$ or $\Phi ,$ from the previous point, we
compute the coefficients $\alpha _{i}$ and $\beta ,$ see (\ref{c1}). This
allows allows us to define $w_{i}$ from the algebraic equations (\ref{e4}).

\item Having computed the coefficient $\gamma $ (\ref{c1}), the
N--connection coefficients $n_{i}$ can be defined after two integrations on $%
y^{4}$ in (\ref{e3}).
\end{enumerate}

We can repeat the steps 2-4 recurrently on shells $s=1,2,3,$ adding
dependencies on $2+2+2$ extra dimension coordinates respectively. In this
way, we involve new classes of generating functions and effective sources.
Solving the systems (\ref{e2aa})-(\ref{e4aa}), (\ref{e4dd}) and (\ref{e5dd})
reccurently, we can compute all d-metric and N--connection coefficients for
a 10-d ansatz (\ref{dm}).

\subsubsection{ Integration of nonholonomic motion equations by generating
functions and effective sources}

The system of nonlinear PDEs (\ref{e1})-- (\ref{e5dd}) with coefficients (%
\ref{c1}) and effective sources (\ref{effsourcp1}) with contributions from
string gravity can be integrated in general forms on any shell up to 10-d.

\paragraph{ 4--d non--vacuum configurations:}

The coefficients $g_{i}=e^{q(x^{k})}$ are defined by solutions of the
corresponding 2-d Poisson equation (\ref{e1}) as we mentioned above (see
point 1 at the end of previous subsection).

We can integrate the system of nonlinear PDEs consisting of the first
equation in (\ref{c1}) and (\ref{e2}) for arbitrary source $\ \Upsilon
(x^{k},y^{4})$ and with generating function $\Phi (x^{k},y^{4})=e^{\varpi }.$
The solutions will be generated for a stationary d--metric when the
coefficients do not depend on time like coordinate $y^{3}=t,$ when $\Phi
^{\ast }\neq 0.$ \ The vertical effective gravitational field equations (\ref%
{e2})-(\ref{e4}) transform respectively into
\begin{eqnarray}
\ \ \Phi ^{\ast }\ h_{3}^{\ast } &=&2\ h_{3}h_{4}\ \Upsilon \Phi ,
\label{rsa1} \\
\sqrt{|h_{3}h_{4}|}\Phi &=&h_{3}^{\ast },  \label{rsa1a} \\
\ n_{i}^{\ast \ast }+\left( \ln |\ h_{3}|^{3/2}/|h_{4}|\right) ^{\ast
}n_{i}^{\ast } &=&0,\   \label{rsa2} \\
\Phi ^{\ast }w_{i}-\partial _{i}\Phi &=&0.\   \label{rsa3}
\end{eqnarray}%
This system can be solved in very general forms by prescribing $\Upsilon $
and $\Phi $ and integrating the equations "step by step". Introducing the
function
\begin{equation}
\rho ^{2}:=-h_{3}h_{4},  \label{qf2}
\end{equation}%
(the sign - is motivated by the pseudo-Euclidean signature), we express (\ref%
{rsa1}) and (\ref{rsa1a}) as
\begin{equation}
\Phi ^{\ast }h_{3}^{\ast }=-2\rho ^{2}\Upsilon \Phi \mbox{ and }\
h_{3}^{\ast }=\rho \Phi .  \label{rsa1b}
\end{equation}%
Using $h_{3}^{\ast }$ from the second equation (\ref{rsa1b}) in the first
equation, we write
\begin{equation}
\rho =-\frac{1}{2}\frac{\ \Phi ^{\ast }}{\Upsilon }.  \label{qf1}
\end{equation}%
This value, together with the second equation of (\ref{rsa1b}) and a further
integration on $y^{4}$, result in
\begin{equation}
\ h_{3}=h_{3}^{[0]}(x^{k})-\frac{1}{4}\int dy^{4}\frac{(\Phi ^{2})^{\ast }}{%
\Upsilon },  \label{h4}
\end{equation}%
where $h_{3}^{[0]}(x^{k})$ is an integration function. Considering (\ref{qf1}%
), (\ref{qf2}) and formula (\ref{h4}), we compute%
\begin{equation}
\ h_{4}=-\frac{1}{4h_{3}}(\frac{\ \Phi ^{\ast }}{\Upsilon })^{2}=-\frac{1}{4}%
(\frac{\ \Phi ^{\ast }}{\Upsilon })^{2}\left( h_{3}^{[0]}-\frac{1}{4}\int
dy^{4}\frac{(\Phi ^{2})^{\ast }}{\Upsilon }\right) ^{-1}.  \label{h3}
\end{equation}%
The first part of the N--connection coefficients are found by integrating
two times on $y^{4}$ in (\ref{rsa2}) written in the form
\begin{equation*}
\ n_{i}^{\ast \ast }=(n_{i}^{\ast })^{\ast }=-\ n_{i}^{\ast }(\ln
|h_{3}|^{3/2}/|h_{4}|)^{\ast }
\end{equation*}%
for the coefficient $\gamma $ defined in (\ref{c1}). \ Integrating two times
on $y^{4}$ for explicit values of (\ref{h3}) and (\ref{h4}), we compute
\begin{eqnarray*}
n_{k}(x^{i},y^{4}) &=&\ _{1}n_{k}+\ _{2}n_{k}\int dy^{4}\ \frac{h_{4}}{|\
h_{3}|^{3/2}}=\ _{1}n_{k}+\ _{2}\widetilde{n}_{k}\int dy^{4}\ \frac{(\ \Phi
^{2})^{\ast }}{|\ h_{3}|^{5/2}\Upsilon ^{2}} \\
&=&\ _{1}n_{k}+\ _{2}n_{k}\int dy^{4}\frac{(\Phi ^{2})^{\ast }}{\ \Upsilon
^{2}}\left\vert h_{3}^{[0]}-\frac{1}{4}\int dy^{4}\frac{(\Phi ^{\ast })^{2}}{%
\Upsilon }\right\vert ^{-5/2},
\end{eqnarray*}%
containing also a second set of integration functions $\ _{1}n_{k}(x^{i})$
and redefined second integration function$\ _{2}\widetilde{n}_{k}(x^{i}).$

We can solve the linear algebraic equations (\ref{rsa3}) and express
\begin{equation*}
\ w_{i}=\partial _{i}\ \Phi /\Phi ^{\ast }.
\end{equation*}

Putting together all the above formulas and writing the effective source (%
\ref{effects}) in explicit form, we obtain the formulas for the coefficients
of a d--metric and a N--connection determining a class of stationary
solutions for the system (\ref{e1})-(\ref{e4}),
\begin{eqnarray}
\ g_{i} &=&g_{i}[\ q,~\ _{h}\Upsilon ,]\ =e^{\ q(x^{k})}%
\mbox{ as
a solution of 2-d Poisson equations (\ref{e1})};  \notag \\
\ h_{3} &=&h_{3}^{[0]}(x^{k})-\frac{1}{4}\int dy^{4}\frac{(\Phi ^{2})^{\ast }%
}{\Upsilon };\   \label{solut1t} \\
h_{4} &=&-\frac{1}{4}(\frac{\ \Phi ^{\ast }}{\Upsilon })^{2}\left(
h_{3}^{[0]}-\frac{1}{4}\int dy^{4}\frac{(\Phi ^{2})^{\ast }}{\Upsilon }%
\right) ^{-1};  \notag \\
n_{k} &=&\ _{1}n_{k}+\ _{2}n_{k}\int dy^{4}\frac{(\Phi ^{2})^{\ast }}{\
\Upsilon ^{2}}\left\vert h_{3}^{[0]}-\frac{1}{4}\int dy^{4}\frac{(\Phi
^{2})^{\ast }}{\Upsilon }\right\vert ^{-5/2};  \notag \\
w_{i} &=&\partial _{i}\ \Phi /\Phi ^{\ast }.\   \notag
\end{eqnarray}

Using coefficients (\ref{solut1t}), we define such a class of quadratic
elements for off--diagonal 4-d stationary configurations in heterotic
supergravity with nonholonomically induced torsion,
\begin{eqnarray}
ds_{K4d}^{2} &=&g_{\alpha \beta }(x^{k},y^{4})du^{\alpha }du^{\beta
}=e^{q}[(dx^{1})^{2}+(dx^{2})^{2}]+  \notag \\
&&\left[ h_{3}^{[0]}(x^{k})-\frac{1}{4}\int dy^{4}\frac{(\Phi ^{2})^{\ast }}{%
\Upsilon }\right] [dt+(\ _{1}n_{k}+\ _{2}n_{k}\int dy^{4}\frac{(\Phi
^{2})^{\ast }}{\ \Upsilon ^{2}}\left\vert h_{3}^{[0]}-\frac{1}{4}\int dy^{4}%
\frac{(\Phi ^{2})^{\ast }}{\Upsilon }\right\vert ^{-5/2})dx^{k}]^{2}  \notag
\\
&&-\frac{1}{4}(\frac{\ \Phi ^{\ast }}{\Upsilon })^{2}\left( h_{3}^{[0]}-%
\frac{1}{4}\int dy^{4}\frac{(\Phi ^{2})^{\ast }}{\Upsilon }\right)
^{-1}[dy^{4}+\frac{\partial _{i}\ \Phi }{\Phi ^{\ast }}dx^{i}]^{2}.
\label{riccisolt}
\end{eqnarray}%
Such a class of metrics also contains exact solutions for the canonical
d--connection $\widehat{\mathbf{D}}$ in $R^{2}$ gravity with effective
scalar field encoded into a nonholonomically polarized vacuum with a special
parametrization of the source $\Upsilon ,$ see details in \cite%
{kehagias,muen01}.

\paragraph{Examples of 4--d - 10- vacuum configurations:}

The configurations with zero source can not be constructed as particular
cases of former off--diagonal solutions with $\ \Upsilon =0$ because these
limits are not smooth for metrics (\ref{riccisolt}). In string heterotic
gravity, such conditions can be satisfied if different effective fields
compensate their mutual contributions and result in an effective vacuum
gravitational configuration.

For the ansatz (\ref{ansk}), we can construct solutions when the nontrivial
coefficients of the Ricci d--tensor (\ref{equ1})--(\ref{equ4d}) are zero but
the Ricci and torsion d--tensor are not trivial. The first equation is a
typical example of 2--d Laplace equation with solutions expressed in the
form $g_{i}=e^{q(x^{k},\Upsilon =0)}.$

In 4-d, there are three general classes of off--diagonal metrics which
result in zero coefficients (\ref{equ2})--(\ref{equ4}). Such constructions
can be generalized for 10-d generic off--diagonals with mixing standard
vacuum configurations for 4-d spacetimes.

\begin{enumerate}
\item If $h_{3}^{\ast }=0$ but $h_{3}\neq 0,$ $h_{4}^{\ast }\neq 0$ and $%
h_{4}\neq 0,$ we obtain only one nontrivial equation (\ref{equ3}),%
\begin{equation*}
n_{k}^{\ast \ast }+n_{k}^{\ast }\ (\ln |h_{4}|)^{\ast }=0,
\end{equation*}%
where $h_{4}(x^{i},y^{4})$ and $w_{k}(x^{i},y^{4})$ are arbitrary generating
functions. Integrating two times on $y^{4},$ we obtain
\begin{equation}
n_{k}=\ _{1}n_{k}+\ _{2}n_{k}\int dy^{4}/h_{4}  \label{wsol}
\end{equation}%
with integration functions $\ _{1}n_{k}(x^{i})$ and $\ _{2}n_{k}(x^{i}).$ In
4-d, this defines a quadratic line element
\begin{eqnarray*}
ds_{v,4-d}^{2} &=&e^{q(x^{k},\Upsilon =0)}(dx^{i})^{2}+\
^{0}h_{3}(x^{k})[dt+(\ _{1}n_{k}(x^{i})+\ _{2}n_{k}(x^{i})\int
dy^{4}/h_{4})dx^{k}]^{2}+ \\
&&h_{4}(x^{i},y^{4})[dy^{4}+w_{i}(x^{k},y^{4})dx^{i}]^{2}.
\end{eqnarray*}%
Recurrently, we can construct effective stationary vacuum configurations in
10-d [with $%
i_{1}=(1,2,3,4);i_{2}=(1,2,3,4,5,6);i_{3}=(1,2,3,4,5,6,7,8);a_{1}=5,6;a_{2}=7,8;a_{3}=9,10],
$
\begin{eqnarray*}
&&{\ ds_{v,10-d}^{2}} =e^{q(x^{k},\Upsilon =0)}(dx^{i})^{2} \\
&&{+\{\ ^{0}h_{3}(x^{k})[dt+(\ _{1}n_{k}(x^{i})+\ _{2}n_{k}(x^{i})\int
dy^{4}/h_{4})dx^{k}]^{2}+h_{4}(x^{i},y^{4})[dy^{4}+w_{i}(x^{k},y^{4})dx^{i}]^{2}\}%
}_{type1,s=0} \\
&&+\ ^{0}h_{5}(x^{k},y^{4})[dy^{5}+(\ _{1}^{1}n_{k_{1}}(x^{i},y^{4})+\
_{2}^{1}n_{k_{1}}(x^{i},y^{6})\int dy^{6}/h_{6})dx^{k_{1}}]^{2}\}_{type1,s=1}
\\
&&+\{h_{6}(x^{i},y^{4},y^{6})[dy^{6}+\
^{1}w_{i_{1}}(x^{k},y^{4},y^{6})dx^{i_{1}}]^{2} \\
&&+\ ^{0}h_{7}(x^{k},y^{4},y^{a_{1}})[dy^{7}+(\
_{1}^{2}n_{k_{2}}(x^{i},y^{4},y^{a_{1}})+\
_{2}^{2}n_{k_{2}}(x^{i},y^{4},y^{a_{1}})\int
dy^{8}/h_{8})dx^{k_{2}}]^{2}\}_{type1,s=2} \\
&&+\{h_{8}(x^{i},y^{4},y^{a_{1}},y^{8})[dy^{8}+\
^{2}w_{i_{2}}(x^{k},y^{4},y^{a_{1}},y^{8})dx^{i_{2}}]^{2} \\
&&+\ ^{0}h_{9}(x^{k},y^{4},y^{a_{1}},y^{a_{2}})[dy^{9}+(\
_{1}^{3}n_{k_{3}}(x^{i},y^{4},y^{a_{1}},y^{a_{2}})+\
_{2}^{3}n_{k_{3}}(x^{i},y^{4},y^{a_{1}},y^{a_{2}})\int
dy^{10}/h_{10})dx^{k_{3}}]^{2} \\
&&+h_{10}(x^{i},y^{4},y^{a_{1}},y^{a_{2}},y^{10})[dy^{10}+\
^{3}w_{i_{3}}(x^{k},y^{4},y^{a_{1}},y^{a_{2}},y^{10})dx^{i_{3}}]^{2}%
\}_{type1,s=3}.
\end{eqnarray*}%
For instance, the bracket $\{...\}_{type1,s=2}$ states that the vacuum
metric is of type 1 on shell 2. The coefficients of such d-metrics do not
depend on variables $(t,y^{9})$ and, on respective shells, the values
\begin{equation*}
\ ^{0}h_{3},\ _{1}n_{k},\ _{2}n_{k},\ ^{0}h_{5},\ _{1}^{1}n_{k_{1}},\
_{2}^{1}n_{k_{1}},\ ^{0}h_{7},\ _{1}^{2}n_{k_{2}},\ _{2}^{2}n_{k_{2}},\
^{0}h_{9},\ _{1}^{3}n_{k_{3}},\ _{2}^{3}n_{k_{3}}
\end{equation*}%
are integration functions and $q,{\ h_{4},w_{i},}h_{6},\
^{1}w_{i_{1}},h_{8},\ ^{2}w_{i_{2}},h_{10},\ ^{3}w_{i_{3}}$ are generating
functions for 10-d nonholonomic effective vacuum heterotic string
configurations. In the above formulas, $h_{5}^{\ast _{1}}=0,$ $h_{5}\neq 0,$
$h_{6}^{\ast _{1}}\neq 0$ and $h_{6}\neq 0;\ h_{7}^{\ast _{2}}=0,$ $%
h_{7}\neq 0,$ $h_{8}^{\ast _{2}}\neq 0$ and $h_{10}\neq 0;\ h_{9}^{\ast
_{3}}=0,$ $h_{9}\neq 0,$ $h_{10}^{\ast _{3}}\neq 0$ and $h_{10}\neq 0.$

\item In such cases, $h_{3}^{\ast }\neq 0$ and $h_{4}^{\ast }\neq 0.$ It is
possible to solve the equation (\ref{equ2}) and (\ref{e2}) in a
self--consistent form for $\ \Upsilon =0$ if $\varpi ^{\ast }=0$ for
respective coefficients in (\ref{c1}). Fixing $\varpi =\varpi _{0}=const,$
we can consider arbitrary functions $w_{i}(x^{k},y^{4})$ because $\beta
=\alpha _{i}=0$ for such configurations. The conditions (\ref{c1}) are
satisfied by any
\begin{equation}
h_{4}=\ ^{0}h_{4}(x^{k})[(\sqrt{|h_{3}|})^{\ast }]^{2},  \label{h34vacuum}
\end{equation}%
where $\ ^{0}h_{3}(x^{k})$ is an integration function and $%
h_{3}(x^{k},y^{4}) $ is any generating function. The coefficients $n_{k}$
should be found from (\ref{equ3}), see (\ref{wsol}). Such a family of 4-d
vacuum generic off--diagonal metrics is described by
\begin{eqnarray}
ds_{v,4d}^{2} &=&e^{q(x^{k},\Upsilon
=0)}(dx^{i})^{2}+h_{3}(x^{i},y^{4})\{dt+(\ _{1}n_{k}(x^{i})+\ _{2}\widetilde{%
n}_{k}(x^{i})\int dy^{4}[(|h_{3}|^{3/4})^{\ast }]^{2}dx^{i}\}^{2}+
\label{vs2} \\
&&\ ^{0}h_{4}(x^{k})[(\sqrt{|h_{3}|})^{\ast
}]^{2}[dy^{4}+w_{i}(x^{k},y^{4})dx^{i}]^{2},  \notag
\end{eqnarray}%
where the integration functions $\ _{2}\widetilde{n}_{k}(x^{i})$ contains
certain integration coefficients. We can extend such metrics on any shell $%
s=1,2,3$ preserving the conditions of zero effective source and adding
respective generating functions {\small
\begin{eqnarray*}
s &=&1:\ ^{1}\varpi ^{\ast _{1}}=0,h_{6}=\ ^{0}h_{6}(x^{k},y^{4})[(\sqrt{%
|h_{5}|})^{\ast _{1}}]^{2},\mbox{ gener. functs. }\left\{
\begin{array}{c}
h_{5}(x^{i},y^{4},y^{6}) \\
\ ^{1}w_{i_{1}}(x^{i},y^{4},y^{6})%
\end{array}%
\right. ; \\
s &=&2:\ ^{2}\varpi ^{\ast _{2}}=0,h_{8}=\ ^{0}h_{8}(x^{k},y^{4},y^{a_{1}})[(%
\sqrt{|h_{7}|})^{\ast _{2}}]^{2},\mbox{ gener. functs. }\left\{
\begin{array}{c}
h_{7}(x^{i},y^{4},y^{a_{1}},y^{8}) \\
\ ^{2}w_{i_{2}}(x^{i},y^{4},y^{a_{1}},y^{8})%
\end{array}%
\right. ; \\
s &=&3:\ ^{3}\varpi ^{\ast _{3}}=0,h_{10}=\
^{0}h_{10}(x^{k},y^{4},y^{a_{1}},y^{a_{2}})[(\sqrt{|h_{9}|})^{\ast
_{3}}]^{2},\mbox{ gener. functs. }\left\{
\begin{array}{c}
h_{9}(x^{i},y^{4},y^{a_{1}},y^{a_{2}},y^{10}) \\
\ ^{3}w_{i_{3}}(x^{i},y^{4},y^{a_{1}},y^{a_{2}},y^{10})%
\end{array}%
\right. .
\end{eqnarray*}%
} In 10-d, the vacuum solutions (\ref{vs2}) are generalized to
\begin{eqnarray*}
ds_{v,10-d}^{2} &=&e^{q}(dx^{i})^{2}+h_{3}\{dt+(\ _{1}n_{k}+\ _{2}\widetilde{%
n}_{k}\int dy^{4}[(|h_{3}|^{3/4})^{\ast }]^{2}dx^{k}\}^{2}+\ ^{0}h_{4}[(%
\sqrt{|h_{3}|})^{\ast }]^{2}[dy^{4}+w_{i}dx^{i}]^{2}+ \\
&&h_{5}\{dy^{5}+(\ _{1}^{1}n_{k_{1}}+\ _{2}^{1}\widetilde{n}_{k_{1}}\int
dy^{6}[(|h_{5}|^{3/4})^{\ast _{1}}]^{2}dx^{k_{1}}\}^{2}+\ ^{0}h_{6}[(\sqrt{%
|h_{5}|})^{\ast _{1}}]^{2}[dy^{6}+\ ^{1}w_{i_{1}}dx^{i_{1}}]^{2}+ \\
&&h_{7}\{dy^{7}+(\ _{1}^{2}n_{k_{2}}+\ _{2}\widetilde{n}_{k_{2}}\int
dy^{8}[(|h_{7}|^{3/4})^{\ast _{2}}]^{2}dx^{k_{2}}\}^{2}+\ ^{0}h_{8}[(\sqrt{%
|h_{7}|})^{\ast _{2}}]^{2}[dy^{8}+\ ^{2}w_{i_{2}}dx^{i_{2}}]^{2}+ \\
&&h_{9}\{dy^{9}+(\ _{1}^{3}n_{k_{3}}+\ _{2}\widetilde{n}_{k_{3}}\int
dy^{10}[(|h_{9}|^{3/4})^{\ast _{3}}]^{2}dx^{k_{3}}\}^{2}+\ ^{0}h_{10}[(\sqrt{%
|h_{9}|})^{\ast _{3}}]^{2}[dy^{10}+\ ^{2}w_{i_{3}}dx^{i_{3}}]^{2},
\end{eqnarray*}%
for integration functions
\begin{eqnarray*}
&&\ \ ^{0}h_{4}(x^{k}),_{1}n_{k}(x^{i}),\ _{2}\widetilde{n}_{k}(x^{i});\
^{0}h_{6}(x^{k},y^{4}),\ _{1}^{1}n_{k_{1}}(x^{i},y^{4}),\ _{2}^{1}\widetilde{%
n}_{k_{1}}(x^{i},y^{4}); \\
&&\ ^{0}h_{8}(x^{k},y^{4},y^{a_{1}}),\
_{1}^{2}n_{k_{2}}(x^{i},y^{4},y^{a_{1}}),\ _{2}^{2}\widetilde{n}%
_{k_{2}}(x^{i},y^{4},y^{a_{1}}); \\
&&\ ^{0}h_{10}(x^{k},y^{4},y^{a_{1}},y^{a_{2}}),\
_{1}^{3}n_{k_{3}}(x^{i},y^{4},y^{a_{1}},y^{a_{2}}),\ _{2}^{3}\widetilde{n}%
_{k_{2}}(x^{i},y^{4},y^{a_{1}},y^{a_{2}}),
\end{eqnarray*}%
In the above shell formulas, $h_{5}^{\ast _{1}}\neq 0$ and $h_{6}^{\ast
_{1}}\neq 0;$ $h_{7}^{\ast _{2}}\neq 0$ and $h_{8}^{\ast _{2}}\neq 0;\
h_{9}^{\ast _{3}}\neq 0$ and $h_{10}^{\ast _{3}}\neq 0.$

\item We consider that $h_{3}^{\ast }\neq 0$ but $h_{4}^{\ast }=0.$ The
equation (\ref{equ2}) transforms into
\begin{equation*}
h_{3}^{\ast \ast }-\frac{\left( h_{3}^{\ast }\right) ^{2}}{2h_{3}}=0,
\end{equation*}%
when the general solution is $h_{3}(x^{k},y^{4})=\left[
c_{1}(x^{k})+c_{2}(x^{k})y^{4}\right] ^{2}$, with generating functions $%
c_{1}(x^{k}),c_{2}(x^{k})$, and $h_{4}=\ ^{0}h_{4}(x^{k}).$ For $\varpi
=\varpi _{0}=const,$ we can take any values $w_{i}(x^{k},y^{4})$ because $%
\beta =\alpha _{i}=0.$ The coefficients $n_{i}$ are found from (\ref{equ3})
and/or, equivalently, to (\ref{e3}) with $\gamma =\frac{3}{2}|h_{3}|^{\ast
}. $ Integrating on $y^{4},$ this subclass of N--coefficients are computed
\begin{equation*}
n_{i}=\ _{1}n_{i}(x^{k})+\ _{2}n_{i}(x^{k})\int dy^{4}|h_{3}|^{-3/2}=\
_{1}n_{i}(x^{k})+\ _{2}\widetilde{n}_{i}(x^{k})\left[
c_{1}(x^{k})+c_{2}(x^{k})y^{4}\right] ^{-2},
\end{equation*}%
with integration functions $\ _{1}n_{i}(x^{k})$ and $\ _{2}n_{i}(x^{k}),$ or
re--defined $\ \ _{2}\widetilde{n}_{i}=-\ _{2}n_{i}/2c_{2}.$ The quadratic
line element for this class of solutions for vacuum metrics is described by
\begin{eqnarray}
ds_{v,4-d}^{2} &=&e^{q}(dx^{i})^{2}+\left[ c_{1}(x^{k})+c_{2}(x^{k})y^{4}%
\right] ^{2}[dt+(\ _{1}n_{i}(x^{k})+\ _{2}\widetilde{n}_{i}(x^{k})\left[
c_{1}(x^{k})+c_{2}(x^{k})y^{4}\right] ^{-2})dx^{i}]^{2}  \notag \\
&&+\ ^{0}h_{4}(x^{k})[dy^{4}+w_{i}(x^{k},y^{4})dx^{i}]^{2}.  \label{vs3}
\end{eqnarray}%
On extra shells, this type of nonholnomic vacuum solutions are given by
quadratic elements
\begin{eqnarray*}
&&ds_{v,10-d3}^{2} =e^{q}(dx^{i})^{2}+ \\
&&\{\left[ c_{1}+c_{2}y^{4}\right] ^{2}[dt+(\ _{1}n_{i}+\ _{2}\widetilde{n}%
_{i}\left[ c_{1}+c_{2}y^{4}\right] ^{-2})dx^{i}]^{2}+\
^{0}h_{4}[dy^{4}+w_{i}dx^{i}]^{2}\}_{type3,s=0}+ \\
&&\{\left[ \ ^{1}c_{1}+\ ^{1}c_{2}y^{6}\right] ^{2}[dy^{5}+(\
_{1}^{1}n_{i_{1}}+\ _{2}^{1}\widetilde{n}_{i_{1}}\left[ \ ^{1}c_{1}+\
^{1}c_{2}y^{6}\right] ^{-2})dx^{i_{1}}]^{2}+\ ^{0}h_{6}[dy^{6}+\
^{1}w_{i_{1}}dx^{i_{1}}]^{2}\}_{type3,s=1}+ \\
&&\{\left[ \ ^{2}c_{1}+\ ^{2}c_{2}y^{8}\right] ^{2}[dy^{7}+(\
_{1}^{2}n_{i_{2}}+\ _{2}^{2}\widetilde{n}_{i_{2}}\left[ \ ^{2}c_{1}+\
^{2}c_{2}y^{8}\right] ^{-2})dx^{i_{2}}]^{2}+\ ^{0}h_{8}[dy^{8}+\
^{2}w_{i_{2}}dx^{i_{2}}]^{2}\}_{type3,s=2}+ \\
&&\{\left[ \ ^{3}c_{1}+\ ^{3}c_{2}y^{10}\right] ^{2}[dy^{9}+(\
_{1}^{3}n_{i_{3}}+\ _{2}^{3}\widetilde{n}_{i_{3}}\left[ \ ^{2}c_{1}+\
^{2}c_{2}y^{10}\right] ^{-2})dx^{i_{3}}]^{2}+\ ^{0}h_{10}[dy^{10}+\
^{3}w_{i_{3}}dx^{i_{3}}]^{2}\}_{type3,s=3}.
\end{eqnarray*}%
The generating functions are
\begin{equation*}
q(x^{k}),w_{i}(x^{k},y^{4}),\ ^{1}w_{i_{1}}(x^{k},y^{4},y^{6}),\
^{2}w_{i_{2}}(x^{k},y^{4},y^{a_{1}},y^{8}),\
^{3}w_{i_{3}}(x^{k},y^{4},y^{a_{1}},y^{a_{2}},y^{10})
\end{equation*}%
and the integration functions are{\small
\begin{eqnarray*}
&&c_{1}(x^{k}),c_{2}(x^{k}),\ _{1}n_{i}(x^{k}),\ _{2}\widetilde{n}%
_{i}(x^{k}),\ ^{0}h_{4}(x^{k}); \\
&&\ ^{1}c_{1}(x^{k},y^{4}),\ ^{1}c_{2}(x^{k},y^{4}),\
_{1}^{1}n_{i_{1}}(x^{k},y^{4}),\ _{2}^{1}\widetilde{n}_{i_{1}}(x^{k},y^{4}),%
\ ^{0}h_{6}(x^{k},y^{4}); \\
&&\ ^{2}c_{1}(x^{k},y^{4},y^{a_{1}}),\ ^{2}c_{2}(x^{k},y^{4},y^{a_{1}}),\
_{1}^{2}n_{i_{2}}(x^{k},y^{4},y^{a_{1}}),\ _{2}^{2}\widetilde{n}%
_{i_{2}}(x^{k},y^{4},y^{a_{1}}),\ ^{0}h_{8}(x^{k},y^{4},y^{a_{1}}); \\
&&\ ^{3}c_{1}(x^{k},y^{4},y^{a_{1}},y^{a_{2}}),\
^{3}c_{2}(x^{k},y^{4},y^{a_{1}},y^{a_{2}}),\
_{1}^{3}n_{i_{2}}(x^{k},y^{4},y^{a_{1}},y^{a_{2}}),\ _{2}^{3}\widetilde{n}%
_{i_{2}}(x^{k},y^{4},y^{a_{1}},y^{a_{2}}),\
^{0}h_{10}(x^{k},y^{4},y^{a_{1}},y^{a_{2}}).
\end{eqnarray*}
}
\end{enumerate}

Finally, we note that we can construct generic off--diagonal vacuum
solutions in heterotic supergravity with different types on different shells%
\begin{equation*}
ds_{vaccum}^{2}=e^{q}+\{...\}_{type,s=0}+\{...\}_{type,s=1}+\{...%
\}_{type,s=2}+\{...\}_{type,s=3}
\end{equation*}
In certain cases, the type may be the same on two or three shells, but the
parametrizations of mixed types of nonholonomic vacuum solutions is given by
different dependencies on shell coordinates of the generating and
integration functions.

\paragraph{Extra dimensional non--vacuum solutions:}

The solutions for extra dimensions can be constructed in certain forms which
are similar to the 4--d ones with (\ref{solut1t}) using new classes of
generating and integration functions with dependencies on extra dimensional
coordinates. For instance, we can generate solutions of the system (\ref%
{e2aa})--(\ref{e4aa}) with coefficients (\ref{c1}) on the shell $s=1$
following a formal analogy when $\partial _{4}\rightarrow \partial _{6},$
i.e. $\ast \rightarrow \ast _{1},\varpi (x^{k},y^{4})\rightarrow \
^{1}\varpi (x^{k_{1}},y^{6}),\ $\ with $\ ^{1}\Phi (x^{k_{1}},y^{4})=e^{\
^{1}\varpi }$ ; $\Upsilon (x^{k},y^{4})\rightarrow \ _{1}\Upsilon
(x^{k_{1}},y^{6})...$ and applying the same procedure as for 4-d but
extending the number of shell coordinates respectively for generating and
integration functions.

The solutions on $s=1$ are also generated as stationary d--metrics when the
coefficients do not depend on time like coordinate $y^{3}=t$ and do not
depend on the coordinate $y^{5}.$ For non-trivial $\ _{1}\Upsilon ,$ we have
to choose $\ ^{1}\Phi ^{\ast _{1}}\neq 0.$ \ The effective gravitational
field equations (\ref{e2aa})-(\ref{e4aa}) and respective equation for $\
^{1}\Phi $ from (\ref{c1}) transform (respectively) into
\begin{eqnarray*}
\ \ \ ^{1}\Phi ^{\ast _{1}}\ h_{5}^{\ast _{1}} &=&2\ h_{5}h_{6}\ \
_{1}\Upsilon \ \ ^{1}\Phi , \\
\sqrt{|h_{5}h_{6}|}\ \ ^{1}\Phi &=&h_{5}^{\ast _{1}}, \\
\ \ ^{1}n_{i_{1}}^{\ast _{1}\ast _{1}}+\left( \ln |\
h_{5}|^{3/2}/|h_{6}|\right) ^{\ast _{1}}\ ^{1}n_{i_{1}}^{\ast _{1}} &=&0,\
\\
\ \ ^{1}\Phi \ ^{1}w_{i_{1}}^{\ast _{1}}-\partial _{i_{1}}\ \ ^{1}\Phi
&=&0.\
\end{eqnarray*}%
Prescribing $\ _{1}\Upsilon $ and $\ ^{1}\Phi$, such equations for 6-d
gravity can be integrated as in the case of $s=0$ with two vertical
coordinates in 4-d gravity, i.e. integrating equations "step by step".
Introducing the function $\ (\ ^{1}\rho )^{2}:=h_{5}h_{6},$ (for extra
shells the sign is different because we work with space like coordinates) we
express the first two equations above as
\begin{equation}
\ ^{1}\Phi ^{\ast _{1}}h_{5}^{\ast _{1}}=2(\ ^{1}\rho )^{2}\ _{1}\Upsilon \
^{1}\Phi \mbox{ and }\ h_{5}^{\ast _{1}}=\ ^{1}\rho \ ^{1}\Phi .
\label{aux5}
\end{equation}%
Using $h_{5}^{\ast _{1}}$ from the second equation in the first equations,
we get $\ ^{1}\rho =\frac{1}{2}\frac{\ \ ^{1}\Phi ^{\ast _{1}}}{\
_{1}\Upsilon }.$ Substituting this value together with the second equation
of (\ref{aux5}) and integrating on $y^{6},$ it is possible to compute
\begin{equation*}
\ h_{5}=h_{5}^{[0]}(x^{k},y^{4})+\frac{1}{4}\int dy^{6}\frac{(\ ^{1}\Phi
^{2})^{\ast _{1}}}{\ _{1}\Upsilon },
\end{equation*}%
where $h_{5}^{[0]}(x^{k},y^{4})$ is an integration function. In result, we
find from the above formula~$(\ ^{1}\rho )^{2}$ the coefficient
\begin{equation*}
\ h_{6}=\frac{1}{4h_{5}}(\frac{\ \Phi ^{\ast _{1}}}{\ _{1}\Upsilon })^{2}=%
\frac{1}{4}(\frac{\ \Phi ^{\ast _{1}}}{\ _{1}\Upsilon })^{2}\left(
h_{5}^{[0]}+\frac{1}{4}\int dy^{6}\frac{(\ ^{1}\Phi ^{2})^{\ast _{1}}}{\
_{1}\Upsilon }\right) ^{-1}.
\end{equation*}

The first part of the N--connection coefficients are found by integrating
two times on $y^{6}$ in (\ref{e3aa}) written in the form
\begin{equation*}
\ \ ^{1}n_{i_{1}}^{\ast _{1}\ast _{1}}=(n_{i_{1}}^{\ast _{1}})^{\ast
_{1}}=-\ n_{i_{1}}^{\ast _{1}}(\ln |h_{5}|^{3/2}/|h_{6}|)^{\ast _{1}}
\end{equation*}%
for the coefficient $\ ^{1}\gamma $ defined in (\ref{c1}). \ This formula
can be integrated two times on $y^{6}$ for explicit values of $h_{a_{1}}$
which results in
\begin{eqnarray*}
\ \ ^{1}n_{k_{1}}(x^{i},y^{4},y^{6}) &=&\ _{1}^{1}n_{k_{1}}+\
_{2}^{1}n_{k_{1}}\int dy^{6}\ \frac{h_{6}}{|\ h_{5}|^{3/2}}=\
_{1}^{1}n_{k_{1}}+\ _{2}^{1}\widetilde{n}_{k_{1}}\int dy^{6}\ \frac{(\ \
^{1}\Phi ^{\ast _{1}})^{2}}{|\ h_{5}|^{5/2}(\ _{1}\Upsilon )^{2}} \\
&=&\ _{1}^{1}n_{k_{1}}+\ _{2}^{1}\widetilde{n}_{k_{1}}\int dy^{6}\frac{(\ \
^{1}\Phi ^{\ast _{1}})^{2}}{(\ _{1}\Upsilon )^{2}}\left\vert h_{5}^{[0]}+%
\frac{1}{4}\int dy^{6}\frac{(\ ^{1}\Phi ^{2})^{\ast _{1}}}{\ _{1}\Upsilon }%
\right\vert ^{-5/2},
\end{eqnarray*}%
also containing a second set of integration functions $\
_{1}^{1}n_{k_{1}}(x^{i},y^{4})$ and redefined second integration function $\
_{2}^{1}\widetilde{n}_{k_{1}}(x^{i},y^{4}).$ The second set of N--connection
coefficients on $s=1$ can be found from the linear algebraic equations (\ref%
{e4aa}) and express $\ \ ^{1}w_{i_{1}}=\partial _{i_{1}}\ \ ^{1}\Phi /\
^{1}\Phi ^{\ast _{1}}.$

Summarizing the above formulas for the d--connection and N--connection
coefficients on $s=1,$ we write the data for generating such generic
off--diagonal metrics as stationary solutions for the system (\ref{e2aa})--(%
\ref{e4aa}),
\begin{eqnarray*}
\ h_{5} &=&h_{5}^{[0]}+\frac{1}{4}\int dy^{6}\frac{(\ ^{1}\Phi ^{2})^{\ast
_{1}}}{\ _{1}\Upsilon }\ ;\  \\
\ h_{6} &=&\frac{1}{4}(\frac{\ \Phi ^{\ast _{1}}}{\ _{1}\Upsilon }%
)^{2}\left( h_{5}^{[0]}+\frac{1}{4}\int dy^{6}\frac{(\ ^{1}\Phi ^{2})^{\ast
_{1}}}{\ _{1}\Upsilon }\right) ^{-1}; \\
\ \ ^{1}n_{k_{1}} &=&\ _{1}^{1}n_{k_{1}}+\ _{2}^{1}\widetilde{n}_{k_{1}}\int
dy^{6}\frac{(\ \ ^{1}\Phi ^{\ast _{1}})^{2}}{(\ _{1}\Upsilon )^{2}}%
\left\vert h_{5}^{[0]}+\frac{1}{4}\int dy^{6}\frac{(\ ^{1}\Phi ^{2})^{\ast
_{1}}}{\ _{1}\Upsilon }\right\vert ^{-5/2}; \\
\ \ ^{1}w_{i_{1}} &=&\partial _{i_{1}}\ \ ^{1}\Phi /\ ^{1}\Phi ^{\ast
_{1}}.\
\end{eqnarray*}%
The quadratic elements for off--diagonal 6-d stationary configurations in
heterotic supergravity with nonholonomically induced torsion are constructed
using above N--adapted coefficients
\begin{eqnarray}
ds_{K6d}^{2} &=&g_{\alpha _{1}\beta _{1}}(x^{k},y^{4},y^{a_{1}})du^{\alpha
_{1}}du^{\beta _{1}}=ds_{K4d}^{2}[\mbox{ see (\ref{riccisolt})}%
]+[h_{5}^{[0]}+\frac{1}{4}\int dy^{6}\frac{(\ ^{1}\Phi ^{2})^{\ast _{1}}}{\
_{1}\Upsilon }]  \label{gensolshell1} \\
&&[dy^{5}+\left( \ _{1}^{1}n_{k_{1}}+\ _{2}^{1}\widetilde{n}_{k_{1}}\int
dy^{6}\frac{(\ \ ^{1}\Phi ^{\ast _{1}})^{2}}{(\ _{1}\Upsilon )^{2}}%
\left\vert h_{5}^{[0]}+\frac{1}{4}\int dy^{6}\frac{(\ ^{1}\Phi ^{2})^{\ast
_{1}}}{\ _{1}\Upsilon }\right\vert ^{-5/2}\right) dx^{k_{1}}]  \notag \\
&&+\frac{1}{4}(\frac{\ \ ^{1}\Phi ^{\ast _{1}}}{\ _{1}\Upsilon })^{2}\left(
h_{5}^{[0]}+\frac{1}{4}\int dy^{6}\frac{(\ ^{1}\Phi ^{2})^{\ast _{1}}}{\
_{1}\Upsilon }\right) ^{-1}[dy^{6}+\frac{\partial _{i_{1}}\ ^{1}\Phi }{\
^{1}\Phi ^{\ast _{1}}}].  \notag
\end{eqnarray}

We can repeat the method of constructing solutions for $s=1$ to next shells $%
s=2$ and $s=3$ and integrate the systems (\ref{e4dd}) and (\ref{e5dd})
respectively. It is necessary to extend the formulas for generating and
integration functions and effective sources recurrently as we provided above
for redefinition of respective values for $s=0$ to $s=1.$ The quadratic
elements are parametrized in corresponding forms:
\begin{eqnarray}
ds_{K8d}^{2} &=&g_{\alpha _{2}\beta
_{2}}(x^{k},y^{4},y^{a_{1}},y^{a_{2}})du^{\alpha _{2}}du^{\beta _{2}}
\label{gensolshell2} \\
&=&ds_{K6d}^{2}[\mbox{ see (\ref{gensolshell1})}]+[h_{6}^{[0]}+\frac{1}{4}%
\int dy^{8}\frac{(\ ^{2}\Phi ^{2})^{\ast _{2}}}{\ _{2}\Upsilon }]  \notag \\
&&[dy^{7}+\left( \ _{1}^{2}n_{k_{2}}+\ _{2}^{2}\widetilde{n}_{k_{2}}\int
dy^{8}\frac{(\ \ ^{2}\Phi ^{\ast _{2}})^{2}}{(\ _{2}\Upsilon )^{2}}%
\left\vert h_{7}^{[0]}+\frac{1}{4}\int dy^{8}\frac{(\ ^{2}\Phi ^{2})^{\ast
_{2}}}{\ _{2}\Upsilon }\right\vert ^{-5/2}\right) dx^{k_{2}}]  \notag \\
&&+\frac{1}{4}(\frac{\ \ ^{2}\Phi ^{\ast _{2}}}{\ _{2}\Upsilon })^{2}\left(
h_{7}^{[0]}+\frac{1}{4}\int dy^{8}\frac{(\ ^{2}\Phi ^{2})^{\ast _{2}}}{\
_{2}\Upsilon }\right) ^{-1}[dy^{8}+\frac{\partial _{i_{2}}\ ^{2}\Phi }{\
^{2}\Phi ^{\ast _{2}}}]  \notag
\end{eqnarray}%
and
\begin{eqnarray}
ds_{K10d}^{2} &=&g_{\alpha _{2}\beta
_{2}}(x^{k},y^{4},y^{a_{1}},y^{a_{2}},y^{a_{2}})du^{\alpha _{3}}du^{\beta
_{3}}=ds_{K8d}^{2}[\mbox{ see (\ref{gensolshell2})}]+[h_{8}^{[0]}+\frac{1}{4}%
\int dy^{10}\frac{(\ ^{3}\Phi ^{2})^{\ast _{3}}}{\ _{3}\Upsilon }]
\label{gensolshell3} \\
&&[dy^{9}+\left( \ _{1}^{3}n_{k_{3}}+\ _{2}^{3}\widetilde{n}_{k_{3}}\int
dy^{10}\frac{(\ \ ^{3}\Phi ^{\ast _{3}})^{2}}{(\ _{3}\Upsilon )^{2}}%
\left\vert h_{9}^{[0]}+\frac{1}{4}\int dy^{10}\frac{(\ ^{3}\Phi ^{2})^{\ast
_{3}}}{\ _{3}\Upsilon }\right\vert ^{-5/2}\right) dx^{k_{3}}]  \notag \\
&&+\frac{1}{4}(\frac{\ \ ^{3}\Phi ^{\ast _{3}}}{\ _{3}\Upsilon })^{2}\left(
h_{9}^{[0]}+\frac{1}{4}\int dy^{10}\frac{(\ ^{3}\Phi ^{2})^{\ast _{3}}}{\
_{3}\Upsilon }\right) ^{-1}[dy^{10}+\frac{\partial _{i_{3}}\ ^{3}\Phi }{\
^{3}\Phi ^{\ast _{3}}}].  \notag
\end{eqnarray}

The generic off--diagonal d--metrics (\ref{gensolshell3}) define exact
stationary solutions with Killing symmetries on $\partial _{t}$ and $%
\partial _{9}$ in effective nonholonomic 10-d gravity with sources (\ref%
{effects}) being determined by the motion equations in heterotic
supergravity (\ref{hs1}).

\subsubsection{A nonlinear symmetry of generating functions and effective
sources}

There is an important nonlinear symmetry relating nontrivial generating
functions and effective sources considered in above classes of solutions,
used in differen forms in \cite{svvvey,tgovsv}. It allows us to re-define
the generating functions and introduce effective cosmological constants
instead of effective sources. Let us study these properties in the case of
heterotic string gravity.

Changing the generating data $(\Phi ,\ \Upsilon )\leftrightarrow (\tilde{\Phi%
},\widetilde{\Lambda }=const),$ where%
\begin{eqnarray}
\frac{(\Phi ^{2})^{\ast }}{\Upsilon } &=&\frac{(\widetilde{\Phi }^{2})^{\ast
}}{\widetilde{\Lambda }},\mbox{ which is equivalent to }  \label{nontransf0}
\\
\widetilde{\Phi }^{2} &=&\widetilde{\Lambda }\int dy^{4}\Upsilon ^{-1}(\Phi
^{2})^{\ast }\mbox{ and/or }\Phi ^{2}=\widetilde{\Lambda }^{-1}\int
dy^{4}\Upsilon (\widetilde{\Phi }^{2})^{\ast },  \notag
\end{eqnarray}%
up to certain classes of integration functions depending on coordinates $%
x^{i},$ we express the solutons (\ref{h4}) and (\ref{h3}) of the system of
nonlinear PDEs (\ref{rsa1}) and (\ref{rsa2}) in two equivalent forms,
\begin{equation*}
\ h_{3}=h_{3}[\Phi ]=h_{3}^{[0]}(x^{k})-\frac{1}{4}\int dy^{4}\frac{(\Phi
^{2})^{\ast }}{\Upsilon }=h_{3}[\widetilde{\Phi }]=h_{3}^{[0]}(x^{k})-\frac{%
\widetilde{\Phi }^{2}}{4\widetilde{\Lambda }},
\end{equation*}%
and
\begin{equation*}
\ h_{4}=\ h_{4}[\Phi ]=-\frac{1}{4h_{3}[\Phi ]}(\frac{\ \Phi ^{\ast }}{%
\Upsilon })^{2}=h_{4}[\widetilde{\Phi }]=-\frac{1}{4h_{3}[\widetilde{\Phi }]}%
\frac{(\ \widetilde{\Phi }^{\ast })^{2}}{\widetilde{\Lambda }^{2}}.
\end{equation*}%
Having defined two equivalent nonlinear formulas for $h_{a}[\Phi ]=h_{a}[%
\widetilde{\Phi }],$ we can express in two equvalent forms the N--adapted
coefficients for the d--metric and N-connection (\ref{solut1t}). In the
second case, the N--connection coefficients are computed using integrals on $%
dy^{4}$ for certain values determined by $(\tilde{\Phi},\widetilde{\Lambda }%
) $ via (\ref{nontransf0}) and integration functions,
\begin{equation*}
n_{k}=\ _{1}n_{k}+\ _{2}n_{k}\int dy^{4}\ \frac{h_{4}[\widetilde{\Phi }]}{|\
h_{3}[\widetilde{\Phi }]|^{3/2}}\mbox{ and }w_{i}=\frac{\partial _{i}\ \Phi
}{\Phi ^{\ast }}=\frac{\partial _{i}\ \Phi ^{2}}{(\Phi ^{2})^{\ast }}=\frac{%
\partial _{i}\ \int dy^{4}\Upsilon \ (\widetilde{\Phi }^{2})^{\ast }}{%
\Upsilon (\widetilde{\Phi }^{2})^{\ast }}.
\end{equation*}%
We observe that if the effective source $\Upsilon =\Upsilon (x^{k})$ does
not depend on $y^{4},$ we have the same expression for $w_{i}$ in terms of
generating functions, $w_{i}=\frac{\partial _{i}\ \Phi }{\Phi ^{\ast }}=%
\frac{\partial _{i}\ \widetilde{\Phi }}{\widetilde{\Phi }^{\ast }}.$

The nonlinear symmetry reflects the property of changing the effective
sources mentioned in (\ref{effects}),
\begin{equation}
\Upsilon (x^{k},y^{4})\rightarrow \ \Lambda =\ \ ^{H}\Lambda +\ ^{F}\Lambda
+\ ^{R}\Lambda ,\mbox{ for }\ ^{\phi }\Lambda =0.  \label{sourc4d}
\end{equation}%
We can indentify $\widetilde{\Lambda }$ with $\Lambda ,$ or any other value $%
\ \ ^{H}\Lambda ,\ ^{F}\Lambda ,\ ^{R}\Lambda ,$ and their sums, depending
on the class of models with effective gauge interactions we consider in our
work.

In a similar form, using recurrent formulas, we can prove the existence of
such nonlinear symmetries generalizing (\ref{nontransf0}), {\small
\begin{eqnarray}
s &=&0:\frac{(\Phi ^{2})^{\ast }}{\Upsilon }=\frac{(\widetilde{\Phi }%
^{2})^{\ast }}{\widetilde{\Lambda }},\mbox{ i.e. }\widetilde{\Phi }^{2}=%
\widetilde{\Lambda }\int dy^{4}\Upsilon ^{-1}(\Phi ^{2})^{\ast }%
\mbox{
and/or }\Phi ^{2}=\widetilde{\Lambda }^{-1}\int dy^{4}\Upsilon (\widetilde{%
\Phi }^{2})^{\ast };  \label{nonltransf} \\
s &=&1:\frac{(\ ^{1}\Phi ^{2})^{\ast _{1}}}{\ _{1}\Upsilon }=\frac{(\ ^{1}%
\widetilde{\Phi }^{2})^{\ast _{1}}}{\ _{1}\widetilde{\Lambda }},\mbox{ i.e. }%
\ ^{1}\widetilde{\Phi }^{2}=\ _{1}\widetilde{\Lambda }\int dy^{6}(\
_{1}\Upsilon )^{-1}(\ ^{1}\Phi ^{2})^{\ast _{1}}\mbox{ and/or }(\ ^{1}\Phi
)^{2}=\ _{1}\widetilde{\Lambda }^{-1}\int dy^{6}\ _{1}\Upsilon (\ ^{1}%
\widetilde{\Phi }^{2})^{\ast _{1}};  \notag \\
s &=&2:\frac{(\ ^{2}\Phi ^{2})^{\ast _{2}}}{\ _{2}\Upsilon }=\frac{(\ ^{2}%
\widetilde{\Phi }^{2})^{\ast _{2}}}{\ _{2}\widetilde{\Lambda }},\mbox{ i.e. }%
\ ^{2}\widetilde{\Phi }^{2}=\ _{2}\widetilde{\Lambda }\int dy^{8}(\
_{2}\Upsilon )^{-1}(\ ^{2}\Phi ^{2})^{\ast _{2}}\mbox{ and/or }(\ ^{2}\Phi
)^{2}=\ _{2}\widetilde{\Lambda }^{-1}\int dy^{8}\ _{2}\Upsilon (\ ^{2}%
\widetilde{\Phi }^{2})^{\ast _{2}};  \notag \\
s &=&3:\frac{(\ ^{3}\Phi ^{2})^{\ast _{3}}}{\ _{3}\Upsilon }=\frac{(\ ^{3}%
\widetilde{\Phi }^{2})^{\ast _{3}}}{\ _{3}\widetilde{\Lambda }},\mbox{ i.e. }%
\ ^{3}\widetilde{\Phi }^{2}=\ _{3}\widetilde{\Lambda }\int dy^{10}(\
_{3}\Upsilon )^{-1}(\ ^{3}\Phi ^{2})^{\ast _{3}}\mbox{ and/or }(\ ^{3}\Phi
)^{2}=\ _{3}\widetilde{\Lambda }^{-1}\int dy^{10}\ _{3}\Upsilon (\ ^{3}%
\widetilde{\Phi }^{2})^{\ast _{3}}.  \notag
\end{eqnarray}
} We consider the convention that $\widehat{\mathbf{\Upsilon }}_{9}^{9}=%
\widehat{\mathbf{\Upsilon }}_{10}^{10}=\ _{3}\Upsilon
(x^{k},y^{a},y^{a_{1}},y^{a_{2}},y^{10})\rightarrow \ _{3}\Lambda =\
_{3}^{\phi }\Lambda +\ _{3}^{H}\Lambda +\ _{3}^{F}\Lambda +\ _{3}^{R}\Lambda
$ and identify $\ _{3}\widetilde{\Lambda }$ with $\ _{3}\Lambda .$ Such
nonlinear transforms can be used for simplifications of formulas for generic
off--diagonal solutions. Prescribing certain effective matter field initial
distributions, we can re--define the constructions equivalently, via new
classes of generating functions, for effective cosmological constants.

Finally, we note that formula (\ref{gensolshell3}) simplifies for $\ \left(
\ ^{3}\Phi ,\ _{3}\Upsilon \right) \rightarrow \left( \ ^{3}\widetilde{\Phi }%
,\ \ _{3}\widetilde{\Lambda }\right) $ if $\ _{3}\Upsilon =\ _{3}\Upsilon
(x^{i},y^{4},y^{a_{1}},y^{a_{2}}),$%
\begin{eqnarray*}
&&ds_{K10d}^{2} =g_{\alpha _{2}\beta
_{2}}(x^{k},y^{4},y^{a_{1}},y^{a_{2}},y^{a_{2}},\ ^{3}\widetilde{\Phi },_{3}%
\widetilde{\Lambda })du^{\alpha _{3}}du^{\beta _{3}} \\
&&=ds_{K8d}^{2}[\mbox{ see (\ref{gensolshell2})}]+[h_{8}^{[0]}+\frac{(\ ^{3}%
\widetilde{\Phi }^{2})}{4\ \ \ _{3}\widetilde{\Lambda }}] [dy^{9}+\left( \
_{1}^{3}n_{k_{3}}+\ _{2}^{3}\widetilde{n}_{k_{3}}\int dy^{10}(\ \ ^{3}%
\widetilde{\Phi }^{\ast _{3}})^{2}\left\vert h_{9}^{[0]}+\frac{(\ ^{3}%
\widetilde{\Phi }^{2})}{4\ \ \ _{3}\widetilde{\Lambda }}\right\vert
^{-5/2}\right) dx^{k_{3}}] \\
&&+\frac{1}{4}\frac{\ (\ ^{3}\widetilde{\Phi }^{\ast _{3}})^{2}}{|\ _{3}%
\widetilde{\Lambda }\ _{3}\Upsilon |}\left( h_{9}^{[0]}+\frac{(\ ^{3}%
\widetilde{\Phi }^{2})}{4\ \ \ _{3}\widetilde{\Lambda }}\right)
^{-1}[dy^{10}+\frac{\partial _{i_{3}}\ ^{3}\widetilde{\Phi }}{\ ^{3}%
\widetilde{\Phi }^{\ast _{3}}}].
\end{eqnarray*}%
Such generic off--diagonal configurations are described by generic
off--diagonal metrics with effective cosmological constants, generalized
generating functions and integration functions. The contributions of $\
_{3}\Upsilon (x^{i},y^{4},y^{a_{1}},y^{a_{2}})$ can be encoded into
generating and integrating functions.

\subsubsection{ The Levi--Civita conditions}

\label{sslc}In general, a solution constructed for a generic off--diagonal
metric (\ref{ansk}) and canonical d--connections $\ ^{s}\widehat{\mathbf{D}}$
is characterized by nonholonomically induced d--torsion coefficients $%
\widehat{\mathbf{T}}_{\ \alpha _{s}\beta _{s}}^{\gamma _{s}}\ $(\ref{dtorss}%
) completely defined by the N--connection and d--metric structure. The zero
torsion conditions (\ref{zerotors}) can be satified by a subclass of
nonholonomic distributions determined by corresponding parametrizations of
the generating and integration functions and sources. By straightforward
computations (see details in Refs. \cite{vex1,vpars,vex2,vex3}), we can
verify that if the coefficients of N--adapted frames and $^{s}v$--components
of d--metrics are subjected to respective conditions,
\begin{eqnarray}
s &=&0:\ w_{i}^{\ast }=\mathbf{e}_{i}\ln \sqrt{|\ h_{4}|},\mathbf{e}_{i}\ln
\sqrt{|\ h_{3}|}=0,\partial _{i}w_{j}=\partial _{j}w_{i}\mbox{ and }%
n_{i}^{\ast }=0;  \notag \\
s &=&1:\ ^{1}w_{i_{1}}^{\ast _{1}}=\ ^{1}\mathbf{e}_{i_{1}}\ln \sqrt{|\
h_{6}|},\ ^{1}\mathbf{e}_{i_{1}}\ln \sqrt{|\ h_{5}|}=0,\partial _{i_{1}}\
^{1}w_{j_{1}}=\partial _{j_{1}}\ ^{1}w_{i_{1}}\mbox{ and }\
^{1}n_{i_{1}}^{\ast _{1}}=0;  \label{zerot} \\
s &=&2:\ ^{2}w_{i_{2}}^{\ast _{2}}=\ ^{2}\mathbf{e}_{i_{2}}\ln \sqrt{|\
h_{8}|},\ ^{2}\mathbf{e}_{i_{2}}\ln \sqrt{|\ h_{7}|}=0,\partial _{i_{2}}\
^{2}w_{j_{2}}=\partial _{j_{2}}\ ^{2}w_{i_{2}}\mbox{ and }\
^{2}n_{i_{2}}^{\ast _{2}}=0;  \notag \\
s &=&3:\ ^{3}w_{i_{3}}^{\ast _{3}}=\ ^{3}\mathbf{e}_{i_{3}}\ln \sqrt{|\
h_{10}|},\ ^{3}\mathbf{e}_{i_{3}}\ln \sqrt{|\ h_{9}|}=0,\partial _{i_{3}}\
^{3}w_{j_{3}}=\partial _{j_{3}}\ ^{3}w_{i_{3}}\mbox{ and }\
^{3}n_{i_{3}}^{\ast _{3}}=0;  \notag
\end{eqnarray}%
all d-torsion coefficients are zero.

The $n$--coefficients solve the conditions (\ref{zerot}) if
\begin{eqnarray}
s &=&0:\ _{2}n_{k}(x^{i})=0\mbox{ and }\partial _{i}\
_{1}n_{j}(x^{k})=\partial _{j}\ _{1}n_{i}(x^{k});  \label{expcondn} \\
s &=&1:\ _{2}^{1}n_{k_{1}}(x^{i_{1}})=0\mbox{ and }\partial _{i_{1}}\
_{1}^{1}n_{j_{1}}(x^{k_{1}})=\partial _{j_{1}}\ _{1}^{1}n_{i_{1}}(x^{k_{1}});
\notag \\
s &=&2:\ _{2}^{2}n_{k_{2}}(x^{i_{2}})=0\mbox{ and }\partial _{i_{2}}\
_{1}^{2}n_{j_{2}}(x^{k_{2}})=\partial _{j_{2}}\ _{1}^{2}n_{i_{2}}(x^{k_{2}});
\notag \\
s &=&3:\ _{2}^{3}n_{k_{3}}(x^{i_{3}})=0\mbox{ and }\partial _{i_{3}}\
_{1}^{3}n_{j_{3}}(x^{k_{3}})=\partial _{j_{3}}\ _{1}^{3}n_{i_{3}}(x^{k_{3}}).
\notag
\end{eqnarray}%
$\ $The explicit form of solutions of constraints on $w_{k}$ derived from (%
\ref{zerot}) depend on the class of vacuum or non--vacuum metrics we try to
construct, see details in \cite{tgovsv}. For instance, if we choose a
generating function $\Phi =\check{\Phi}(x^{i},y^{4}),$ for which $(\partial
_{i}\check{\Phi})^{\ast }=\partial _{i}\check{\Phi}^{\ast },$we solve the
conditions for $w_{i}$ in (\ref{zerot}) in explicit form if $\Upsilon
=const, $ or if $\ $such an effective source can be expressed as a
functional $\ \Upsilon (x^{i},y^{4})=\Upsilon \lbrack \check{\Phi}].$ The
third condition for $s=0$, $\partial _{i}w_{j}=\partial _{j}w_{i},$ can be
satisfied for any generating function $\check{A}=\check{A}(x^{k},y^{4})$ for
which $\ w_{i}=\check{w}_{i}=\partial _{i}\check{\Phi}/\partial _{4}\check{%
\Phi}=\partial _{i}\check{A}.$ Following similar considerations for shells $%
s=1,2,3,$ we formulate the LC-conditions for generating functions
\begin{eqnarray}
s=0: &&\Phi =\check{\Phi}(x^{i},y^{4}),(\partial _{i}\check{\Phi})^{\ast
}=\partial _{i}\check{\Phi}^{\ast },\check{w}_{i}=\partial _{i}\check{\Phi}%
/\partial _{4}\check{\Phi}=\partial _{i}\check{A};  \label{expconda} \\
&&\Upsilon (x^{i},y^{4})=\Upsilon \lbrack \check{\Phi}],\mbox{ or }\Upsilon
=const;  \notag \\
s=1: &&\ ^{1}\Phi =\ ^{1}\check{\Phi}(u^{\tau },y^{6}),(\partial _{i_{1}}\
^{1}\check{\Phi})^{\ast _{1}}=\partial _{i_{1}}\ ^{1}\check{\Phi}^{\ast
_{1}},\partial _{\alpha }\ ^{1}\check{\Phi}/\partial _{6}\ ^{1}\check{\Phi}%
=\partial _{\alpha }\ ^{1}\check{A};\ _{1}^{1}n_{\tau }=\partial _{\tau }\
^{1}n(u^{\beta });  \notag \\
&&\Upsilon (x^{i},y^{4})=\Upsilon \lbrack \check{\Phi}],\mbox{ or }\Upsilon
=const;  \notag \\
s=2: &&\ ^{2}\Phi =\ ^{2}\check{\Phi}(u^{\tau _{1}},y^{8}),\partial
_{8}\partial _{\tau _{1}}\ ^{2}\check{\Phi}=\partial _{\tau _{1}}\partial
_{8}\ ^{2}\check{\Phi};\partial _{\alpha _{1}}\ ^{2}\check{\Phi}/\partial
_{8}\ ^{2}\check{\Phi}=\partial _{\alpha _{2}}\ ^{2}\check{A};\
_{1}^{2}n_{\tau _{1}}=\partial _{\tau _{1}}\ ^{2}n(u^{\beta _{1}});  \notag
\\
&&\Upsilon (x^{i},y^{4})=\Upsilon \lbrack \check{\Phi}],\mbox{ or }\Upsilon
=const;  \notag \\
s=3: &&\ ^{2}\Phi =\ ^{2}\check{\Phi}(u^{\tau _{1}},y^{8}),\partial
_{8}\partial _{\tau _{1}}\ ^{2}\check{\Phi}=\partial _{\tau _{1}}\partial
_{8}\ ^{2}\check{\Phi};\partial _{\alpha _{1}};\ ^{2}\check{\Phi}/\partial
_{8}\ ^{2}\check{\Phi}=\partial _{\alpha _{2}}\ ^{2}\check{A};\
_{1}^{2}n_{\tau _{1}}=\partial _{\tau _{1}}\ ^{2}n(u^{\beta _{1}});  \notag
\\
&&\Upsilon (x^{i},y^{4})=\Upsilon \lbrack \check{\Phi}],\mbox{ or }\Upsilon
=const;  \notag
\end{eqnarray}

Imposing respective conditions from (\ref{expcondn}) and (\ref{expconda}) on
coefficients of (\ref{solut1t}), we define such a class of quadratic
elements for off--diagonal 4-d stationary configurations in heterotic
supergravity with zero induced torsion,
\begin{eqnarray}
ds_{K4d}^{2}&=&\check{g}_{\alpha \beta }(x^{k},y^{4})du^{\alpha }du^{\beta
}=e^{q}[(dx^{1})^{2}+(dx^{2})^{2}]+ [h_{3}^{[0]}(x^{k})-\frac{1}{4}\int
dy^{4}\frac{(\check{\Phi}^{2})^{\ast }}{\Upsilon \lbrack \check{\Phi}]}%
][dt+(\partial _{k}\ n)dx^{k}]^{2}  \notag \\
&&-\frac{1}{4}(\frac{\check{\Phi}^{\ast }}{\Upsilon \lbrack \check{\Phi}]}%
)^{2}\left( h_{3}^{[0]}-\frac{1}{4}\int dy^{4}\frac{(\check{\Phi}^{2})^{\ast
}}{\Upsilon \lbrack \check{\Phi}]}\right) ^{-1}[dy^{4}+(\partial _{i}\check{A%
})dx^{i}]^{2}.  \label{4dlc}
\end{eqnarray}
We can impose similar conditions and generate exact off--diagonal solutions
with respective data $(\ ^{s}\mathbf{\check{g},}\ ^{s}\mathbf{\check{N},}\
^{s}\mathbf{\check{\nabla}})$ for which all nonholonomic torsions are zero,
but this will impose very strong restrictions on the dynamics of effective
matter fields on extra shells in heterotic supergravity. In order to
consider realistic solutions in string gravity with 6-d interior almost-K%
\"{a}hler configurations, the torsion is positively not zero. Such 10-d
solutions are conventionally parameterized as $(s=0:\check{g},N,\nabla
;s=1,2,3:\ ^{s}\mathbf{g,}\ ^{s}\mathbf{N,}\ ^{s}\widehat{\mathbf{D}}),$ for
stationary configurations with Killing symmetry on $\partial _{t}$ and $%
\partial _{9}.$ Hereafter we shall work with solutions with nontrivial
nonholonomically induced torsions considering that it is always possible to
state additional constraints resulting in LC-configurations in 4-d or extra
dimensions.

\subsection{ Small N--adapted nonholonomic stationary deformations}

\label{ssedef}We can construct very general classes of generic off--diagonal
stationary solutions in heterotic supergravity. It is not clear what
physical meaning these configurations may have for generating and
integration functions with arbitrary data. Using small off--diagonal
deformations of some known physically important solutons we can understand
physical properties of such solutions characterized by locally anisotropic
polarization/ running of constants and nonlinear off-diagonal gravitational
interactions determined by (super) string corrections.

We consider a "prime" pseudo--Riemannian metric of type $\mathbf{\mathring{g}%
}=[\mathring{g}_{i},\mathring{h}_{a_{s}},\mathring{N}_{b_{s}}^{j_{s}}]$ when
\begin{eqnarray}
ds^{2} &=&\mathring{g}_{i}(x^{k})(dx^{i})^{2}+\mathring{h}%
_{a}(x^{k},y^{4})(dy^{a})^{2}(\mathbf{\mathring{e}}^{a})^{2}+\mathring{g}%
_{a_{1}}(x^{k},y^{4},y^{6})\left( \mathbf{\mathring{e}}^{a_{1}}\right) ^{2}
\label{pmc1c} \\
&&+\mathring{g}_{a_{2}}(x^{k},y^{4},y^{a_{1}},y^{8})\left( \mathbf{\mathring{%
e}}^{a_{2}}\right) ^{2}+\mathring{g}%
_{a_{3}}(x^{k},y^{4},y^{a_{1}},y^{a_{2}},y^{10})\left( \mathbf{\mathring{e}}%
^{a_{3}}\right) ^{2},  \notag
\end{eqnarray}%
where $\mathbf{\mathring{e}}^{a_{s}}$ as N--elongated differentials. Such a
metric is diagonalizable if there is a coordinate transform $u^{\alpha
_{s}^{\prime }}=u^{\alpha _{s}^{\prime }}(u^{\alpha _{s}})$ when  $ds^{2}=%
\mathring{g}_{i^{\prime }}(x^{k\prime })(dx^{i^{\prime }})^{2}+\mathring{h}%
_{a_{s}^{\prime }}(x^{k\prime })(dy^{a_{s}^{\prime }})^{2}$,  with $\ ^{s}%
\mathring{w}_{i_{s}}=\ ^{s}\mathring{n}_{i_{s}}=0.$ To construct exact
solutions with non--singular coordinates it is important to work with
"formal" off--diagonal parameterizations when the coefficients $\ ^{s}%
\mathring{w}_{i_{s}}$ and/or $^{s}\mathring{n}_{i_{s}}$ are not zero but the
anholonomy coefficients $\mathring{W}_{\beta _{s}\gamma _{s}}^{\alpha
_{s}}(u^{\mu _{s}})=0,$ see (\ref{anhrel}). We suppose that some data $(%
\mathring{g}_{i^{\prime }},\mathring{h}_{a_{s}^{\prime }})$ may define a
known physically important diagonal exact solution in GR or heterotic string
gravity (for instance, a black hole, BH, configuration of Kerr ors
Schwarzshild type). Our goal is to study certain small generic off--diagonal
parametric deformations of the prime d--metric and N--connection
coefficients (\ref{pmc1c}) into certain target metrics
\begin{eqnarray}
ds^{2} &=&\eta _{i}\mathring{g}_{i}(dx^{i})^{2}+\eta _{a_{s}}\mathring{g}%
_{a_{s}}(\mathbf{e}^{a_{s}})^{2},  \label{targm} \\
\mathbf{e}^{3} &=&dt+\ ^{n}\eta _{i}\mathring{n}_{i}dx^{i},\mathbf{e}%
^{4}=dy^{4}+\ ^{w}\eta _{i}\mathring{w}_{i}dx^{i},\ \mathbf{e}^{5}=dy^{5}+\
^{n}\eta _{i_{1}}\mathring{n}_{i_{1}}dx^{i_{1}},\mathbf{e}^{6}=dy^{6}+\
^{w}\eta _{i_{1}}\mathring{w}_{i_{1}}dx^{i_{1}},  \notag \\
\mathbf{e}^{7} &=&dy^{7}+\ ^{n}\eta _{i_{2}}\mathring{n}_{i_{2}}dx^{i_{2}},%
\mathbf{e}^{8}=dy^{8}+\ ^{w}\eta _{i_{2}}\mathring{w}_{i_{s}}dx^{i_{s}},\
\mathbf{e}^{9}=dy^{9}+\ ^{n}\eta _{i_{3}}\mathring{n}_{i_{3}}dx^{i_{3}},%
\mathbf{e}^{10}=dy^{10}+\ ^{w}\eta _{i_{3}}\mathring{w}_{i_{3}}dx^{i_{3}},
\notag
\end{eqnarray}%
where the coefficients $(g_{\alpha _{s}}=\eta _{\alpha _{s}}\mathring{g}%
_{\alpha _{s}},^{w}\eta _{i_{s}}\mathring{w}_{i_{s}},\ ^{n}\eta
_{i_{s}}n_{i_{s}})$ define, for instance, a d--metric $^{s}\mathbf{g}$%
\textbf{\ }(\ref{ansk}) as a solution of nonholonomic motion equations in
heterotic string gravity (\ref{hs1}) rewritten as a nonlinear system of PDEs
in nonholonomic 10-d gravity (\ref{e1})-(\ref{e5dd}).

Let us construct exact solutions, for instance, of type (\ref{gensolshell3}%
). For certain well--defined conditions, we can express using d-metric and
N--connection deformations stated in explicit form on all shells for the
d--metric and $n$- and $w$-coefficients:
\begin{eqnarray}
&&\mbox{For the coefficients of d-metrics,  }  \label{smpolariz} \\
\eta _{i} &=&1+\varepsilon \chi _{i}(x^{k}),\eta _{a}=1+\varepsilon \chi
_{a}(x^{k},y^{4}),\eta _{a_{1}}=1+\varepsilon \chi
_{a_{1}}(x^{k},y^{4},y^{6}),  \notag \\
\eta _{a_{2}} &=&1+\varepsilon \chi
_{a_{2}}(x^{k},y^{4},y^{a_{1}},y^{8}),\eta _{a_{3}}=1+\varepsilon \chi
_{a_{3}}(x^{k},y^{4},y^{a_{1}},y^{a_{2}},y^{10});  \notag \\
&&\mbox{ and  for the coeffisients of N-connection,}  \notag \\
\ ^{n}\eta _{i} &=&1+\varepsilon \ \ ^{n}\eta _{i}(x^{k},y^{4}),\ ^{w}\eta
_{i}=1+\varepsilon \ ^{w}\chi _{i}(x^{k},y^{4}),  \notag \\
\ ^{n}\eta _{i_{1}} &=&1+\varepsilon \ \ ^{n}\eta
_{i_{1}}(x^{k},y^{4},y^{6}),\ ^{w}\eta _{i_{1}}=1+\varepsilon \ ^{w}\chi
_{i_{1}}(x^{k},y^{4},y^{6}),  \notag \\
\ ^{n}\eta _{i_{2}} &=&1+\varepsilon \ \ ^{n}\eta
_{i_{2}}(x^{k},y^{4},y^{a_{1}},y^{8}),\ ^{w}\eta _{i_{2}}=1+\varepsilon \
^{w}\chi _{i_{2}}(x^{k},y^{4},y^{a_{1}},y^{8}),  \notag \\
\ ^{n}\eta _{i_{3}} &=&1+\varepsilon \ \ ^{n}\eta
_{i_{3}}(x^{k},y^{4},y^{a_{1}},y^{a_{2}},y^{10}),\ ^{w}\eta
_{i_{3}}=1+\varepsilon \ ^{w}\chi
_{i_{3}}(x^{k},y^{4},y^{a_{1}},y^{a_{2}},y^{10}),  \notag
\end{eqnarray}%
for a small parameter $0\leq \varepsilon \ll 1,$ when (\ref{targm})
transforms into (\ref{pmc1c}) for $\varepsilon \rightarrow 0$ (which in
turn, can be diagonalized). In general, there are no smooth limits from such
nonholonomic deformations which can be satisfied for arbitrary generation
and integration functions, integration constants and general (effective)
sources on corresponding shells. The goal of this subsection is to analyze
such conditions when $\varepsilon $-deformations with nontrivial
N--connection coefficients for the prime and target d--metrics can be
related to new classes of solutions of motion heterotic string equations.

We denote nonholonomic $\varepsilon $--deformations of certain prime
d--metric (\ref{pmc1c}) into a target one (\ref{targm}) with polarizations (%
\ref{smpolariz}) in the form $\mathbf{\mathring{g}}\rightarrow \
_{\varepsilon }\mathbf{g=(\ }_{\varepsilon }g_{i},\ _{\varepsilon
}h_{a_{s}},\ _{\varepsilon }N_{b_{s}}^{a_{s-1}}).$ The goal of this
subsection is to compute the formulas for $\varepsilon $--deformations of
prime d--metrics resulting in stationary solutions in equivalent
nonholonomic 10-d gravity with Killing symmetries on $\partial _{t}$ and $%
\partial _{9}.$

The geometric constructions will be provided in detalis for 4-d
configurations and then extended to higher dimensional shells. Deformations
of $h$-components are characterized by $\ _{\varepsilon }g_{i}=\mathring{g}%
_{i}(1+\varepsilon \chi _{i})=e^{q(x^{k})}$ resulting in a solution of the
2-d Laplace equation (\ref{e1}). For $\ q=\ \ ^{0}q(x^{k})+\varepsilon \
^{1}q(x^{k})$ and $\ _{h}\Upsilon =\ _{h}^{0}\Upsilon (x^{k})+\ _{h}^{\tilde{%
1}}\overline{\Upsilon }(x^{k}),$ we compute the deformation polarization
functions $\chi _{i}=e^{\ ^{0}q}\ ^{1}q/\mathring{g}_{i}\ \ _{h}^{0}\Upsilon
.$ In this formula, we use certain generating and source functions as
solutions of $\ ^{0}q^{\bullet \bullet }+\ ^{0}q^{\prime \prime }=\
^{0}\Upsilon$ and $\ ^{1}q^{\bullet \bullet }+\ ^{1}q^{\prime \prime }=\ \
^{1}\Upsilon $.

At the next step, we compute $\varepsilon $--deformations of $v$--components
on the $s=0$ shell,
\begin{eqnarray}
\ _{\varepsilon }h_{3} &=&h_{3}^{[0]}(x^{k})-\frac{1}{4}\int dy^{4}\frac{%
(\Phi ^{2})^{\ast }}{\ \Upsilon }=(1+\varepsilon \chi _{3})\mathring{g}_{3},
\label{h3b} \\
\ _{\varepsilon }h_{4} &=&-\frac{1}{4}\frac{(\ \Phi ^{\ast })^{2}}{\
\Upsilon ^{2}}\left( h_{3}^{[0]}-\frac{1}{4}\int dy^{4}\frac{(\Phi
^{2})^{\ast }}{\Upsilon }\right) ^{-1}=(1+\varepsilon \chi _{4})\mathring{g}%
_{4};  \label{h4b}
\end{eqnarray}%
$\ $Parametrizing the generation function
\begin{equation}
\ \Phi \rightarrow \ _{\varepsilon }\Phi =\mathring{\Phi}(x^{k},y^{4})[1+%
\varepsilon \chi (x^{k},y^{4})],  \label{epsdefgf}
\end{equation}%
and introducing this value in (\ref{h4b}), one obtains
\begin{equation}
\chi _{3}=-\frac{1}{4\mathring{g}_{3}}\int dy^{4}\frac{(\mathring{\Phi}%
^{2}\chi )^{\ast }}{\Upsilon }\mbox{ and }\int dy^{4}\frac{(\mathring{\Phi}%
^{2})^{\ast }}{\Upsilon }=4(h_{3}^{[0]}-\mathring{g}_{3}).  \label{cond2a}
\end{equation}%
In result, we can compute $\chi _{3}$ for any deformation $\chi $ from a
2-hypersurface $y^{4}=y^{4}(x^{k}).$ Such a hypersurface, in general, is
defined in non-explicit form from $\mathring{\Phi}=\mathring{\Phi}%
(x^{k},y^{4})$ when the integration function $h_{3}^{[0]}(x^{k}),$ the prime
value $\mathring{g}_{3}(x^{k})$ and the fraction $(\mathring{\Phi}%
^{2})^{\ast }/\Upsilon $ satisfy the condition (\ref{cond2a}).

We can find the formula for hypersurface $\mathring{\Phi}(x^{k},y^{4})$
prescribing a value of $\Upsilon .$ Introducing (\ref{aux5}) into (\ref{h4b}%
), one obtains%
\begin{equation*}
\chi _{4}=2(\chi +\frac{\mathring{\Phi}}{\mathring{\Phi}^{\ast }}\chi ^{\ast
})-\chi _{3}=2(\chi +\frac{\mathring{\Phi}}{\mathring{\Phi}^{\ast }}\chi
^{\ast })+\frac{1}{4\mathring{g}_{3}}\int dy^{4}\frac{(\mathring{\Phi}%
^{2}\chi )^{\ast }}{\ \Upsilon },
\end{equation*}%
i.e. we compute $\chi _{4}$ for any data $\left( \mathring{\Phi},\mathring{g}%
_{3},\chi \right) .$ The formula for a compatible source is $\ _{h}\Upsilon
=\pm \mathring{\Phi}^{\ast }/2\sqrt{|\mathring{g}_{4}h_{3}^{[0]}|}.$ It
transforms (\ref{cond2a}) into a 2-d hypersurface formula $%
y^{4}=y^{4}(x^{k}) $ defined in non-explicit form from
\begin{equation}
\int dy^{4}\mathring{\Phi}=\pm (h_{3}^{[0]}-\mathring{g}_{3})/\sqrt{|%
\mathring{g}_{4}h_{3}^{[0]}|}.  \label{cond2b}
\end{equation}%
The $\varepsilon $--deformations of N--connection coefficients $%
w_{i}=\partial _{i}\Phi /\Phi ^{\ast }$ for nontrivial $\ \mathring{w}%
_{i}=\partial _{i}\mathring{\Phi}/\mathring{\Phi}^{\ast }$ are found
following formulas (\ref{aux5}) and (\ref{smpolariz}), $\ ^{w}\chi _{i}=%
\frac{\partial _{i}(\chi \ \mathring{\Phi})}{\partial _{i}\ \mathring{\Phi}}-%
\frac{(\chi \ \mathring{\Phi})^{\ast }}{\mathring{\Phi}^{\ast }},$ where
there is no summation on index $i.$ We can compute the deformations of the $n
$--coefficients (we omit such details, see below necessary formulas).

In a similar way, the $\varepsilon $--deformations of $\ ^{1}v$--components
on the $s=1$ shell are computed,
\begin{equation*}
\ _{\varepsilon }h_{5} = h_{5}^{[0]}(x^{k},y^{4})+ \int dy^{6}\frac{(\
^{1}\Phi ^{2})^{\ast _{1}}}{4\ _{1}\Upsilon }=(1+\varepsilon \chi _{5})%
\mathring{g}_{5}, \ _{\varepsilon }h_{6} = -\frac{(\ ^{1}\Phi ^{\ast
_{1}})^{2}}{4\ _{1}\Upsilon ^{2}}\left( h_{5}^{[0]}+\frac{1}{4}\int dy^{6}%
\frac{(\ ^{1}\Phi ^{2})^{\ast _{1}}}{\ _{1}\Upsilon }\right)
^{-1}=(1+\varepsilon \chi _{6})\mathring{g}_{6};
\end{equation*}%
The first shell generation function is parameterized $\ \ ^{1}\Phi
\rightarrow \ _{\varepsilon }^{1}\Phi =\ ^{1}\mathring{\Phi}%
(x^{k},y^{4},y^{6})[1+\varepsilon \ ^{1}\chi (x^{k},y^{4},y^{6})],$
resulting in
\begin{equation*}
\chi _{5}=-\frac{1}{4\mathring{g}_{5}}\int dy^{6}\frac{(\ ^{1}\mathring{\Phi}%
^{2}\ ^{1}\chi )^{\ast _{1}}}{\ _{1}\Upsilon }\mbox{ and }\int dy^{6}\frac{%
(\ ^{1}\mathring{\Phi}^{2})^{\ast _{1}}}{\ _{1}\Upsilon }=4(h_{5}^{[0]}-%
\mathring{g}_{5}).
\end{equation*}%
It is possible to compute $\chi _{5}$ for any deformation $\ ^{1}\chi $ from
a 3-hypersurface $y^{6}=y^{6}(x^{k},y^{4})$ (for stationary solutions with
Killing symmetries on $\partial _{t}$ and $\partial _{5}).$ We note that, in
general, such a hypersurface is defined in non-explicit form from $\ ^{1}%
\mathring{\Phi}=\ ^{1}\mathring{\Phi}(x^{k},y^{4},y^{6})$ when the
integration function $h_{5}^{[0]}(x^{k},y^{4}),$ the prime value $\mathring{g%
}_{5}(x^{k},y^{4})$ and the fraction $(\ ^{1}\mathring{\Phi}^{2})^{\ast
_{1}}/\ _{1}\Upsilon $ satisfy a condition similar to (\ref{cond2a}).

We can find the formula for hypersurface $\ ^{1}\mathring{\Phi}%
(x^{k},y^{4},y^{6})$ by prescribing a first shell value of effective source $%
\ _{1}\Upsilon .$ Similarly to (\ref{aux5}) and (\ref{h4b}), it is possible
to generalize and compute%
\begin{equation*}
\chi _{6}=2(\ ^{1}\chi +\frac{\ ^{1}\mathring{\Phi}}{\ ^{1}\mathring{\Phi}%
^{\ast _{1}}}\ ^{1}\chi ^{\ast _{1}})-\chi _{5}=2(\ ^{1}\chi +\frac{\ ^{1}%
\mathring{\Phi}}{\ ^{1}\mathring{\Phi}^{\ast _{1}}}\ ^{1}\chi ^{\ast _{1}})+%
\frac{1}{4\mathring{g}_{5}}\int dy^{6}\frac{(\ ^{1}\mathring{\Phi}^{2}\
^{1}\chi )^{\ast _{1}}}{\ _{1}\Upsilon },
\end{equation*}%
i.e. we compute $\chi _{6}$ for any data $\left( \ ^{1}\mathring{\Phi},%
\mathring{g}_{5},^{1}\chi \right) .$ $\ $One defines a first shell
compatible source if $\ \ _{1}\Upsilon =\pm \ ^{1}\mathring{\Phi}^{\ast
_{1}}/2\sqrt{|\mathring{g}_{6}h_{5}^{[0]}|}.$ It generalizes (\ref{cond2a})
into a 2-d hypersurface formula $y^{6}=y^{6}(x^{k},y^{5})$ which has to be
computed in non-explicit form from $\int dy^{6}\ ^{1}\mathring{\Phi}=\pm
(h_{5}^{[0]}-\mathring{g}_{5})/\sqrt{|\mathring{g}_{6}h_{5}^{[0]}|}.$ On the
first shell, the $\varepsilon $--deformations of N--connection coefficients $%
w_{i_{1}}=\partial _{i_{1}}\ ^{1}\Phi /\ ^{1}\Phi ^{\ast _{1}}$ for
nontrivial $\ \mathring{w}_{i_{1}}=\partial _{i_{1}}\ ^{1}\mathring{\Phi}/\
^{1}\mathring{\Phi}^{\ast _{1}}$ are $\ ^{w}\chi _{i_{1}}=\frac{\partial
_{i_{1}}(\ ^{1}\chi \ \ ^{1}\mathring{\Phi})}{\partial _{i_{1}}\ \ ^{1}%
\mathring{\Phi}}-\frac{(\ ^{1}\chi \ \ ^{1}\mathring{\Phi})^{\ast _{1}}}{\
^{1}\mathring{\Phi}^{\ast _{1}}},$ where there is no summation on index $%
i_{1}.$ We omit computations of deformations of the $n$--coefficients but we
shall present necessary formulas below, see similar details in \cite{tgovsv}.

Summarizing the above formulas and extending for all shells $s=0,1,2,3,$ we
obtain such coefficients for $\varepsilon $--deformations of a prime metric (%
\ref{pmc1c}) into a target stationary metric satisfying the motion equations
in heterotic supergravity:
\begin{eqnarray}
\ ^{\varepsilon }g_{i} &=&\mathring{g}_{i}[1+\varepsilon \chi
_{i}]=[1+\varepsilon e^{\ ^{0}q}\ ^{1}q/\mathring{g}_{i}\ \ _{h}^{0}\Upsilon
]\mathring{g}_{i}\mbox{ as
a solution of 2-d Poisson equations (\ref{e1})};  \notag \\
\ \ _{\varepsilon }h_{3} &=&[1+\varepsilon \ \chi _{3}]\mathring{g}_{3}=%
\left[ 1-\varepsilon \frac{1}{4\mathring{g}_{3}}\int dy^{4}\frac{(\mathring{%
\Phi}^{2}\chi )^{\ast }}{\ \Upsilon }\right] \mathring{g}_{3}\ ;\   \notag \\
\ _{\varepsilon }h_{4} &=&[1+\varepsilon \ \chi _{4}]\mathring{g}_{4}=\left[
1+\varepsilon \ \left( 2(\chi +\frac{\mathring{\Phi}}{\mathring{\Phi}^{\ast }%
}\chi ^{\ast })+\frac{1}{4\mathring{g}_{3}}\int dy^{4}\frac{(\mathring{\Phi}%
^{2}\chi )^{\ast }}{\Upsilon }\right) \right] \mathring{g}_{4};
\label{ersdef} \\
\ \ _{\varepsilon }n_{i} &=&[1+\varepsilon \ ^{n}\chi _{i}]\mathring{n}_{i}=%
\left[ 1+\varepsilon \ \widetilde{n}_{i}\int dy^{4}\ \frac{1}{\Upsilon ^{2}}%
\left( \chi +\frac{\mathring{\Phi}}{\mathring{\Phi}^{\ast }}\chi ^{\ast }+%
\frac{5}{8\mathring{g}_{3}}\frac{(\mathring{\Phi}^{2}\chi )^{\ast }}{%
\Upsilon }\right) \right] \mathring{n}_{i};  \notag \\
\ _{\varepsilon }w_{i} &=&[1+\varepsilon \ ^{w}\chi _{i}]\mathring{w}_{i}=%
\left[ 1+\varepsilon (\frac{\partial _{i}(\chi \ \mathring{\Phi})}{\partial
_{i}\ \mathring{\Phi}}-\frac{(\chi \ \mathring{\Phi})^{\ast }}{\mathring{\Phi%
}^{\ast }})\right] \mathring{w}_{i},  \notag
\end{eqnarray}%
where $\widetilde{n}_{i}(x^{k})$ is a re-defined integration function
including contributions from the prime metric. On a shell $s,$ these
formulas are defined recurrently (we omit parameterizations of functions on
shell coordinates because they can be determined in compatible form with
indices and labels for shells),
\begin{eqnarray*}
\ \ _{\varepsilon }h_{3+2s} &=&[1+\varepsilon \ \chi _{3+2s}]\mathring{g}%
_{3+2s}=\left[ 1-\varepsilon \frac{1}{4\mathring{g}_{3+2s}}\int dy^{4+2s}%
\frac{(\ ^{s}\mathring{\Phi}^{2}\ ^{s}\chi )^{\ast _{s}}}{\ _{s}\Upsilon }%
\right] \mathring{g}_{3+2s}\ ;\  \\
\ _{\varepsilon }h_{4+2s} &=&[1+\varepsilon \ \chi _{4+2s}]\mathring{g}%
_{4+2s}=\left[ 1+\varepsilon \ \left( 2(\ ^{s}\chi +\frac{\ ^{s}\mathring{%
\Phi}}{\ ^{s}\mathring{\Phi}^{\ast _{s}}}\ ^{s}\chi ^{\ast _{s}})+\frac{1}{4%
\mathring{g}_{3+2s}}\int dy^{4+2s}\frac{(\ ^{s}\mathring{\Phi}^{2}\ ^{s}\chi
)^{\ast _{1}}}{_{s}\Upsilon }\right) \right] \mathring{g}_{4+2s}; \\
\ \ _{\varepsilon }n_{i_{s}} &=&[1+\varepsilon \ ^{n}\chi _{i_{s}}]\mathring{%
n}_{i_{s}}=\left[ 1+\varepsilon \ \widetilde{n}_{i_{s}}\int dy^{5+2s}\ \frac{%
1}{\ _{s}\Upsilon ^{2}}\left( \ \ ^{s}\chi +\frac{\ ^{s}\mathring{\Phi}}{\
^{s}\mathring{\Phi}^{\ast }}\ ^{s}\chi ^{\ast _{s}}+\frac{5}{8\mathring{g}%
_{3+2s}}\frac{(\ ^{s}\mathring{\Phi}^{2}\ ^{s}\chi )^{\ast _{s}}}{\
_{s}\Upsilon }\right) \right] \mathring{n}_{i_{s}}; \\
\ _{\varepsilon }w_{i_{s}} &=&[1+\varepsilon \ ^{w}\chi _{i_{s}}]\mathring{w}%
_{i_{s}}=\left[ 1+\varepsilon (\frac{\partial _{i_{s}}(\ ^{s}\chi \ \ ^{s}%
\mathring{\Phi})}{\partial _{i_{s}}\ \ ^{s}\mathring{\Phi}}-\frac{(\
^{s}\chi \ \ ^{s}\mathring{\Phi})^{\ast _{s}}}{\ ^{s}\mathring{\Phi}^{\ast
_{s}}})\right] \mathring{w}_{i_{s}}.
\end{eqnarray*}%
These formulas with $s=1,2,3$ allow us to parametrize all coefficients of
vertical components of d--metrics and N--connections. For small parametric
deformations, the values $\chi ,\ ^{s}\chi $ and $\Upsilon ,\ _{s}\Upsilon $
have to be considered as generating functions. The values with a "circle"
are prescribed by a chosen prime solution (in our case, we can chose the 4-d
Kerr metric embedded into a 10-d gravity string spacetime). Fixing a small
value $\varepsilon $, we compute such deformations and prove their stability
(see \cite{vex1} and references therein) for any stable prime solution.

The $\varepsilon $--deformed quadratic elements are written {\small
\begin{eqnarray}
ds_{\varepsilon t}^{2} &=& \ _{\varepsilon }g_{\alpha _{s}\beta
_{s}}(x^{k},y^{4},y^{a_{1}},y^{a_{2}},y^{10})du^{\alpha _{s}}du^{\beta _{s}}
\label{edeformedsol} \\
&=&\ _{\varepsilon }g_{i}\left( x^{k}\right) [(dx^{1})^{2}+(dx^{2})^{2}]+\
_{\varepsilon }h_{3}(x^{k},y^{4})[dt+\ _{\varepsilon }n_{k}\
(x^{k},y^{4})dx^{k}]^{2}+\ _{\varepsilon }h_{4}(x^{k},y^{4})\ [dy^{4}+\
_{\varepsilon }w_{i}(x^{k},y^{4})dx^{i}]^{2}+  \notag \\
&&\ _{\varepsilon }h_{5}(x^{k},y^{4},y^{6})[dy^{5}+\ _{\varepsilon
}n_{k_{1}}\ (x^{k},y^{4},y^{6})dx^{k_{1}}]^{2}+\ _{\varepsilon
}h_{6}(x^{k},y^{4},y^{6})\ [dy^{6}+\ _{\varepsilon
}w_{i_{1}}(x^{k},y^{4},y^{6})dx^{i_{1}}]^{2}+  \notag \\
&&\ _{\varepsilon }h_{7}(x^{k},y^{4},y^{a_{1}},y^{8})[dy^{7}+\ _{\varepsilon
}n_{k_{2}}\ (x^{k},y^{4},y^{a_{1}},y^{8})dx^{k_{2}}]^{2}+  \notag \\
&&\ _{\varepsilon }h_{8}(x^{k},y^{4},y^{a_{1}},y^{8})\ [dy^{8}+\
_{\varepsilon }w_{i_{2}}(x^{k},y^{4},y^{a_{1}},y^{8})dx^{i_{2}}+  \notag \\
&&\ _{\varepsilon }h_{9}(x^{k},y^{4},y^{a_{1}},y^{a_{2}},y^{10})[dy^{9}+\
_{\varepsilon }n_{k_{3}}\
(x^{k},y^{4},y^{a_{1}},y^{a_{2}},y^{10})dx^{k_{3}}]^{2}+  \notag \\
&&\ _{\varepsilon }h_{10}(x^{k},y^{4},y^{a_{1}},y^{a_{2}},y^{10})\
[dy^{10}+\ _{\varepsilon
}w_{i_{3}}(x^{k},y^{4},y^{a_{1}},y^{a_{2}},y^{10})dx^{i_{3}}].  \notag
\end{eqnarray}%
} We can impose additional constraints in order to extract
LC--configurations as we considered in section (\ref{sslc}).


\section{Nonholonomic Heterotic String Deformations of the Kerr Metric}

\label{s3} In this section, we study generic off--diagonal deformations and
generalizations of the 4-d Kerr metric to new classes of exact solutions of
motion equations in heterotic string theory, see \cite%
{kramer,griff,misner,heusler}. We prove that using the AFDM extended to
models with almost-K\"{a}hler internal spaces, the Kerr solution can be
constructed as a particular case by prescribing a corresponding class of
generating and integration functions. In general, such solutions are with
nontrivial cosmological constants and nonholonomically induced torsions.
Imposing additional nonholonomic constraints, we can generate effective
vacuum solutions and extract Levi Civita configurations.A series of new
classes of small parametric solutions when the Kerr metrics are
nonholonomically deformed into general or ellipsoidal stationary
configurations in four dimensional gravity and/or extra dimensions are
considered. We provide and study examples of generic off--diagonal metrics
encoding nonlinear interactions with 3-form and gauge like fields,
nonholonomically induced torsion effects and instanton configurations.

\subsection{Preliminaries on the Kerr vacuum solution and nonholonomic
variables}

A 4-d ansatz
\begin{equation*}
ds_{[0]}^{2}=Y^{-1}e^{2h}(d\rho ^{2}+dz^{2})-\rho
^{2}Y^{-1}dt^{2}+Y(d\varphi +Adt)^{2}
\end{equation*}%
written in terms of three functions $(h,Y,A)$ on coordinates $x^{i}=(\rho
,z),$ defines the Kerr solution of the vacuum Einstein equations (for
rotating black holes) if we choose
\begin{eqnarray*}
Y &=&\frac{1-(p\widehat{x}_{1})^{2}-(q\widehat{x}_{2})^{2}}{(1+p\widehat{x}%
_{1})^{2}+(q\widehat{x}_{2})^{2}},\ A=2M\frac{q}{p}\frac{(1-\widehat{x}%
_{2})(1+p\widehat{x}_{1})}{1-(p\widehat{x}_{1})-(q\widehat{x}_{2})}, \\
e^{2h} &=&\frac{1-(p\widehat{x}_{1})^{2}-(q\widehat{x}_{2})^{2}}{p^{2}[(%
\widehat{x}_{1})^{2}+(\widehat{x}_{2})^{2}]},\ \rho ^{2}=M^{2}(\widehat{x}%
_{1}^{2}-1)(1-\widehat{x}_{2}^{2}),\ z=M\widehat{x}_{1}\widehat{x}_{2}.
\end{eqnarray*}%
Some values $M=const$ and $\rho =0$ result in a horizon $\widehat{x}_{1}=0$
and the "north / south" segments of the rotation axis, $\widehat{x}%
_{2}=+1/-1.$ For further applications of the AFDM, we can write this prime
solution in the form
\begin{equation}
ds_{[0]}^{2}=(dx^{1})^{2}+(dx^{2})^{2}-\rho ^{2}Y^{-1}(\mathbf{e}^{3})^{2}+Y(%
\mathbf{e}^{4})^{2}.  \label{kerr1}
\end{equation}%
This is possible if the coordinates $x^{1}(\widehat{x}_{1},\widehat{x}_{2})$
and $x^{2}(\widehat{x}_{1},\widehat{x}_{2})$ are defined for any%
\begin{equation*}
(dx^{1})^{2}+(dx^{2})^{2}=M^{2}e^{2h}(\widehat{x}_{1}^{2}-\widehat{x}%
_{2}^{2})Y^{-1}\left( \frac{d\widehat{x}_{1}^{2}}{\widehat{x}_{1}^{2}-1}+%
\frac{d\widehat{x}_{2}^{2}}{1-\widehat{x}_{2}^{2}}\right)
\end{equation*}%
and $y^{3}=t+\widehat{y}^{3}(x^{1},x^{2}),y^{4}=\varphi +\widehat{y}%
^{4}(x^{1},x^{2},t).$ We can consider an N-adapted basis  $\mathbf{e}%
^{3}=dt+(\partial _{i}\widehat{y}^{3})dx^{i},\mathbf{e}^{4}=dy^{4}+(\partial
_{i}\widehat{y}^{4})dx^{i}$,  for some functions $\widehat{y}^{a},$ $a=3,4,$
with $\partial _{t}\widehat{y}^{4}=-A(x^{k}).$

The Kerr metric was intensively studied in the so--called Boyer--Linquist
coordinates $(r,\vartheta ,\varphi ,t),$ for $r=m_{0}(1+p\widehat{x}_{1}),%
\widehat{x}_{2}=\cos \vartheta ,$ which can be considered for applications
of the AFDM. Such coordinates are expressed via parameters $p,q$ which are
related to the total black hole mass, $m_{0}$ and the total angular
momentum, $am_{0},$ for the asymptotically flat, stationary and
anti-symmetric Kerr spacetime. The formulas $m_{0}=Mp^{-1}$ and $a=Mqp^{-1}$
when $p^{2}+q^{2}=1$ implies $m_{0}^{2}-a^{2}=M^{2}$. In these variables,
the metric (\ref{kerr1}) can be written%
\begin{eqnarray}
ds_{[0]}^{2} &=&(dx^{1^{\prime }})^{2}+(dx^{2^{\prime }})^{2}+\overline{A}(%
\mathbf{e}^{3^{\prime }})^{2}+(\overline{C}-\overline{B}^{2}/\overline{A})(%
\mathbf{e}^{4^{\prime }})^{2},  \label{kerrbl} \\
\mathbf{e}^{3^{\prime }} &=&dt+d\varphi \overline{B}/\overline{A}%
=dy^{3^{\prime }}-\partial _{i^{\prime }}(\widehat{y}^{3^{\prime }}+\varphi
\overline{B}/\overline{A})dx^{i^{\prime }},\mathbf{e}^{4^{\prime
}}=dy^{4^{\prime }}=d\varphi ,  \notag
\end{eqnarray}%
for any coordinate functions $\ x^{1^{\prime }}(r,\vartheta ),\ x^{2^{\prime
}}(r,\vartheta ),\ y^{3^{\prime }}=t+\widehat{y}^{3^{\prime }}(r,\vartheta
,\varphi )+\varphi \overline{B}/\overline{A},y^{4^{\prime }}=\varphi ,\
\partial _{\varphi }\widehat{y}^{3^{\prime }}=-\overline{B}/\overline{A},$
for which $(dx^{1^{\prime }})^{2}+(dx^{2^{\prime }})^{2}=\Xi \left( \Delta
^{-1}dr^{2}+d\vartheta ^{2}\right) $, and the coefficients are%
\begin{eqnarray}
\overline{A} &=&-\Xi ^{-1}(\Delta -a^{2}\sin ^{2}\vartheta ),\overline{B}%
=\Xi ^{-1}a\sin ^{2}\vartheta \left[ \Delta -(r^{2}+a^{2})\right] ,\overline{%
C}=\Xi ^{-1}\sin ^{2}\vartheta \left[ (r^{2}+a^{2})^{2}-\Delta a^{2}\sin
^{2}\vartheta \right] ,\mbox{ and }  \notag \\
\Delta &=&r^{2}-2m_{0}+a^{2},\ \Xi =r^{2}+a^{2}\cos ^{2}\vartheta .
\label{kerrcoef}
\end{eqnarray}%
We send readers to \cite{heusler,kramer,misner} for main results and methods
for stationary black hole solutions (the coordinates $\widehat{x}_{1},%
\widehat{x}_{2}$ introduced above correspond to $x,y$ respectively from
chapter 4 of the first book).

The primed quadratic linear elements (\ref{kerr1}) (or (\ref{kerrbl}))
\begin{eqnarray}
\mathring{g}_{1} &=&1,\mathring{g}_{2}=1,\mathring{h}_{3}=-\rho ^{2}Y^{-1},%
\mathring{h}_{4}=Y,\mathring{N}_{i}^{a}=\partial _{i}\widehat{y}^{a},%
\mbox{
or \ }  \label{dkerr} \\
\mathring{g}_{1^{\prime }} &=&1,\mathring{g}_{2^{\prime }}=1,\mathring{h}%
_{3^{\prime }}=\overline{A},\mathring{h}_{4^{\prime }}=\overline{C}-%
\overline{B}^{2}/\overline{A},\ \mathring{N}_{i^{\prime }}^{3}=\mathring{n}%
_{i^{\prime }}=-\partial _{i^{\prime }}(\widehat{y}^{3^{\prime }}+\varphi
\overline{B}/\overline{A}),\mathring{N}_{i^{\prime }}^{4}=\mathring{w}%
_{i^{\prime }}=0  \notag
\end{eqnarray}%
define solutions of vacuum Einstein equations parametrized in the form (\ref%
{cdeinst}) and (\ref{lcconstr}) with zero sources. A straightforward
application of the AFDM is possible if we consider a correspondingly
N--adapted system of coordinates instead of the "standard" prolate
spherical, or Boyer--Linquist system. Parametrizations (\ref{dkerr}) are
most convenient for a straightforward application of the AFDM. This way we
can generalize the solutions for coefficients depending on more than two
coordinates, with non--Killing configurations and/or extra dimensions.

Working with general classes of stationary solutions generated by the AFDM,
the Kerr vacuum solution in GR can be considered as a "degenerated" case of
4--d off--diagonal vacuum solutions determined by primary metrics with data (%
\ref{dkerr}) when the diagonal coefficients depend only on two "horizontal"
N--adapted coordinates. Such a metric contains off--diagonal terms induced
by rotation frames in a form when the nonolonomically induced torsion is
zero. In N-adapted frames, further generic off--diagonal and extra dimension
generalizations can be performed following standard geometric methods (see
the following sections).

\subsection{Off--diagonal deformations of 4-d Kerr metrics by heterotic
string sources}

Let us consider the coefficients (\ref{dkerr}) for the Kerr metric as the
prime metric $\mathbf{\mathring{g}}$ (in general, a prime metric may or may
not be an exact solution of the Einstein or other modified gravitational
equations). Our goal is to construct nonholonomic deformations,
\begin{equation*}
(\mathbf{\mathring{g}},\mathbf{\mathring{N},\ }^{v}\mathring{\Upsilon}=0,%
\mathring{\Upsilon}=0)\rightarrow (\mathbf{g},\mathbf{N,\ }_{h}\Upsilon
(x^{k})\rightarrow \mathbf{\ }_{h}\ \Lambda =const\ \neq 0,\Upsilon
\rightarrow \Lambda =const\neq 0),
\end{equation*}%
see sources (\ref{effects}), (\ref{effsourcp1}), and (\ref{es}) respectively
\ with (\ref{es2}), when $\mathbf{\ }_{h}\Upsilon (x^{k})\rightarrow \
\mathbf{\ }_{h}\Lambda =\ \ _{h}^{H}\Lambda +\ _{h}^{F}\Lambda +\
_{h}^{R}\Lambda ,$ $\Upsilon (x^{k},y^{4})\rightarrow \ \Lambda =\ \
^{H}\Lambda +\ ^{F}\Lambda +\ ^{R}\Lambda $ and$\ \ _{h}^{\phi }\Lambda =\
^{\phi }\Lambda =0,$ as in (\ref{sourc4d}). The main condition is that the
target metric $\mathbf{g}$ is of type (\ref{4dlc}) which positively defines
a torsionless off--diagonal solution of field equations in the 4--d gravity
sector with sources determined from the heterotic string theory (\ref{hs1})
with ansatz for sources (\ref{ansatzsourc}). The N--adapted deformations of
coefficients of metrics, frames and sources are parametrized in the form
\begin{eqnarray}
&&[\mathring{g}_{i},\mathring{h}_{a},\mathring{w}_{i},\mathring{n}%
_{i}]\rightarrow \lbrack g_{i}=\eta _{i}\mathring{g}_{i},h_{3}=\eta _{3}%
\mathring{h}_{3},h_{4}=\eta _{4}\mathring{h}_{4},w_{i}=\mathring{w}_{i}+\
^{\eta }w_{i},n_{i}=\mathring{n}_{i}+\ ^{\eta }n_{i},\mathring{w}_{i}=0],
\notag \\
\mathbf{\ }_{h}\Upsilon &=&\Upsilon =\ _{K}\Lambda =-5[(\ ^{H}s)^{2}+n_{F}(\
^{F}s)^{2}(\ ^{R}s)^{2}],\check{\Phi}^{2}=\exp [\check{\varpi}(x^{k^{\prime
}},y^{4})],\ \mathring{h}_{3}=h_{3}^{(0)},\ \mathring{h}_{4}=-\check{\Upsilon%
}^{2}/\ _{K}\Lambda ^{2}  \label{ndefbm}
\end{eqnarray}%
The primes $\mathring{g}_{i},\mathring{h}_{a},\mathring{w}_{i},\mathring{n}%
_{i}$ (\ref{dkerr}) are given by coefficients depending only on $(x^{k\prime
}).$ The general deformations of the Kerr solution determined by generating
function and extra dimensional string sources with ansatz resulting in
cosmological constants described in terms of polarization functions, where $%
|\eta _{4^{\prime }}|$, may have not a smooth limit to $1,$
\begin{equation*}
\eta _{3^{\prime }}=1-(\check{\Phi}^{2})/\ _{K}\Lambda \mathring{h}_{3},\eta
_{4^{\prime }}=(\check{\Phi}^{\ast })^{2}/(\ _{K}\Lambda )^{2}\eta
_{3^{\prime }}\mathring{h}_{3},\ ^{\eta }w_{i^{\prime }}=\partial
_{i^{\prime }}\check{A},\ ^{\eta }n_{k^{\prime }}=\partial _{k^{\prime }}\
^{\eta }n(x^{i^{\prime }}),
\end{equation*}%
where the coefficient $1/4$ was introduced in $\check{\Phi}$ and the
function $\check{A}$ is any value for which $\partial _{i^{\prime }}\check{%
\Phi}/\check{\Phi}^{\ast }=\partial _{i^{\prime }}\check{A}.$

Summarizing the above formulas, we obtain the quadratic element%
\begin{eqnarray}
ds_{K\eta 4d}^{2} &=&\check{g}_{\alpha \beta }(x^{k},y^{4})du^{\alpha
}du^{\beta }=\eta _{i}\mathring{g}_{i}(dx^{i})^{2}+\eta _{a}\mathring{g}_{a}(%
\mathbf{e}^{a})^{2}  \label{nvlcmgs} \\
&=&e^{q[\ _{K}\Lambda ]}[(dx^{1\prime })^{2}+(dx^{2\prime })^{2}]+\left(
\overline{A}-\frac{\check{\Phi}^{2}}{\ _{K}\Lambda }\right) [dy^{3^{\prime
}}+\left( \partial _{k^{\prime }}\ ^{\eta }n(x^{i^{\prime }})-\partial
_{k^{\prime }}(\widehat{y}^{3^{\prime }}+\varphi \frac{\overline{B}}{%
\overline{A}})\right) dx^{k^{\prime }}]^{2}+  \notag \\
&&\frac{(\check{\Phi}^{\ast })^{2}}{\ _{K}\Lambda \left( \ _{K}\Lambda
\overline{A}-\check{\Phi}^{2}\right) }(\overline{C}-\frac{\overline{B}^{2}}{%
\overline{A}})[d\varphi +\partial _{i^{\prime }}\check{A})dx^{i^{\prime
}}]^{2},  \notag
\end{eqnarray}%
where $\ w_{i^{\prime }}=\mathring{w}_{i^{\prime }}+\ ^{\eta }w_{i^{\prime
}}=\partial _{i^{\prime }}(\ ^{\eta }\widetilde{A}[\varpi ]),\ n_{k^{\prime
}}=\mathring{n}_{k^{\prime }}+\ ^{\eta }n_{k^{\prime }}=\partial _{k^{\prime
}}(-\widehat{y}^{3^{\prime }}+\varphi \overline{B}/\overline{A}+\ ^{\eta }n)$
and $q$ is a solution of
\begin{equation*}
q^{\bullet \bullet }+q^{\prime \prime }=2\ \ _{K}\Lambda .
\end{equation*}%
We used an important relation $\mathring{h}_{3^{\prime }}\mathring{h}%
_{4^{\prime }}=\overline{A}\overline{C}-\overline{B}^{2}$ and emphasize that
it is possible to take any function $\ ^{\eta }n(x^{k}).$

The solutions (\ref{nvlcmgs}) are for stationary LC--configurations
determined by off--diagonal heterotic string gravity effects on Kerr black
holes when the new class of spacetimes are with Killing symmetry on $%
\partial /\partial y^{3^{\prime }}$ and generic dependence on three (from
maximally four) coordinates, $(x^{i^{\prime }}(r,\vartheta ),\varphi ).$
Similar solutions were constructed and studied in massive gravity with extra
dimensions \cite{tgovsv}. Off--diagonal modifications are possible for any
nontrivial values of $\check{\Phi}$ and any small constants $\ \ ^{H}\Lambda
,\ ^{F}\Lambda ,\ ^{R}\Lambda .$ The solutions depend on the type of
generating function $\check{\Phi}(x^{i^{\prime }},\varphi )$ we fix in order
to suit certain experimental/observational data for fixed systems of
reference/coordinates. These can be re--parameterized for an effective $%
_{K}\Lambda ,$ which should also be compatible with experimental data. In
such variables, we can mimic stationary heterotic string gravity effects by
off--diagonal configurations in GR with integration parameters which should
also be fixed by additional assumptions on symmetries of interactions. See
section \ref{4dellipsc} for ellipsoid configurations and details on
parametric Killing symmetries in Refs. \cite{ger1,ger2,vpars}.

\subsubsection{Nonholonomically string induced torsion for Kerr metrics in
the 4-d sector}

If we do not impose the LC--conditions (\ref{lcconstr}), a nontrivial source
$\Upsilon \rightarrow \Lambda $ from heterotic string gravity induces
stationary configurations with nontrivial d--torsion (\ref{dtorss}). The
torsion coefficients are determined by metrics of type (\ref{riccisolt})
with $\mathbf{\ }_{h}\Upsilon =\Upsilon =\ _{K}\Lambda $ as in (\ref{ndefbm}%
) and parametrizations of coefficients and coordinates distinguishing the
prime data for a Kerr metric (\ref{dkerr}). Such solutions can be written in
the form
\begin{eqnarray}
ds^{2} &=&e^{q}[(dx^{1^{\prime }})^{2}+(dx^{2^{\prime }})^{2}]+  \notag \\
&&\left( \overline{A}-\frac{\Phi ^{2}}{\ _{K}\Lambda }\right) [dy^{3^{\prime
}}+\left( \ _{1}n_{k^{\prime }}(x^{i^{\prime }})+\ _{2}n_{k^{\prime
}}(x^{i^{\prime }})\int dy^{4}(\Phi ^{2})^{\ast }(_{K}\Lambda \overline{A}%
-\Phi ^{2})^{-5/2}-\partial _{k^{\prime }}(\widehat{y}^{3^{\prime }}+\varphi
\frac{\overline{B}}{\overline{A}})\right) dx^{k^{\prime }}]^{2}  \notag \\
&&-\frac{\ (\Phi ^{\ast })^{2}}{\ _{K}\Lambda }\frac{\overline{A}\overline{C}%
-\overline{B}^{2}}{_{K}\Lambda \overline{A}-\Phi ^{2}}[d\varphi +\frac{%
\partial _{i^{\prime }}\Phi }{\partial _{\varphi }\Phi }dx^{i^{\prime
}}]^{2},  \label{ofindtmg}
\end{eqnarray}
where the generating function $\Phi (x^{i^{\prime }},\varphi )$ is not
subjected to any integrability conditions. Nontrivial stationary
off--diagonal torsion effects may result in additional effective rotations
if the integration function $\ _{2}n_{k}\neq 0.$ Considering two different
classes of off--diagonal solutions (\ref{ofindtmg}) and (\ref{nvlcmgs}), we
can study if heterotic string corrections to GR can have nonholonomically
induced torsion or if such effects are characterized by additional
nonholonomic constraints as in GR (for zero torsion).

It should be noted that configurations of type (\ref{ofindtmg}) can be
constructed in various theories with noncommutative and commutative
variables. We can consider warped and trapped brane type variables in
string, Finsler-like and/or Ho\v{r}ava--Lifshitz theories \cite%
{vex1,vp,vt,vgrg,vbranef} when nonholonomically induced torsion effects play
a substantial role.

\subsubsection{Small modifications of Kerr metrics and effective string
sources}

It is not clear what physical meaning general deformations of the Kerr
metric described by metrics of type (\ref{ofindtmg}) may have. We can choose
certain subclasses of nonholonomic distributions describing stationary $%
\varepsilon $-deformations described by formulas (\ref{ersdef}). Using the
Kerr solution (\ref{dkerr}) as a primary metric with assumptions for the
string sources, we compute small deformations into d--metric and
N--connection coefficients,%
\begin{equation*}
ds_{K\varepsilon 4d}^{2}=\check{g}_{\alpha \beta }(x^{k},y^{4})du^{\alpha
}du^{\beta }=[1+\varepsilon \chi _{i}(x^{k})]\mathring{g}%
_{i}(dx^{i})^{2}+[1+\varepsilon \chi _{a}(x^{k},y^{4})]\mathring{g}_{a}(%
\mathbf{e}^{a})^{2}.
\end{equation*}%
The $\varepsilon $--deformations are computed
\begin{eqnarray*}
\ ^{\varepsilon }g_{i} &=&\mathring{g}_{i}[1+\varepsilon \chi
_{i}]=[1+\varepsilon e^{\ ^{0}q}\ ^{1}q/_{K}\Lambda ]\mathring{g}_{i}%
\mbox{ as
a solution of 2-d Poisson equations (\ref{e1})}; \\
\ \ _{\varepsilon }h_{3} &=&[1+\varepsilon \ \chi _{3}]\mathring{g}_{3}=%
\left[ 1-\varepsilon \frac{\mathring{\Phi}^{2}\chi }{4\overline{A}\
_{K}\Lambda }\right] \mathring{g}_{3}\ ;\  \\
\ _{\varepsilon }h_{4} &=&[1+\varepsilon \ \chi _{4}]\mathring{g}_{4}=\left[
1+\varepsilon \ \left( 2(\chi +\frac{\mathring{\Phi}}{\mathring{\Phi}^{\ast }%
}\chi ^{\ast })+\frac{\mathring{\Phi}^{2}\chi }{4\overline{A}\ _{K}\Lambda }%
\right) \right] \mathring{g}_{4}; \\
\ \ _{\varepsilon }n_{i} &=&[1+\varepsilon \ ^{n}\chi _{i}]\mathring{n}_{i}=%
\left[ 1+\varepsilon \ \widetilde{n}_{i}\int dy^{4}\ \left( \chi +\frac{%
\mathring{\Phi}}{\mathring{\Phi}^{\ast }}\chi ^{\ast }+\frac{5(\mathring{\Phi%
}^{2}\chi )^{\ast }}{8\overline{A}_{K}\Lambda }\right) \right] \mathring{n}%
_{i}; \\
\ _{\varepsilon }w_{i} &=&[1+\varepsilon \ ^{w}\chi _{i}]\mathring{w}_{i}=%
\left[ 1+\varepsilon (\frac{\partial _{i}(\chi \ \mathring{\Phi})}{\partial
_{i}\ \mathring{\Phi}}-\frac{(\chi \ \mathring{\Phi})^{\ast }}{\mathring{\Phi%
}^{\ast }})\right] \mathring{w}_{i},
\end{eqnarray*}%
where $\mathring{g}_{1^{\prime }}=1,\mathring{g}_{2^{\prime }}=1,\mathring{h}%
_{3^{\prime }}=\overline{A},\mathring{h}_{4^{\prime }}=\overline{C}-%
\overline{B}^{2}/\overline{A},\ \mathring{N}_{i^{\prime }}^{3}=\mathring{n}%
_{i^{\prime }}=-\partial _{i^{\prime }}(\widehat{y}^{3^{\prime }}+\varphi
\overline{B}/\overline{A})$ are kept as above but a coordinate transform is
performed in order to have $\mathring{N}_{i^{\prime }}^{4}=\mathring{w}%
_{i^{\prime }}\neq 0.$ We use polarization functions linearized on $%
\varepsilon $,
\begin{eqnarray}
\eta _{i} &=&1+\varepsilon \chi _{i}(x^{k}),\eta _{a}=1+\varepsilon \chi
_{a}(x^{k},y^{4});  \notag \\
&&\mbox{ and  for the coeffisients of N-connection,}  \notag \\
\ ^{n}\eta _{i} &=&1+\varepsilon \ \ ^{n}\eta _{i}(x^{k},y^{4}),\ ^{w}\eta
_{i}=1+\varepsilon \ ^{w}\chi _{i}(x^{k},y^{4}).  \notag
\end{eqnarray}

Summarizing the above for 4--d $\varepsilon -$configurations, we obtain the
quadratic element
\begin{eqnarray*}
ds_{K\varepsilon 4d}^{2} &=&(1+\varepsilon e^{\ ^{0}q}\frac{\ ^{1}q}{%
_{K}\Lambda })[(dx^{1\prime })^{2}+(dx^{2\prime })^{2}]+ \\
&&(\overline{A}-\varepsilon \frac{\mathring{\Phi}^{2}\chi }{4\ _{K}\Lambda })%
\left[ dy^{3^{\prime }}-[1+\varepsilon \ \widetilde{n}_{i}\int dy^{4}\
\left( \chi +\frac{\mathring{\Phi}}{\mathring{\Phi}^{\ast }}\chi ^{\ast }+%
\frac{5(\mathring{\Phi}^{2}\chi )^{\ast }}{8\overline{A}_{K}\Lambda }\right)
]\partial _{k^{\prime }}(\widehat{y}^{3^{\prime }}+\varphi \frac{\overline{B}%
}{\overline{A}})dx^{k^{\prime }}\right] ^{2} \\
&&+\left[ 1+\varepsilon \ \left( 2(\chi +\frac{\mathring{\Phi}}{\mathring{%
\Phi}^{\ast }}\chi ^{\ast })+\frac{\mathring{\Phi}^{2}\chi }{4\overline{A}\
_{K}\Lambda }\right) \right] \left( \overline{C}-\frac{\overline{B}^{2}}{%
\overline{A}}\right) \left[ d\varphi +[1+\varepsilon (\frac{\partial
_{i}(\chi \ \mathring{\Phi})}{\partial _{i}\ \mathring{\Phi}}-\frac{(\chi \
\mathring{\Phi})^{\ast }}{\mathring{\Phi}^{\ast }})]\mathring{w}_{i}\right]
^{2}
\end{eqnarray*}%
In general, these heterotic string deformations of the Kerr metric are
nonholonomically induced torsion coefficients, linear on $\varepsilon $.

We can impose additional constraints on $\chi ,\,\widetilde{n}_{i}$ and $%
\mathring{\Phi}$ which allows us to extract LC--configurations. The
corresponding $\varepsilon $--deformed analogue of the metric (\ref{nvlcmgs}%
) can be written \
\begin{eqnarray}
ds_{K\eta 4d}^{2} &=&(1+\varepsilon e^{\ ^{0}q}\frac{\ ^{1}q}{_{K}\Lambda }%
)[(dx^{1\prime })^{2}+(dx^{2\prime })^{2}]+\left( \overline{A}-\varepsilon
\frac{\mathring{\Phi}^{2}\chi }{4\ _{K}\Lambda }\right) [dy^{3^{\prime
}}+\left( \varepsilon \partial _{k^{\prime }}\ ^{\chi }n(x^{i^{\prime
}})-\partial _{k^{\prime }}(\widehat{y}^{3^{\prime }}+\varphi \frac{%
\overline{B}}{\overline{A}})\right) dx^{k^{\prime }}]^{2}  \notag \\
&&+\left[ 1+\varepsilon \ \left( 2(\chi +\frac{\mathring{\Phi}}{\mathring{%
\Phi}^{\ast }}\chi ^{\ast })+\frac{\mathring{\Phi}^{2}\chi }{4\overline{A}\
_{K}\Lambda }\right) \right] (\overline{C}-\frac{\overline{B}^{2}}{\overline{%
A}})[d\varphi +\varepsilon \partial _{i^{\prime }}\check{A})dx^{i^{\prime
}}]^{2},  \label{epsdeflc}
\end{eqnarray}%
where $\chi $ defines a deformed generating function $\ ^{\varepsilon }\Phi =%
\mathring{\Phi}(x^{k},y^{4})[1+\varepsilon \chi (x^{k},y^{4})]$ as in
formula (\ref{epsdefgf}) but subjected to the condition
\begin{equation*}
\varepsilon \partial _{i^{\prime }}\check{A}=\partial _{i^{\prime }}(\
^{\varepsilon }\Phi )/(\ ^{\varepsilon }\Phi )^{\ast }
\end{equation*}%
which together with $\ ^{\chi }n(x^{i^{\prime }})$ are chosen to result in
zero nonholonomically induced torsion.

\subsection{String induced ellipsoidal 4--d deformations of the Kerr metric}

\label{4dellipsc}We provide some examples how the Kerr primary data (\ref%
{dkerr}) are nonholonomically deformed by heterotic string interactions into
target generic off--diagonal solutions of vacuum and non--vacuum Einstein
equations for the canonical d--connection and/or the Levi--Civita
connection. Generic off--diagonal metrics of type (\ref{epsdeflc}) can be
parameterized as ellipsoidal deformations of the Kerr metric on a small
eccentricity parameter $\varepsilon .$

\subsubsection{ Ellipsoidal configurations with string induced cosmological
constant}

Let us construct solutions for $\varepsilon $--deformations of type (\ref%
{epsdeflc}) \ with ellipsoidal configurations. We choose a generating
function $\chi _{3^{\prime }},$ when the constraint $h_{3^{\prime }}=0$
defines a stationary rotoid configuration (different from the ergo sphere
for the Kerr solutions): \ We prescribe
\begin{equation}
\chi _{3^{\prime }}=\frac{\mathring{\Phi}^{2}\chi }{4\overline{A}\
_{K}\Lambda }=2\underline{\zeta }\sin (\omega _{0}\varphi +\varphi _{0}),
\label{chi3prim}
\end{equation}%
for constant parameters $\underline{\zeta },\omega _{0}$ and $\varphi _{0},$
when the values
\begin{equation*}
\overline{A}(r,\vartheta )[1+\varepsilon \chi _{3^{\prime }}(r,\vartheta
,\varphi )]=\widehat{A}(r,\vartheta ,\varphi )=-\Xi ^{-1}(\widehat{\Delta }%
-a^{2}\sin ^{2}\vartheta )\mbox{ and }\widehat{\Delta }(r,\varphi
)=r^{2}-2m(\varphi )+a^{2},
\end{equation*}%
are considered as $\varepsilon $--deformations of Kerr coefficients (\ref%
{kerrcoef}). We get an effective "anisotropically polarized" mass
\begin{equation}
m(\varphi )=m_{0}/\left( 1+\varepsilon \underline{\zeta }\sin (\omega
_{0}\varphi +\varphi _{0})\right) .  \label{polarm}
\end{equation}%
In result, the condition $h_{3}=0,$ i.e. $\ ^{\varphi }\Delta (r,\varphi
,\varepsilon )=a^{2}\sin ^{2}\vartheta ,$ states an ellipsoidal "deformed
horizon"
\begin{equation*}
r(\vartheta ,\varphi )=m(\varphi )+\left( m^{2}(\varphi )-a^{2}\sin
^{2}\vartheta \right) ^{1/2}.
\end{equation*}%
If $a=0$, we obtain the parametric formula for an ellipse with eccentricity $%
\varepsilon ,$ $r_{+}=\frac{2m_{0}}{1+\varepsilon \underline{\zeta }\sin
(\omega _{0}\varphi +\varphi _{0})}.$ Such configurations correspond to the
generating function%
\begin{equation}
\chi =\frac{8\widetilde{\zeta }\overline{A}\ _{K}\Lambda }{\mathring{\Phi}%
^{2}}\sin (\omega _{0}\varphi +\varphi _{0})  \label{ellipsoidgf}
\end{equation}%
determined by effective heterotic string source $_{K}\Lambda ,$ as follows
from (\ref{chi3prim}).

If the anholonomy coefficients (\ref{anhrel}) computed for (\ref{epsdeflc})
are not trivial for $w_{i}$ and $\ n_{k}=\ _{1}n_{k},$ the generated
solutions can not be diagonalized via coordinate transforms.

The corresponding nonholonomically deformed 4--d spacetimes have one Killing
symmetry on $\partial /\partial y^{3^{\prime }}.$ For small $\varepsilon ,$
the singularity at $\Xi =0$ is "hidden" under ellipsoidal deformed horizons
if $m_{0}\geq a.$ Similarly to the Kerr solution, there are $\varphi $%
--deformed both the event horizon,
\begin{equation*}
r_{+}=m(\varphi )+\left( m^{2}(\varphi )-a^{2}\sin ^{2}\vartheta \right)
^{1/2},
\end{equation*}%
and the Cauchy horizon,
\begin{equation*}
r_{-}=m(\varphi )-\left( m^{2}(\varphi )-a^{2}\sin ^{2}\vartheta \right)
^{1/2},
\end{equation*}%
which are effectively imbedded into an off--diagonal background determined
by N--coefficients. Using an ellipsoid type generating function (\ref%
{ellipsoidgf}) in (\ref{epsdeflc}), we construct a class of generic
off--diagonal solutions of effective Einstein equations with heterotic
string gravity effective cosmological constant $_{K}\Lambda, $ which in its
turn can be related to arbitrary sources via a re-definition of generating
functions (see formulas (\ref{nonltransf}) and (\ref{effsourcp1})) adapted
to $\varepsilon $--deformations. The corresponding quadratic line elements
are {\small
\begin{eqnarray}
ds_{K\eta 4d}^{2} &=&(1+\varepsilon e^{\ ^{0}q}\frac{\ ^{1}q}{\ _{K}\Lambda }%
)[(dx^{1\prime })^{2}+(dx^{2\prime })^{2}]+\overline{A}[1-2\varepsilon
\widetilde{\zeta }\sin (\omega _{0}\varphi +\varphi _{0})][dy^{3^{\prime
}}+\left( \varepsilon \partial _{k^{\prime }}\ ^{\chi }n(x^{i^{\prime
}})-\partial _{k^{\prime }}(\widehat{y}^{3^{\prime }}+\varphi \frac{%
\overline{B}}{\overline{A}})\right) dx^{k^{\prime }}]^{2}  \notag \\
&+&\left[ 1+\varepsilon \ \left( \frac{(8\overline{A}\ _{K}\Lambda +%
\mathring{\Phi}^{2})\widetilde{\zeta }}{4\ _{K}\Lambda \mathring{\Phi}^{2}}%
\sin (\omega _{0}\varphi +\varphi _{0})-\frac{16\widetilde{\zeta }\omega _{0}%
\overline{A}\ _{K}\Lambda }{3\mathring{\Phi}^{2}}\cos (\omega _{0}\varphi
+\varphi _{0})\right) \right] (\overline{C}-\frac{\overline{B}^{2}}{%
\overline{A}})[d\varphi +\varepsilon \partial _{i^{\prime }}\check{A}%
)dx^{i^{\prime }}]^{2}.  \label{ellips01}
\end{eqnarray}%
}

The new classes of $\varepsilon $--deformed solutions determine Kerr-like
black hole solutions with additional dependencies on variable $\varphi $ of
certain diagonal and off--diagonal coefficients of metric. There is an
obvious anisotropy in angle $\varphi .$ The values $\widetilde{\zeta }$ and $%
\omega _{0}$ have to be chosen in accordance with experimental data. The
function $\mathring{\Phi}$ depends on corresponding frame distributions for
the prime metric. Fixing $a=0$ for a $\varepsilon \neq 0,$ we get
ellipsoidal deformations of the Schwarzschild black holes. We studied these
constructions in details in \cite{vex1}, see also references inside on
stability and interpretation of such solutions with both commutative and/or
noncommutative deformation parameters. In general, a black hole/ellipsoid
interpretation is not possible for "non-small" $N$--deformations of the Kerr
metric. For certain embeddings, we can generate black hole-like
configurations with deformed horizons and locally anisotropic polarized
physical constants.

\subsubsection{Ellipsoid Kerr -- de Sitter configurations in $R^{2}$ and
heterotic string gravity}

The first examples of generic off--diagonal ellipsoid--solitonic
deformations of similar Kerr-Sen black holes were constructed in \cite%
{vkerrads}. \ Recently, asymptotically de Sitter solutions with spherical
symmetry for $R^{2}$ gravity were studied in \cite{kehagias} and the
nonholonomic geometric off--diagonal evolution of such metrics was analyzed
in \cite{muen01}. In this section, we show that those constructions can be
related to the 4-d part of heterotic string MGTs.

We consider a prime 4-d metric
\begin{equation}
d\overline{s}^{2}=\frac{3\lambda }{2\varsigma ^{2}}\left\{ (1-\frac{M}{r}%
-\lambda r^{2})^{-1}dr^{2}+r^{2}d\theta ^{2}+r^{2}\sin \theta d\varphi
^{2}-(1-\frac{M}{r}-\lambda r^{2})dt^{2}\right\}  \label{ads}
\end{equation}%
which for
\begin{equation*}
e^{\sqrt{1/3}\phi }=\frac{3\lambda }{2\varsigma ^{2}}=\frac{1}{8\varsigma
^{2}}R\text{\mbox{ and }}\overline{g}_{\mu \nu }=e^{\sqrt{1/3}\phi }g_{\mu
\nu }=\frac{R}{8\varsigma ^{2}}g_{\mu \nu },R\neq 0,
\end{equation*}%
defines exact solutions with spherical symmetry in $R^{2}$ gravity, for
equations $\ \overline{R}_{\mu \nu }=2\varsigma ^{2}\overline{g}_{\mu \nu }.
$ The effective cosmological constant $\varsigma ^{2}$ is usually related to
nonlinear scalar fields/ dilaton like interactions in effective Einstein
gravity resulting from $R^{2}$ gravity. In our model of heterotic string
gravity, we can choose%
\begin{equation}
2\varsigma ^{2}=\ _{K}\Lambda  \label{stringquadr}
\end{equation}%
and study quadratic gravity 4-d models determined by heterotic string
effective sources.\ The metric (\ref{ads}) describes asymptotically de
Sitter solutions with $\lambda >0$ and $R\neq 0.$ Introducing new 4-d
coordinates,
\begin{eqnarray}
\widetilde{x}^{1^{^{\prime }}}(r) &=&\sqrt{\left\vert \frac{3\lambda }{2}%
\right\vert }\frac{1}{\varsigma }\int dr(1-\frac{M}{r}-\lambda r^{2})^{-1/2},%
\widetilde{x}^{2^{\prime }}=\theta ,y^{3^{\prime }}=\varphi ,y^{4^{\prime
}}=t;  \label{newcoord} \\
\underline{\mathring{g}}_{1^{\prime }} &=&1,\underline{\mathring{g}}%
_{2^{\prime }}(\widetilde{x}^{1^{\prime }})=r^{2}(\widetilde{x}^{1^{\prime
}}),\mathring{h}_{3^{\prime }}=r^{2}(\widetilde{x}^{1^{\prime }})\sin
(x^{2^{\prime }}),\mathring{h}_{4^{\prime }}=-(1-\frac{M}{r(\widetilde{x}%
^{1^{\prime }})}+\lambda r^{2}(\widetilde{x}^{1^{\prime }})),  \notag
\end{eqnarray}%
the metric (\ref{ads}) is written as a "prime" metric
\begin{equation*}
ds^{2}=\underline{\mathring{g}}_{\alpha ^{\prime }\beta ^{\prime }}(%
\widetilde{x}^{k^{\prime }})du^{\alpha ^{\prime }}du^{\beta ^{\prime }}=%
\underline{\mathring{g}}_{1^{\prime }}(d\widetilde{x}^{1^{\prime }})^{2}+%
\underline{\mathring{g}}_{2^{\prime }}(\widetilde{x}^{1^{^{\prime }}})(d%
\widetilde{x}^{2^{\prime }})^{2}+\underline{\mathring{h}}_{3^{\prime }}(%
\widetilde{x}^{1^{\prime }},\widetilde{x}^{2^{\prime }})(dy^{3^{\prime
}})^{2}+\underline{\mathring{h}}_{4^{\prime }}(\widetilde{x}^{1^{\prime
}})(dy^{4^{\prime }})^{2},
\end{equation*}%
for some constants $M$, $\lambda $ and $u^{\alpha }=(\widetilde{x}%
^{k^{\prime }},y^{a}).$ In order to work with "formal" off--diagonal metric
of type (\ref{pmc1c}) with nontrivial values $\mathring{h}_{a}^{\ast },%
\mathring{w}_{i}$ and $\mathring{n}_{i},$ but $\mathring{W}_{\beta \gamma
}^{\alpha }(\widetilde{u}^{\mu })=0,$ see (\ref{anhrel}), we consider a
coordinate transform $u^{\alpha ^{\prime }}=u^{\alpha ^{\prime }}(u^{\alpha
})$ with $\varphi =\varphi (y^{4},\widetilde{x}^{k})$ and $t=t(y^{3},%
\widetilde{x}^{k}),$ when
\begin{equation*}
dt=\frac{\partial t}{\partial y^{3}}[dy^{3}+(\partial _{3}t)^{-1}(\widetilde{%
\partial }_{k}t)\widetilde{d}x^{k}]\mbox{ and }d\varphi =\frac{\partial
\varphi }{\partial y^{4}}[dy^{4}+(\partial _{4}\varphi )^{-1}(\widetilde{%
\partial }_{k}\varphi )\widetilde{d}x^{k}]
\end{equation*}%
for $\widetilde{\partial }_{i}\varphi =\partial \varphi /\partial \widetilde{%
x}^{i}$ and $\partial _{a}\varphi =\partial \varphi /\partial y^{a}.$
Choosing
\begin{equation*}
\underline{\mathring{n}}_{i}=\widetilde{\partial }_{i}n(x^{k})=(\partial
_{3}t)^{-1}(\widetilde{\partial }_{i}t),\mbox{ and }\underline{\mathring{w}}%
_{i}=\widetilde{\partial }_{i}\mathring{\Phi}\ /\mathring{\Phi}^{\ast
}=(\partial _{4}\varphi )^{-1}(\widetilde{\partial }_{i}\varphi ),
\end{equation*}%
we express (\ref{ads}) as
\begin{eqnarray}
ds^{2} &=&\underline{\mathring{g}}_{1^{\prime }}(d\widetilde{x}^{1^{\prime
}})^{2}+\underline{\mathring{g}}_{2^{\prime }}(\widetilde{x}^{1^{^{\prime
}}})(d\widetilde{x}^{2^{\prime }})^{2}+\underline{\mathring{g}}_{3}(x^{k}(%
\widetilde{x}^{k^{\prime }}))[dy^{3}+\underline{\mathring{n}}_{i}(\widetilde{%
x}^{k})d\widetilde{x}^{i}]^{2}+\underline{\mathring{g}}_{4}[dy^{4}+%
\underline{\mathring{w}}_{i}(\widetilde{x}^{k})dx^{i}]^{2},  \label{pm2a} \\
\underline{\mathring{g}}_{4}(\widetilde{x}^{k}(\widetilde{x}^{k^{\prime }}))
&=&(\partial _{4}\varphi )^{2}r^{2}(\widetilde{x}^{1^{\prime }})\sin (%
\widetilde{x}^{2^{\prime }})\mbox{ and }\underline{\mathring{g}}_{3}(%
\widetilde{x}^{k}(\widetilde{x}^{k^{\prime }}))=-(\partial _{3}t)^{2}(1-%
\frac{M}{r}+\lambda r^{2}).  \notag
\end{eqnarray}%
The prime d--metric (\ref{pm2a}) allows us to apply the AFDM and construct $%
\varepsilon $--deformations of geometric/ physical objects and physical
parameters as was shown in section \ref{ssedef} .

For $\underline{\mathring{g}}_{3}=\underline{\mathring{h}}_{3}(\widetilde{x}%
^{1^{\prime }})=(1-\frac{M}{r}+\lambda r^{2})$ and $(\partial _{3}t)^{2}=1$
and anisotropically polarized mass $\ \widetilde{M}(\varphi
)=M[1+\varepsilon \cos (\omega _{0}\varphi +\varphi _{0})],$ we obtain
\begin{eqnarray*}
\ ^{s=0}h_{3} &=&-(1-\frac{M}{r}+\lambda r^{2})[1-\varepsilon \frac{M}{r}%
\frac{\cos (\omega _{0}\varphi +\varphi _{0})}{1-\frac{M}{r}+\lambda r^{2}}]
\\
&=&\underline{\mathring{h}}_{3}(\widetilde{x}^{1^{\prime }})\left[
1-\varepsilon \frac{M}{r}(\underline{\mathring{h}}_{3})^{-1}\cos (\omega
_{0}\varphi +\varphi _{0})\right] \simeq -\left[ 1-\frac{\ \widetilde{M}%
(\varphi )}{r}+\lambda r^{2}\right]
\end{eqnarray*}%
The parametric equation of an ellipse with radial parameter $\mathring{r}%
_{+}=M$ and eccentricity $\varepsilon ,$ $r_{+}\simeq \frac{M}{1-\varepsilon
\cos (\omega _{0}\varphi +\varphi _{0})},$ can be determined in a simple way
for $\lambda =0.$ We have to find solutions of a third order algebraic
equation in order to determine possible horizons for nontrivial $\lambda .$

We construct ellipsoidal deformations of d--metric (\ref{pm2a}) if $\ \chi
=\ _{\varsigma }\chi =8\frac{M}{r}\varsigma ^{2}\ \mathring{\Phi}^{-2}\cos
(\omega _{0}\varphi +\varphi _{0}),$ with identification (\ref{stringquadr}%
). \ Following the same method as in the previous subsection but for $\
_{\varsigma }\chi $ used for d--metric coefficients (\ref{ersdef}), we
compute
\begin{eqnarray}
\ _{\varsigma }g_{i} &=&\underline{\mathring{g}}_{i}[1+\varepsilon \chi
_{i}]=[1+\varepsilon e^{\ ^{0}q}\ ^{1}q/\ \ 2\varsigma ^{2}]\underline{%
\mathring{g}}_{i}\mbox{ solution of 2-d Poisson equations (\ref{e1})};
\notag \\
\ \ _{\varsigma }h_{3} &=&[1+\varepsilon \ \ \ _{\varsigma }\chi _{3}]%
\mathring{g}_{3}=\left[ 1-\varepsilon \frac{1}{8\varsigma ^{2}\underline{%
\mathring{g}}_{3}}\mathring{\Phi}^{2}\ ^{\varsigma }\chi \right] \underline{%
\mathring{g}}_{3};  \notag \\
\ \ \ _{\varsigma }h_{4} &=&[1+\varepsilon \ \ _{\varsigma }\chi _{3}]%
\mathring{g}_{3}=\left[ 1+\varepsilon \ \left( 2(\ ^{\varsigma }\chi +\frac{%
\mathring{\Phi}}{\mathring{\Phi}^{\ast }}\ ^{\varsigma }\chi ^{\ast })+\frac{%
1}{8\varsigma ^{2}\underline{\mathring{g}}_{4}}\mathring{\Psi}^{2}\
^{\varsigma }\chi \right) \right] \underline{\mathring{g}}_{3};
\label{scoeff} \\
\ _{\varsigma }n_{i} &=&[1+\varepsilon \ _{\varsigma }^{n}\chi _{i}]%
\mathring{n}_{i}=\left[ 1+\varepsilon \ \widetilde{n}_{i}\int dy^{4}\left( \
^{\varsigma }\chi +\frac{\mathring{\Phi}}{\mathring{\Phi}^{\ast }}\
^{\varsigma }\chi ^{\ast }+\frac{5}{16\varsigma ^{2}}\frac{1}{\underline{%
\mathring{g}}_{4}}(\mathring{\Phi}^{2}\ ^{\varsigma }\chi )^{\ast }\right) %
\right] \underline{\mathring{n}}_{i},\ \   \notag \\
\ \ \ _{\varsigma }w_{i} &=&[1+\varepsilon \ \ _{\varsigma }^{w}\chi _{i}]%
\mathring{w}_{i}=\left[ 1+\varepsilon (\frac{\partial _{i}(\ ^{\varsigma
}\chi \ \mathring{\Phi})}{\partial _{i}\ \mathring{\Phi}}-\frac{(\
^{\varsigma }\chi \ \mathring{\Phi})^{\ast }}{\mathring{\Phi}^{\ast }})%
\right] \underline{\mathring{w}}_{i};  \notag
\end{eqnarray}%
where $\ \widetilde{n}_{i}(x^{k})$ is a re-defined integration function
including contributions from the prime metric (\ref{pm2a}). The generating
functions $\ ^{\varsigma }\chi $ and $\ \ ^{0}q$ can be determined for an
ellipsoid configuration induced by the effective cosmological constant \ $%
\varsigma ^{2}$ in $R^{2}$ gravity.

Finally, we generate a class of generic off--diagonal metrics for ellipsoid
Kerr -- de Sitter configurations related to the cosmological constant in
heterotic string gravity, {\small
\begin{eqnarray}
&&ds^{2}=[1+\varepsilon e^{\ ^{0}q}\ ^{1}q/\ \ 2\varsigma ^{2}][\underline{%
\mathring{g}}_{1^{\prime }}(d\widetilde{x}^{1^{\prime }})^{2}+\underline{%
\mathring{g}}_{2^{\prime }}(\widetilde{x}^{1^{^{\prime }}})(d\widetilde{x}%
^{2^{\prime }})^{2}]-  \notag \\
&&\left[ 1-\varepsilon \frac{1}{8\varsigma ^{2}\underline{\mathring{g}}_{3}}%
\mathring{\Phi}^{2}\ ^{\varsigma }\chi \right] \underline{\mathring{g}}%
_{3}(x^{k}(\widetilde{x}^{k^{\prime }}))\left[ dy^{3}+[1+\varepsilon \
\widetilde{n}_{i}\int dy^{4}\left( \ ^{\varsigma }\chi +\frac{\mathring{\Phi}%
}{\mathring{\Phi}^{\ast }}\ ^{\varsigma }\chi ^{\ast }+\frac{5}{16\varsigma
^{2}}\frac{1}{\underline{\mathring{g}}_{4}}(\mathring{\Phi}^{2}\ ^{\varsigma
}\chi )^{\ast }\right) ]\underline{\mathring{n}}_{i}d\widetilde{x}^{i}\right]
^{2}+  \label{elkdscon} \\
&&\left[ 1+\varepsilon \ \left( 2(\ ^{\varsigma }\chi +\frac{\mathring{\Phi}%
}{\mathring{\Phi}^{\ast }}\ ^{\varsigma }\chi ^{\ast })+\frac{1}{8\varsigma
^{2}\underline{\mathring{g}}_{4}}\mathring{\Phi}^{2}\ ^{\varsigma }\chi
\right) \right] \underline{\mathring{g}}_{4}\left[ d\varphi +[1+\varepsilon (%
\frac{\partial _{i}(\ ^{\varsigma }\chi \ \mathring{\Phi})}{\partial _{i}\
\mathring{\Phi}}-\frac{(\ ^{\varsigma }\chi \ \mathring{\Phi})^{\ast }}{%
\mathring{\Phi}^{\ast }})]\underline{\mathring{w}}_{i}d\widetilde{x}^{i}%
\right] ^{2}.  \notag
\end{eqnarray}%
}

Such metrics have Killing symmetry on $\partial /\partial y^{3}$ and are
completely defined by generating functions $\ ^{1}q$ and $\ ^{\varsigma
}\chi $ and effective source $2\varsigma ^{2}=\ _{K}\Lambda \ $\ induced
from sting theory. They define $\varepsilon $--deformations of Kerr -- de
Sitter black holes into ellipsoid configurations with effective (polarized)
cosmological constants determined by constants in string theory and
equivalent MGTs. If the LC--conditions are satisfied, such metrics can be
modelled in GR with nontrivial cosmological constant.

\subsection{Extra dimensional off--diagonal string modifications of the Kerr
solutions}

Various classes of exact solutions in heterotic string gravity can be
constructed, which depend on which type of effective sources (\ref{es}) and (%
\ref{ansatzsourc}) are chosen. As a result, various classes of generic
off--diagonal deformations of the Kerr metric into higher dimensional exact
solutions can be constructed. In this subsection we shall construct and
analyze a series of 6--d and 10--d solutions encoding possible higher
dimension interactions with effective cosmological constants, warping
configurations, $f$--modifications and certain analogies to almost-K\"{a}%
hler gravity models.

\subsubsection{6--d deformations with nontrivial cosmological constant}

Solutions are determined by certain configurations of the NS-fields $%
\widehat{\mathbf{H}}_{\alpha _{1}\beta _{1}\mu _{1}}$ which are nontrivial
on the first shell $s=1,$ see formulas (\ref{es2}) and respective term in (%
\ref{ansatzsourc})
\begin{equation*}
\ ^{H}\mathbf{\Upsilon }_{\mu _{1}\nu _{1}}=\frac{1}{4}\widehat{\mathbf{H}}%
_{\alpha _{1}\beta _{1}\mu _{1}}\widehat{\mathbf{H}}_{\nu _{1}}^{\quad
\alpha _{1}\beta _{1}}\mbox{ with effective constant }\ ^{H}\Lambda ;\ ^{H}%
\mathbf{\Upsilon }_{\mu _{1}\nu _{1}}=-\frac{6}{2}(\ ^{H}s)^{2}\mathbf{g}%
_{\beta _{1}\mu _{1}},\mbox{ for }\ ^{H}\Lambda =-3(\ ^{H}s)^{2},
\end{equation*}%
where the coefficient $3$ is used for an effective 6-d space time with
trivial extension on 4 other internal space coordinates. We introduce an
effective source in the quadratic element (\ref{ellips01}) extended with one
nontrivial extra 2-d shell. Such a family of generic off-diagonal stationary
solutions is described by {\small
\begin{eqnarray}
&&ds_{K\eta 6d}^{2}=(1-\frac{e^{\ ^{0}q}\ ^{1}q}{\ 12(\ ^{H}s)^{2}}%
)[(dx^{1^{\prime }})^{2}+(dx^{2^{\prime }})^{2}]+ \overline{A}%
[1-2\varepsilon \widetilde{\zeta }\sin (\omega _{0}\varphi +\varphi
_{0})][dy^{3^{\prime }}+(\varepsilon \partial _{k^{\prime }}\ ^{\chi
}n(x^{i^{\prime }})-\partial _{k^{\prime }}(\widehat{y}^{3^{\prime
}}+\varphi \frac{\overline{B}}{\overline{A}})) dx^{k^{\prime }}]^{2}  \notag
\\
&&+\left[ 1+\varepsilon \left( -\frac{[-24\overline{A}\ (\ ^{H}s)^{2}+%
\mathring{\Phi}^{2}]\widetilde{\zeta }}{12(\ ^{H}s)^{2}\mathring{\Phi}^{2}}%
\sin (\omega _{0}\varphi +\varphi _{0})+\frac{16\widetilde{\zeta }\omega _{0}%
\overline{A}\ (\ ^{H}s)^{2}}{\mathring{\Phi}^{2}}\cos (\omega _{0}\varphi
+\varphi _{0})\right) \right] (\overline{C}-\frac{\overline{B}^{2}}{%
\overline{A}})[d\varphi +\varepsilon \partial _{i^{\prime }}\check{A}%
)dx^{i^{\prime }}]^{2}  \notag \\
&&+[1+\varepsilon \frac{\ ^{1}\mathring{\Phi}^{2}\ _{1}^{H}\chi }{12(\
^{H}s)^{2}\underline{\mathring{g}}_{5}}]\underline{\mathring{g}}_{5}(x^{k}(%
\widetilde{x}^{k^{\prime }}))[dy^{5}+[1+\varepsilon \ \ ^{1}\widetilde{n}%
_{i_{1}}\int dy^{6}(\ _{1}^{H}\chi +\frac{\ ^{1}\mathring{\Phi}}{\ ^{1}%
\mathring{\Phi}^{\ast _{1}}}\ _{1}^{H}\chi ^{\ast _{1}}+\frac{5}{24(\
^{H}s)^{2}\underline{\mathring{g}}_{6}}(\ ^{1}\mathring{\Phi}^{2}\
_{1}^{H}\chi )^{\ast _{1}})]\underline{\mathring{n}}_{i_{1}}d\widetilde{x}%
^{i_{1}}]^{2}  \notag \\
&&+[1+\varepsilon \ \left( 2(\ _{1}^{H}\chi +\frac{\ ^{1}\mathring{\Phi}}{\
^{1}\mathring{\Phi}^{\ast _{1}}}\ _{1}^{H}\chi ^{\ast _{1}})+\frac{\ ^{1}%
\mathring{\Phi}}{12(\ ^{H}s)^{2}\underline{\mathring{g}}_{6}}\ _{1}^{H}\chi
\right) ]\underline{\mathring{g}}_{6}\left[ dy^{6}+[1+\varepsilon (\frac{%
\partial _{i_{1}}(\ _{1}^{H}\chi \ \ ^{1}\mathring{\Phi})}{\partial
_{i_{1}}\ \ ^{1}\mathring{\Phi}}-\frac{(\ _{1}^{H}\chi \ \ ^{1}\mathring{\Phi%
})^{\ast _{1}}}{\ ^{1}\mathring{\Phi}^{\ast _{1}}})]\widetilde{\mathring{w}}%
_{i_{1}}d\widetilde{x}^{i_{1}}\right] ^{2}.  \label{ellips026d}
\end{eqnarray}%
}In the above formulas, we consider a generating function $\ _{1}^{H}\chi (%
\widetilde{x}^{i^{\prime }},y^{4},y^{6})$ which results, in general, in
nontrivial nonholonomic torsions encoding contributions of $\widehat{\mathbf{%
H}}_{\alpha _{1}\beta _{1}\mu _{1}}$ via an effective cosmological constant $%
(\ ^{H}s)^{2}.$ One considers a summation on $i_{1}^{\prime }=1,2,4$ for
terms like $\frac{\partial _{i_{1}}(\ _{1}^{H}\chi \ \ ^{1}\mathring{\Phi})}{%
\partial _{i_{1}}\ \ ^{1}\mathring{\Phi}}\widetilde{\mathring{w}}_{i_{1}}d%
\widetilde{x}^{i_{1}}.$

Off--diagonal extra dimensional gravitational interactions modify a Kerr
metric for any nontrivial cosmological constant determined by a
corresponding ansatz for $\widehat{\mathbf{H}}_{\alpha _{1}\beta _{1}\mu
_{1}}$ in 6--d. \ In a similar form we can generalize the constructions in
8--d and 10-d with nontrivial extra shell components $\widehat{\mathbf{H}}%
_{\alpha _{s}\beta _{s}\mu _{s}}$ and $\ ^{R}\Lambda =0$ in (\ref{es}).

\subsubsection{10--d deformations with NS 3-form and 6--d almost K\"{a}hler
internal spaces}

The class of solutions (\ref{ellips026d}) can be generalized for 10-d
spacetimes with indices of fields running 10-d values and nontrivial sources
(\ref{effects}) re-defined in (\ref{ansatzsourc}), with nontrivial $\
^{H}\Lambda =-5(\ ^{H}s)^{2},\ ^{F}\Lambda =-5n_{F}(\ ^{F}s)^{2}$ and$\
^{R}\Lambda =-5trn_{R}(\ ^{R}s)^{2}$ in $\Lambda =\ ^{H}\Lambda +\
^{F}\Lambda +\ ^{R}\Lambda .$ The corresponding quadratic line element is
{\small
\begin{eqnarray}
&&ds_{K\eta 10d}^{2}=(1+e^{\ ^{0}q}\frac{\ ^{1}q}{\ 4\Lambda }%
)[(dx^{1^{\prime }})^{2}+(dx^{2^{\prime }})^{2}]+  \label{10dsol} \\
&&\overline{A}[1-2\varepsilon \widetilde{\zeta }\sin (\omega _{0}\varphi
+\varphi _{0})][dy^{3^{\prime }}+\left( \varepsilon \partial _{k^{\prime }}\
^{\chi }n(x^{i^{\prime }})-\partial _{k^{\prime }}(\widehat{y}^{3^{\prime
}}+\varphi \frac{\overline{B}}{\overline{A}})\right) dx^{k^{\prime }}]^{2}+
\notag \\
&&\left[ 1+\varepsilon \left( \frac{\lbrack 8\overline{A}\Lambda +\mathring{%
\Phi}^{2}]\widetilde{\zeta }}{4\Lambda \mathring{\Phi}^{2}}\sin (\omega
_{0}\varphi +\varphi _{0})-\frac{4\widetilde{\zeta }\omega _{0}\overline{A}\
\Lambda }{\mathring{\Phi}^{2}}\cos (\omega _{0}\varphi +\varphi _{0})\right) %
\right] (\overline{C}-\frac{\overline{B}^{2}}{\overline{A}})[d\varphi
+\varepsilon (\partial _{i^{\prime }}\check{A})dx^{i^{\prime }}]^{2}+  \notag
\\
&&[1-\varepsilon \frac{1}{4\Lambda \underline{\mathring{g}}_{5}}\ ^{1}%
\mathring{\Phi}^{2}\ _{1}^{\Lambda }\chi ]\underline{\mathring{g}}_{5}(x^{k}(%
\widetilde{x}^{k^{\prime }}))[dy^{5}+[1+\varepsilon \ \ ^{1}\widetilde{n}%
_{i_{1}}\int dy^{6}(\ _{1}^{\Lambda }\chi +\frac{\ ^{1}\mathring{\Phi}}{\
^{1}\mathring{\Phi}^{\ast _{1}}}\ _{1}^{\Lambda }\chi ^{\ast _{1}}-\frac{5}{%
8\Lambda }\frac{1}{\underline{\mathring{g}}_{6}}(\ ^{1}\mathring{\Phi}^{2}\
_{1}^{\Lambda }\chi )^{\ast _{1}})]\underline{\mathring{n}}_{i_{1}}d%
\widetilde{x}^{i_{1}}]^{2}  \notag \\
&&+[1+\varepsilon \ \left( 2(\ _{1}^{\Lambda }\chi +\frac{\ ^{1}\mathring{%
\Phi}}{\ ^{1}\mathring{\Phi}^{\ast _{1}}}\ _{1}^{\Lambda }\chi ^{\ast _{1}})-%
\frac{\ ^{1}\mathring{\Phi}}{4\Lambda \underline{\mathring{g}}_{6}}\
_{1}^{\Lambda }\chi \right) ]\underline{\mathring{g}}_{6}\left[
dy^{6}+[1+\varepsilon (\frac{\partial _{i_{1}}(\ _{1}^{\Lambda }\chi \ \ ^{1}%
\mathring{\Phi})}{\partial _{i_{1}}\ \ ^{1}\mathring{\Phi}}-\frac{(\
_{1}^{\Lambda }\chi \ \ ^{1}\mathring{\Phi})^{\ast _{1}}}{\ ^{1}\mathring{%
\Phi}^{\ast _{1}}})]\widetilde{\mathring{w}}_{i_{1}}d\widetilde{x}^{i_{1}}%
\right] ^{2}+  \notag \\
&&[1-\varepsilon \frac{1}{4\Lambda \underline{\mathring{g}}_{7}}\ ^{2}%
\mathring{\Phi}^{2}\ _{2}^{\Lambda }\chi ]\underline{\mathring{g}}_{7}(x^{k}(%
\widetilde{x}^{k^{\prime }}))[dy^{7}+[1+\varepsilon \ \ ^{2}\widetilde{n}%
_{i_{2}}\int dy^{8}(\ _{2}^{\Lambda }\chi +\frac{\ ^{2}\mathring{\Phi}}{\
^{2}\mathring{\Phi}^{\ast _{2}}}\ _{2}^{\Lambda }\chi ^{\ast _{2}}-\frac{5}{%
8\Lambda }\frac{1}{\underline{\mathring{g}}_{8}}(\ ^{2}\mathring{\Phi}^{2}\
_{2}^{\Lambda }\chi )^{\ast _{2}})]\underline{\mathring{n}}_{i_{2}}d%
\widetilde{x}^{i_{2}}]^{2}  \notag \\
&&+[1+\varepsilon \ \left( 2(\ _{2}^{\Lambda }\chi +\frac{\ ^{2}\mathring{%
\Phi}}{\ ^{2}\mathring{\Phi}^{\ast _{2}}}\ _{2}^{\Lambda }\chi ^{\ast _{2}})-%
\frac{\ ^{2}\mathring{\Phi}}{4\Lambda \underline{\mathring{g}}_{8}}\
_{2}^{\Lambda }\chi \right) ]\underline{\mathring{g}}_{8}\left[
dy^{8}+[1+\varepsilon (\frac{\partial _{i_{2}}(\ _{2}^{\Lambda }\chi \ \ ^{2}%
\mathring{\Phi})}{\partial _{i_{2}}\ \ ^{2}\mathring{\Phi}}-\frac{(\
_{2}^{\Lambda }\chi \ \ ^{2}\mathring{\Phi})^{\ast _{2}}}{\ ^{2}\mathring{%
\Phi}^{\ast _{2}}})]\widetilde{\mathring{w}}_{i_{2}}d\widetilde{x}^{i_{2}}%
\right] ^{2}+  \notag
\end{eqnarray}%
\begin{eqnarray*}
&&\lbrack 1-\varepsilon \frac{1}{4\Lambda \underline{\mathring{g}}_{9}}\ ^{3}%
\mathring{\Phi}^{2}\ _{3}^{\Lambda }\chi ]\underline{\mathring{g}}_{9}(x^{k}(%
\widetilde{x}^{k^{\prime }}))[dy^{9}+[1+\varepsilon \ \ ^{3}\widetilde{n}%
_{i_{3}}\int dy^{10}(\ _{3}^{\Lambda }\chi +\frac{\ ^{3}\mathring{\Phi}}{\
^{3}\mathring{\Phi}^{\ast _{3}}}\ _{3}^{\Lambda }\chi ^{\ast _{3}}-\frac{5}{%
8\Lambda }\frac{1}{\underline{\mathring{g}}_{10}}(\ ^{3}\mathring{\Phi}^{2}\
_{3}^{\Lambda }\chi )^{\ast _{3}})]\underline{\mathring{n}}_{i_{3}}d%
\widetilde{x}^{i_{3}}]^{2} \\
&&+[1+\varepsilon \ \left( 2(\ _{3}^{\Lambda }\chi +\frac{\ 3\mathring{\Phi}%
}{\ ^{3}\mathring{\Phi}^{\ast _{3}}}\ _{3}^{\Lambda }\chi ^{\ast _{3}})-%
\frac{\ ^{3}\mathring{\Phi}}{4\Lambda \underline{\mathring{g}}_{10}}\
_{3}^{\Lambda }\chi \right) ]\underline{\mathring{g}}_{10}\left[
dy^{10}+[1+\varepsilon (\frac{\partial _{i_{3}}(\ _{3}^{\Lambda }\chi \ \
^{3}\mathring{\Phi})}{\partial _{i_{3}}\ \ ^{3}\mathring{\Phi}}-\frac{(\
_{3}^{\Lambda }\chi \ \ ^{3}\mathring{\Phi})^{\ast _{3}}}{\ ^{3}\mathring{%
\Phi}^{\ast _{3}}})]\widetilde{\mathring{w}}_{i_{3}}d\widetilde{x}^{i_{3}}%
\right] ^{2}.
\end{eqnarray*}%
} This generic off--diagonal stationary metric for 10-d spacetimes defines a
very general class of stationary solutions of the nonholonomic motion
equations in heterotic gravity (\ref{hs1})-(\ref{hs4}) with effective scalar
field encoded into the N--connection structure. It is important to use
"shell by shell" nonholonomic variables for the quadratic element (\ref%
{10dsol}). Only in such cases, we can understand the nonlinear symmetries
and classify the types of generating and integration functions corresponding
to horizontal and higher order vertical conventional sources. Such
properties can not be encoded in a minimal form if notations for indices
running coordinate values from 0 to 9 are used as in standard papers on
heterotic supergravity. For small $\varepsilon $-deformations, the 4-d
component of such metrics describes Kerr type black ellipsoid with
eccentricity $\varepsilon $ and $\varphi $-anisotropic polarization of
physical constants and horizons. The nonholonomic deformations also encode
sources from all gauge like and interior space gravitational fields via
re-definition of generating functions.

Similar classes of non--vacuum solutions can also be modelled for
Einstein--Finsler spaces if extra dimensional coordinates are treated as
velocity/momentum coordinates \cite%
{vjgp,vgrg,vcosmsol1,vnpfins,vapfins,tgovsv,vtamsuper}. The metrics possess
a respective Killing symmetry on $\partial _{t}$ and $\partial /\partial
y^{9}.$ Using the AFDM, we can construct solutions depending on all 10-d
coordinates which may describe geometric evolution and time propogating Kerr
black holes determined by heterotic sting gravity effects. They define $%
\varepsilon $--deformations of Kerr -- de Sitter black holes into ellipsoid
configurations with effective cosmological constants determined by constants
in GR, possible $f$--modifications and extra dimension contributions \cite%
{vkerrads}. By nonholonomic frame transforms and distortions of the
canonical d--connection, the above metric can be rewritten in canonical
almost-K\"{a}hler variables on 6-d internal space, as we prove in \cite%
{partner1}. We omit such constructions (being very important in various
models of deformation and brane quantization of MGTs and geometric evolution
theories \cite{vjgp,vwitten,vmedit,muen01}) in this work.

\subsubsection{Off-diagonal solutions in standard 10-d heterotic string
coordinates}

We can re-write the solution (\ref{10dsol}) in standard variables used in
\cite{lecht1,harl,gran,wecht, doug,lust, samt} with coordinates having prime
Greek indices re--defined in spherical 4-d coordinates (\ref{newcoord}) and
extra dimensions,
\begin{equation*}
x^{\mu }=(x^{0}=x^{0^{\prime }}=t,x^{1}=x^{1^{\prime }}=r,x^{2}=x^{2^{\prime
}}=\vartheta ,x^{3}=x^{3^{\prime }}=\varphi ,x^{4}=u^{5^{\prime
}}=u^{5},...,x^{9}=u^{10^{\prime }}=u^{10}),
\end{equation*}%
considering respective partial derivatives and explicit values for the
functions $\overline{A},\overline{B},\overline{C}$ \ (\ref{kerrcoef})\
determined by the 4-d Kerr black hole solution and explicit
parameterizations of integration and generating functions,{\small
\begin{eqnarray}
ds_{K\eta 10d}^{2} &=&(1+e^{\ ^{0}q(r)}\frac{\ ^{1}q(r)}{\ 4\Lambda }%
)[(dr)^{2}+\vartheta ^{2}(dr)^{2}]-\frac{r^{2}-2m_{0}+a^{2}-a^{2}\sin
^{2}\vartheta }{r^{2}+a^{2}\cos ^{2}\vartheta }[1-2\varepsilon \widetilde{%
\zeta }\sin (\omega _{0}\varphi +\varphi _{0})]  \notag \\
&&[dt+\left( \varepsilon \partial _{r}\ ^{\chi }n(r,\vartheta )-\partial
_{r}[\widehat{y}^{3^{\prime }}(r,\vartheta ,\varphi )+\varphi \frac{%
2m_{0}a\sin ^{2}\vartheta }{(r^{2}-2m_{0}+a^{2}-a^{2}\sin ^{2}\vartheta )}%
]\right) dr+  \notag \\
&&\left( \varepsilon \partial _{\vartheta }\ ^{\chi }n(r,\vartheta
)-\partial _{\vartheta }[\widehat{y}^{3^{\prime }}(r,\vartheta ,\varphi
)+\varphi \frac{2m_{0}a\sin ^{2}\vartheta }{(r^{2}-2m_{0}+a^{2}-a^{2}\sin
^{2}\vartheta )}]\right) d\vartheta ]^{2}+  \label{detailed10d} \\
&&\left[ 1+\varepsilon \left( \frac{\lbrack \mathring{\Phi}^{2}(r,\vartheta
,\varphi )-8\overline{A}\Lambda \frac{r^{2}-2m_{0}+a^{2}-a^{2}\sin
^{2}\vartheta }{r^{2}+a^{2}\cos ^{2}\vartheta }]\widetilde{\zeta }}{4\Lambda
\mathring{\Phi}^{2}(r,\vartheta ,\varphi )}\sin (\omega _{0}\varphi +\varphi
_{0})-\frac{4\widetilde{\zeta }\omega _{0}\overline{A}\ \Lambda }{\mathring{%
\Phi}^{2}}\cos (\omega _{0}\varphi +\varphi _{0})\right) \right]  \notag \\
&&(\frac{\sin ^{2}\vartheta \left[ (r^{2}+a^{2})^{2}-\Delta a^{2}\sin
^{2}\vartheta \right] }{r^{2}+a^{2}\cos ^{2}\vartheta }+\frac{%
4(m_{0})^{2}a^{2}\sin ^{4}\vartheta }{(r^{2}+a^{2}\cos ^{2}\vartheta
)(r^{2}-2m_{0}+a^{2}-a^{2}\sin ^{2}\vartheta )})  \notag \\
&&[d\varphi +\varepsilon (\partial _{r}\check{A}(r,\vartheta ,\varphi
))dr+\varepsilon (\partial _{\vartheta }\check{A}(r,\vartheta ,\varphi
))d\vartheta ]^{2}+  \notag
\end{eqnarray}%
\begin{eqnarray*}
&&[1-\varepsilon \frac{1}{4\Lambda \underline{\mathring{g}}_{5}(r,\vartheta
,\varphi )}\ ^{1}\mathring{\Phi}^{2}(r,\vartheta ,\varphi ,x^{5})\
_{1}^{\Lambda }\chi (r,\vartheta ,\varphi ,x^{5})]\underline{\mathring{g}}%
_{4}(r,\vartheta ,\varphi )[dx^{4}+  \notag \\
&&[1+\varepsilon \ \ ^{1}\widetilde{n}_{1}(r,\vartheta ,\varphi )\int
dx^{5}(\ _{1}^{\Lambda }\chi (r,\vartheta ,\varphi ,x^{5})+\frac{\ ^{1}%
\mathring{\Phi}(r,\vartheta ,\varphi ,x^{5})}{\frac{\partial }{\partial x^{5}%
}\ ^{1}\mathring{\Phi}(r,\vartheta ,\varphi ,x^{5})}\frac{\partial }{%
\partial x^{5}}\ _{1}^{\Lambda }\chi (r,\vartheta ,\varphi ,x^{5})  \notag \\
&&-\frac{5}{8\Lambda }\frac{1}{\underline{\mathring{g}}_{5}(r,\vartheta
,\varphi )}\frac{\partial }{\partial x^{5}}(\ ^{1}\mathring{\Phi}%
^{2}(r,\vartheta ,\varphi ,x^{5})\ _{1}^{\Lambda }\chi (r,\vartheta ,\varphi
,x^{5})))]\ ^{1}\underline{\mathring{n}}_{1}(r,\vartheta ,\varphi )dr+
\notag \\
&&\varepsilon \ \ ^{1}\widetilde{n}_{2}(r,\vartheta ,\varphi )\int dx^{5}(\
_{1}^{\Lambda }\chi (r,\vartheta ,\varphi ,x^{5})+\frac{\ ^{1}\mathring{\Phi}%
(r,\vartheta ,\varphi ,x^{5})}{\frac{\partial }{\partial x^{5}}\ ^{1}%
\mathring{\Phi}(r,\vartheta ,\varphi ,x^{5})}\frac{\partial }{\partial x^{5}}%
\ _{1}^{\Lambda }\chi (r,\vartheta ,\varphi ,x^{5})  \notag \\
&&-\frac{5}{8\Lambda }\frac{1}{\underline{\mathring{g}}_{5}(r,\vartheta
,\varphi )}\frac{\partial }{\partial x^{5}}(\ ^{1}\mathring{\Phi}%
^{2}(r,\vartheta ,\varphi ,x^{5})\ _{1}^{\Lambda }\chi (r,\vartheta ,\varphi
,x^{5})))]\ ^{1}\underline{\mathring{n}}_{2}(r,\vartheta ,\varphi
)d\vartheta +  \notag \\
&&\varepsilon \ \ ^{1}\widetilde{n}_{3}(r,\vartheta ,\varphi )\int dx^{5}(\
_{1}^{\Lambda }\chi (r,\vartheta ,\varphi ,x^{5})+\frac{\ ^{1}\mathring{\Phi}%
(r,\vartheta ,\varphi ,x^{5})}{\frac{\partial }{\partial x^{5}}\ ^{1}%
\mathring{\Phi}(r,\vartheta ,\varphi ,x^{5})}\frac{\partial }{\partial x^{5}}%
\ _{1}^{\Lambda }\chi (r,\vartheta ,\varphi ,x^{5})  \notag \\
&&-\frac{5}{8\Lambda }\frac{1}{\underline{\mathring{g}}_{5}(r,\vartheta
,\varphi )}\frac{\partial }{\partial x^{5}}(\ ^{1}\mathring{\Phi}%
^{2}(r,\vartheta ,\varphi ,x^{5})\ _{1}^{\Lambda }\chi (r,\vartheta ,\varphi
,x^{5})))]\ ^{1}\underline{\mathring{n}}_{3}(r,\vartheta ,\varphi )d\varphi
]^{2}+  \notag \\
&&[1+\varepsilon \ (2(\ _{1}^{\Lambda }\chi (r,\vartheta ,\varphi ,x^{5})+%
\frac{\ ^{1}\mathring{\Phi}(r,\vartheta ,\varphi ,x^{5})}{\frac{\partial }{%
\partial x^{5}}\ ^{1}\mathring{\Phi}(r,\vartheta ,\varphi ,x^{5})}\frac{%
\partial }{\partial x^{5}}\ _{1}^{\Lambda }\chi (r,\vartheta ,\varphi
,x^{5})) \\
&& -\frac{\ ^{1}\mathring{\Phi}(r,\vartheta ,\varphi ,x^{5})}{4\Lambda
\underline{\mathring{g}}_{5}(r,\vartheta ,\varphi )}\ _{1}^{\Lambda }\chi
(r,\vartheta ,\varphi ,x^{5}))]\underline{\mathring{g}}_{5}(r,\vartheta
,\varphi ) [dx^{5}+[1+\varepsilon (\frac{\partial _{r}(\ _{1}^{\Lambda }\chi
(r,\vartheta ,\varphi ,x^{5})\ \ ^{1}\mathring{\Phi}(r,\vartheta ,\varphi
,x^{5}))}{\partial _{r}\ \ ^{1}\mathring{\Phi}(r,\vartheta ,\varphi ,x^{5})}
\\
&&-\frac{\frac{\partial }{\partial x^{5}}(\ _{1}^{\Lambda }\chi (r,\vartheta
,\varphi ,x^{5})\ \ ^{1}\mathring{\Phi}(r,\vartheta ,\varphi ,x^{5}))}{\frac{%
\partial }{\partial x^{5}}\ ^{1}\mathring{\Phi}(r,\vartheta ,\varphi ,x^{5})}%
)]\ ^{1}\widetilde{\mathring{w}}_{1}(r,\vartheta ,\varphi )dr+ \\
&&[1+\varepsilon (\frac{\partial _{\vartheta }(\ _{1}^{\Lambda }\chi
(r,\vartheta ,\varphi ,x^{5})\ \ ^{1}\mathring{\Phi}(r,\vartheta ,\varphi
,x^{5}))}{\partial _{\vartheta }\ \ ^{1}\mathring{\Phi}(r,\vartheta ,\varphi
,x^{5})} -\frac{\frac{\partial }{\partial x^{5}}(\ _{1}^{\Lambda }\chi
(r,\vartheta ,\varphi ,x^{5})\ \ ^{1}\mathring{\Phi}(r,\vartheta ,\varphi
,x^{5}))}{\frac{\partial }{\partial x^{5}}\ ^{1}\mathring{\Phi}(r,\vartheta
,\varphi ,x^{5})})]\ ^{1}\widetilde{\mathring{w}}_{2}(r,\vartheta ,\varphi
)d\vartheta + \\
&&[1+\varepsilon (\frac{\partial _{\varphi }(\ _{1}^{\Lambda }\chi
(r,\vartheta ,\varphi ,x^{5})\ \ ^{1}\mathring{\Phi}(r,\vartheta ,\varphi
,x^{5}))}{\partial _{\varphi }\ \ ^{1}\mathring{\Phi}(r,\vartheta ,\varphi
,x^{5})} -\frac{\frac{\partial }{\partial x^{5}}(\ _{1}^{\Lambda }\chi
(r,\vartheta ,\varphi ,x^{5})\ \ ^{1}\mathring{\Phi}(r,\vartheta ,\varphi
,x^{5}))}{\frac{\partial }{\partial x^{5}}\ ^{1}\mathring{\Phi}(r,\vartheta
,\varphi ,x^{5})})]\ ^{1}\widetilde{\mathring{w}}_{3}(r,\vartheta ,\varphi
)d\varphi ]^{2}+
\end{eqnarray*}%
\begin{eqnarray*}
&&[1-\varepsilon \frac{1}{4\Lambda \underline{\mathring{g}}_{6}(r,\vartheta
,\varphi ,x^{5})}\ ^{2}\mathring{\Phi}^{2}(r,\vartheta ,\varphi
,x^{5},x^{7})\ _{2}^{\Lambda }\chi (r,\vartheta ,\varphi ,x^{5},x^{7})]%
\underline{\mathring{g}}_{6}(r,\vartheta ,\varphi ,x^{5}) \\
&&[dx^{6}+[1+\varepsilon \ \ ^{2}\widetilde{n}_{r}(r,\vartheta ,\varphi
,x^{5})\int dx^{7}(\ _{2}^{\Lambda }\chi (r,\vartheta ,\varphi ,x^{5},x^{7})+%
\frac{\ ^{2}\mathring{\Phi}(r,\vartheta ,\varphi ,x^{5},x^{7})}{\frac{%
\partial }{\partial x^{7}}\ ^{2}\mathring{\Phi}(r,\vartheta ,\varphi
,x^{5},x^{7})}\frac{\partial }{\partial x^{7}}\ _{2}^{\Lambda }\chi
(r,\vartheta ,\varphi ,x^{5},x^{7}) \\
&&-\frac{5}{8\Lambda }\frac{1}{\underline{\mathring{g}}_{7}(r,\vartheta
,\varphi ,x^{5})}\frac{\partial }{\partial x^{7}}(\ ^{2}\mathring{\Phi}%
^{2}(r,\vartheta ,\varphi ,x^{5},x^{7})\ _{2}^{\Lambda }\chi (r,\vartheta
,\varphi ,x^{5},x^{7})))]\ ^{2}\underline{\mathring{n}}_{r}(r,\vartheta
,\varphi ,x^{5})dr+ \\
&&[1+\varepsilon \ \ ^{2}\widetilde{n}_{\vartheta }(r,\vartheta ,\varphi
,x^{5})\int dx^{7}(\ _{2}^{\Lambda }\chi (r,\vartheta ,\varphi ,x^{5},x^{7})+%
\frac{\ ^{2}\mathring{\Phi}(r,\vartheta ,\varphi ,x^{5},x^{7})}{\frac{%
\partial }{\partial x^{7}}\ ^{2}\mathring{\Phi}(r,\vartheta ,\varphi
,x^{5},x^{7})}\frac{\partial }{\partial x^{7}}\ _{2}^{\Lambda }\chi
(r,\vartheta ,\varphi ,x^{5},x^{7}) \\
&&-\frac{5}{8\Lambda }\frac{1}{\underline{\mathring{g}}_{7}(r,\vartheta
,\varphi ,x^{5})}\frac{\partial }{\partial x^{7}}(\ ^{2}\mathring{\Phi}%
^{2}(r,\vartheta ,\varphi ,x^{5},x^{7})\ _{2}^{\Lambda }\chi (r,\vartheta
,\varphi ,x^{5},x^{7})))]\ ^{2}\underline{\mathring{n}}_{\vartheta
}(r,\vartheta ,\varphi ,x^{5})d\vartheta + \\
&&[1+\varepsilon \ \ ^{2}\widetilde{n}_{\varphi }(r,\vartheta ,\varphi
,x^{5})\int dx^{7}(\ _{2}^{\Lambda }\chi (r,\vartheta ,\varphi ,x^{5},x^{7})+%
\frac{\ ^{2}\mathring{\Phi}(r,\vartheta ,\varphi ,x^{5},x^{7})}{\frac{%
\partial }{\partial x^{7}}\ ^{2}\mathring{\Phi}(r,\vartheta ,\varphi
,x^{5},x^{7})}\frac{\partial }{\partial x^{7}}\ _{2}^{\Lambda }\chi
(r,\vartheta ,\varphi ,x^{5},x^{7}) \\
&&-\frac{5}{8\Lambda }\frac{1}{\underline{\mathring{g}}_{7}(r,\vartheta
,\varphi ,x^{5})}\frac{\partial }{\partial x^{7}}(\ ^{2}\mathring{\Phi}%
^{2}(r,\vartheta ,\varphi ,x^{5},x^{7})\ _{2}^{\Lambda }\chi (r,\vartheta
,\varphi ,x^{5},x^{7})))]\ ^{2}\underline{\mathring{n}}_{\varphi
}(r,\vartheta ,\varphi ,x^{5})d\varphi + \\
&&[1+\varepsilon \ \ ^{2}\widetilde{n}_{5}(r,\vartheta ,\varphi ,x^{5})\int
dx^{7}(\ _{2}^{\Lambda }\chi (r,\vartheta ,\varphi ,x^{5},x^{7})+\frac{\ ^{2}%
\mathring{\Phi}(r,\vartheta ,\varphi ,x^{5},x^{7})}{\frac{\partial }{%
\partial x^{7}}\ ^{2}\mathring{\Phi}(r,\vartheta ,\varphi ,x^{5},x^{7})}%
\frac{\partial }{\partial x^{7}}\ _{2}^{\Lambda }\chi (r,\vartheta ,\varphi
,x^{5},x^{7}) \\
&&-\frac{5}{8\Lambda }\frac{1}{\underline{\mathring{g}}_{7}(r,\vartheta
,\varphi ,x^{5})}\frac{\partial }{\partial x^{7}}(\ ^{2}\mathring{\Phi}%
^{2}(r,\vartheta ,\varphi ,x^{5},x^{7})\ _{2}^{\Lambda }\chi (r,\vartheta
,\varphi ,x^{5},x^{7})))]\ ^{2}\underline{\mathring{n}}_{5}(r,\vartheta
,\varphi ,x^{5})dx^{5}]^{2}+
\end{eqnarray*}%
\begin{eqnarray*}
&&[1+\varepsilon \ (2(\ _{2}^{\Lambda }\chi (r,\vartheta ,\varphi
,x^{5},x^{7})+\frac{\ ^{2}\mathring{\Phi}(r,\vartheta ,\varphi ,x^{5},x^{7})%
}{\frac{\partial }{\partial x^{7}}\ ^{2}\mathring{\Phi}(r,\vartheta ,\varphi
,x^{5},x^{7})}\frac{\partial }{\partial x^{7}}\ _{2}^{\Lambda }\chi
(r,\vartheta ,\varphi ,x^{5},x^{7})) \\
&&-\frac{\ ^{2}\mathring{\Phi}(r,\vartheta ,\varphi ,x^{5},x^{7})}{4\Lambda
\underline{\mathring{g}}_{7}(r,\vartheta ,\varphi ,x^{5},x^{7})}\
_{7}^{\Lambda }\chi (r,\vartheta ,\varphi ,x^{5},x^{7}))]\underline{%
\mathring{g}}_{7}(r,\vartheta ,\varphi ,x^{5}) \\
&&[dx^{7}+[1+\varepsilon (\frac{\partial _{r}(\ _{2}^{\Lambda }\chi
(r,\vartheta ,\varphi ,x^{5},x^{7})\ \ ^{2}\mathring{\Phi}(r,\vartheta
,\varphi ,x^{5},x^{7}))}{\partial _{r}\ \ ^{2}\mathring{\Phi}(r,\vartheta
,\varphi ,x^{5},x^{7})} \\
&&-\frac{\frac{\partial }{\partial x^{7}}(\ _{2}^{\Lambda }\chi (r,\vartheta
,\varphi ,x^{5},x^{7})\ \ ^{2}\mathring{\Phi}(r,\vartheta ,\varphi
,x^{5},x^{7}))}{\frac{\partial }{\partial x^{7}}\ ^{2}\mathring{\Phi}%
(r,\vartheta ,\varphi ,x^{5},x^{7})})]\ ^{2}\widetilde{\mathring{w}}%
_{r}(r,\vartheta ,\varphi ,x^{5})dr+ \\
&&[1+\varepsilon (\frac{\partial _{\vartheta }(\ _{2}^{\Lambda }\chi
(r,\vartheta ,\varphi ,x^{5},x^{7})\ \ ^{2}\mathring{\Phi}(r,\vartheta
,\varphi ,x^{5},x^{7}))}{\partial _{\vartheta }\ \ ^{2}\mathring{\Phi}%
(r,\vartheta ,\varphi ,x^{5},x^{7})} \\
&&-\frac{\frac{\partial }{\partial x^{7}}(\ _{2}^{\Lambda }\chi (r,\vartheta
,\varphi ,x^{5},x^{7})\ \ ^{2}\mathring{\Phi}(r,\vartheta ,\varphi
,x^{5},x^{7}))}{\frac{\partial }{\partial x^{7}}\ ^{2}\mathring{\Phi}%
(r,\vartheta ,\varphi ,x^{5},x^{7})})]\ ^{2}\widetilde{\mathring{w}}%
_{\vartheta }(r,\vartheta ,\varphi ,x^{5})d\vartheta + \\
&&[1+\varepsilon (\frac{\partial _{\varphi }(\ _{2}^{\Lambda }\chi
(r,\vartheta ,\varphi ,x^{5},x^{7})\ \ ^{2}\mathring{\Phi}(r,\vartheta
,\varphi ,x^{5},x^{7}))}{\partial _{\varphi }\ \ ^{2}\mathring{\Phi}%
(r,\vartheta ,\varphi ,x^{5},x^{7})} \\
&&-\frac{\frac{\partial }{\partial x^{7}}(\ _{2}^{\Lambda }\chi (r,\vartheta
,\varphi ,x^{5},x^{7})\ \ ^{2}\mathring{\Phi}(r,\vartheta ,\varphi
,x^{5},x^{7}))}{\frac{\partial }{\partial x^{7}}\ ^{2}\mathring{\Phi}%
(r,\vartheta ,\varphi ,x^{5},x^{7})})]\ ^{2}\widetilde{\mathring{w}}%
_{\varphi }(r,\vartheta ,\varphi ,x^{5})d\varphi + \\
&&[1+\varepsilon (\frac{\frac{\partial }{\partial x^{5}}(\ _{2}^{\Lambda
}\chi (r,\vartheta ,\varphi ,x^{5},x^{7})\ \ ^{2}\mathring{\Phi}(r,\vartheta
,\varphi ,x^{5},x^{7}))}{\frac{\partial }{\partial x^{5}}\ \ ^{2}\mathring{%
\Phi}(r,\vartheta ,\varphi ,x^{5},x^{7})} \\
&&-\frac{\frac{\partial }{\partial x^{7}}(\ _{2}^{\Lambda }\chi (r,\vartheta
,\varphi ,x^{5},x^{7})\ \ ^{2}\mathring{\Phi}(r,\vartheta ,\varphi
,x^{5},x^{7}))}{\frac{\partial }{\partial x^{7}}\ ^{2}\mathring{\Phi}%
(r,\vartheta ,\varphi ,x^{5},x^{7})})]\ ^{2}\widetilde{\mathring{w}}%
_{5}(r,\vartheta ,\varphi ,x^{5})dx^{5}]^{2}+
\end{eqnarray*}%
\begin{eqnarray*}
&&[1-\varepsilon \frac{1}{4\Lambda \underline{\mathring{g}}_{8}(r,\vartheta
,\varphi ,x^{5},x^{6},x^{7})}\ ^{3}\mathring{\Phi}^{2}(r,\vartheta ,\varphi
,x^{5},x^{6},x^{7},x^{9})\ _{3}^{\Lambda }\chi (r,\vartheta ,\varphi
,x^{5},x^{6},x^{7},x^{9})]\underline{\mathring{g}}_{8}(r,\vartheta ,\varphi
,x^{5},x^{6},x^{7}) \\
&&[dx^{8}+[1+\varepsilon \ \ ^{3}\widetilde{n}_{r}(r,\vartheta ,\varphi
,x^{5},x^{6},x^{7})\int dx^{9}(\ _{3}^{\Lambda }\chi (r,\vartheta ,\varphi
,x^{5},x^{6},x^{7},x^{9})+ \\
&&\frac{\ ^{3}\mathring{\Phi}(r,\vartheta ,\varphi ,x^{5},x^{6},x^{7},x^{9})%
}{\frac{\partial }{\partial x^{9}}\ ^{3}\mathring{\Phi}(r,\vartheta ,\varphi
,x^{5},x^{6},x^{7},x^{9})}\frac{\partial }{\partial x^{9}}\ _{3}^{\Lambda
}\chi (r,\vartheta ,\varphi ,x^{5},x^{6},x^{7},x^{9})- \\
&&\frac{5}{8\Lambda }\frac{1}{\underline{\mathring{g}}_{9}(r,\vartheta
,\varphi ,x^{5},x^{6},x^{7})}\frac{\partial }{\partial x^{9}}(\ ^{3}%
\mathring{\Phi}^{2}\ (r,\vartheta ,\varphi ,x^{5},x^{6},x^{7},x^{9})\
_{3}^{\Lambda }\chi (r,\vartheta ,\varphi ,x^{5},x^{6},x^{7},x^{9})))]\  \\
&&\ ^{3}\underline{\mathring{n}}_{r}(r,\vartheta ,\varphi
,x^{5},x^{6},x^{7})dr+[1+\varepsilon \ \ ^{3}\widetilde{n}_{\vartheta
}(r,\vartheta ,\varphi ,x^{5},x^{6},x^{7})\int dx^{9}(\ _{3}^{\Lambda }\chi
(r,\vartheta ,\varphi ,x^{5},x^{6},x^{7},x^{9})+ \\
&&\frac{\ ^{3}\mathring{\Phi}(r,\vartheta ,\varphi ,x^{5},x^{6},x^{7},x^{9})%
}{\frac{\partial }{\partial x^{9}}\ ^{3}\mathring{\Phi}(r,\vartheta ,\varphi
,x^{5},x^{6},x^{7},x^{9})}\frac{\partial }{\partial x^{9}}\ _{3}^{\Lambda
}\chi (r,\vartheta ,\varphi ,x^{5},x^{6},x^{7},x^{9})- \\
&&\frac{5}{8\Lambda }\frac{1}{\underline{\mathring{g}}_{9}(r,\vartheta
,\varphi ,x^{5},x^{6},x^{7})}\frac{\partial }{\partial x^{9}}(\ ^{3}%
\mathring{\Phi}^{2}\ (r,\vartheta ,\varphi ,x^{5},x^{6},x^{7},x^{9})\
_{3}^{\Lambda }\chi (r,\vartheta ,\varphi ,x^{5},x^{6},x^{7},x^{9})))] \\
&&\ ^{3}\underline{\mathring{n}}_{\vartheta }(r,\vartheta ,\varphi
,x^{5},x^{6},x^{7})d\vartheta +[1+\varepsilon \ \ ^{3}\widetilde{n}_{\varphi
}(r,\vartheta ,\varphi ,x^{5},x^{6},x^{7})\int dx^{9}(\ _{3}^{\Lambda }\chi
(r,\vartheta ,\varphi ,x^{5},x^{6},x^{7},x^{9})+ \\
&&\frac{\ ^{3}\mathring{\Phi}(r,\vartheta ,\varphi ,x^{5},x^{6},x^{7},x^{9})%
}{\frac{\partial }{\partial x^{9}}\ ^{3}\mathring{\Phi}(r,\vartheta ,\varphi
,x^{5},x^{6},x^{7},x^{9})}\frac{\partial }{\partial x^{9}}\ _{3}^{\Lambda
}\chi (r,\vartheta ,\varphi ,x^{5},x^{6},x^{7},x^{9})- \\
&&\frac{5}{8\Lambda }\frac{1}{\underline{\mathring{g}}_{9}(r,\vartheta
,\varphi ,x^{5},x^{6},x^{7})}\frac{\partial }{\partial x^{9}}(\ ^{3}%
\mathring{\Phi}^{2}\ (r,\vartheta ,\varphi ,x^{5},x^{6},x^{7},x^{9})\
_{3}^{\Lambda }\chi (r,\vartheta ,\varphi ,x^{5},x^{6},x^{7},x^{9})))]\ \
^{3}\underline{\mathring{n}}_{\varphi }(r,\vartheta ,\varphi
,x^{5},x^{6},x^{7})d\varphi +
\end{eqnarray*}%
\begin{eqnarray*}
&&[1+\varepsilon \ \ ^{3}\widetilde{n}_{5}(r,\vartheta ,\varphi
,x^{5},x^{6},x^{7})\int dx^{9}(\ _{3}^{\Lambda }\chi (r,\vartheta ,\varphi
,x^{5},x^{6},x^{7},x^{9})+ \\
&&\frac{\ ^{3}\mathring{\Phi}(r,\vartheta ,\varphi ,x^{5},x^{6},x^{7},x^{9})%
}{\frac{\partial }{\partial x^{9}}\ ^{3}\mathring{\Phi}(r,\vartheta ,\varphi
,x^{5},x^{6},x^{7},x^{9})}\frac{\partial }{\partial x^{9}}\ _{3}^{\Lambda
}\chi (r,\vartheta ,\varphi ,x^{5},x^{6},x^{7},x^{9})- \\
&&\frac{5}{8\Lambda }\frac{1}{\underline{\mathring{g}}_{9}(r,\vartheta
,\varphi ,x^{5},x^{6},x^{7})}\frac{\partial }{\partial x^{9}}(\ ^{3}%
\mathring{\Phi}^{2}\ (r,\vartheta ,\varphi ,x^{5},x^{6},x^{7},x^{9})\
_{3}^{\Lambda }\chi (r,\vartheta ,\varphi ,x^{5},x^{6},x^{7},x^{9})))]\ \
^{3}\underline{\mathring{n}}_{5}(r,\vartheta ,\varphi
,x^{5},x^{6},x^{7})dx^{5}+ \\
&&[1+\varepsilon \ \ ^{3}\widetilde{n}_{6}(r,\vartheta ,\varphi
,x^{5},x^{6},x^{7})\int dx^{9}(\ _{3}^{\Lambda }\chi (r,\vartheta ,\varphi
,x^{5},x^{6},x^{7},x^{9})+ \\
&& \frac{\ ^{3}\mathring{\Phi}(r,\vartheta ,\varphi ,x^{5},x^{6},x^{7},x^{9})%
}{\frac{\partial }{\partial x^{9}}\ ^{3}\mathring{\Phi}(r,\vartheta ,\varphi
,x^{5},x^{6},x^{7},x^{9})}\frac{\partial }{\partial x^{9}}\ _{3}^{\Lambda
}\chi (r,\vartheta ,\varphi ,x^{5},x^{6},x^{7},x^{9})- \\
&&\frac{5}{8\Lambda }\frac{1}{\underline{\mathring{g}}_{9}(r,\vartheta
,\varphi ,x^{5},x^{6},x^{7})}\frac{\partial }{\partial x^{9}}(\ ^{3}%
\mathring{\Phi}^{2}\ (r,\vartheta ,\varphi ,x^{5},x^{6},x^{7},x^{9})\
_{3}^{\Lambda }\chi (r,\vartheta ,\varphi ,x^{5},x^{6},x^{7},x^{9})))]\  \\
&&\ ^{3}\underline{\mathring{n}}_{6}(r,\vartheta ,\varphi
,x^{5},x^{6},x^{7})dx^{6}+[1+\varepsilon \ \ ^{3}\widetilde{n}%
_{7}(r,\vartheta ,\varphi ,x^{5},x^{6},x^{7})\int dx^{9}(\ _{3}^{\Lambda
}\chi (r,\vartheta ,\varphi ,x^{5},x^{6},x^{7},x^{9})+ \\
&&\frac{\ ^{3}\mathring{\Phi}(r,\vartheta ,\varphi ,x^{5},x^{6},x^{7},x^{9})%
}{\frac{\partial }{\partial x^{9}}\ ^{3}\mathring{\Phi}(r,\vartheta ,\varphi
,x^{5},x^{6},x^{7},x^{9})}\frac{\partial }{\partial x^{9}}\ _{3}^{\Lambda
}\chi (r,\vartheta ,\varphi ,x^{5},x^{6},x^{7},x^{9})- \\
&&\frac{5}{8\Lambda }\frac{1}{\underline{\mathring{g}}_{9}(r,\vartheta
,\varphi ,x^{5},x^{6},x^{7})}\frac{\partial }{\partial x^{9}}(\ ^{3}%
\mathring{\Phi}^{2}\ (r,\vartheta ,\varphi ,x^{5},x^{6},x^{7},x^{9})\
_{3}^{\Lambda }\chi (r,\vartheta ,\varphi ,x^{5},x^{6},x^{7},x^{9})))] \\
&&\ ^{3}\underline{\mathring{n}}_{7}(r,\vartheta ,\varphi
,x^{5},x^{6},x^{7})dx^{7}]^{2}+[1+\varepsilon \ (2(\ _{3}^{\Lambda }\chi
(r,\vartheta ,\varphi ,x^{5},x^{6},x^{7},x^{9})+\frac{\ 3\mathring{\Phi}%
(r,\vartheta ,\varphi ,x^{5},x^{6},x^{7},x^{9})}{\frac{\partial }{\partial
x^{9}}\ ^{3}\mathring{\Phi}(r,\vartheta ,\varphi ,x^{5},x^{6},x^{7},x^{9})}
\\
&&\frac{\partial }{\partial x^{9}}\ _{3}^{\Lambda }\chi (r,\vartheta
,\varphi ,x^{5},x^{6},x^{7},x^{9}))-\frac{\ ^{3}\mathring{\Phi}(r,\vartheta
,\varphi ,x^{5},x^{6},x^{7},x^{9})}{4\Lambda \underline{\mathring{g}}%
_{9}(r,\vartheta ,\varphi ,x^{5},x^{6},x^{7},x^{9})}\ _{3}^{\Lambda }\chi
(r,\vartheta ,\varphi ,x^{5},x^{6},x^{7},x^{9}))]\underline{\mathring{g}}%
_{9}(r,\vartheta ,\varphi ,x^{5},x^{6},x^{7}) \\
&&[dx^{9}+[1+\varepsilon (\frac{\partial _{r}(\ _{3}^{\Lambda }\chi
(r,\vartheta ,\varphi ,x^{5},x^{6},x^{7},x^{9})\ \ ^{3}\mathring{\Phi}%
(r,\vartheta ,\varphi ,x^{5},x^{6},x^{7},x^{9}))}{\partial _{i_{3}}\ \ ^{3}%
\mathring{\Phi}(r,\vartheta ,\varphi ,x^{5},x^{6},x^{7},x^{9})} \\
&&-\frac{\frac{\partial }{\partial x^{9}}(\ _{3}^{\Lambda }\chi (r,\vartheta
,\varphi ,x^{5},x^{6},x^{7},x^{9})\ \ ^{3}\mathring{\Phi}(r,\vartheta
,\varphi ,x^{5},x^{6},x^{7},x^{9}))}{\frac{\partial }{\partial x^{9}}\ ^{3}%
\mathring{\Phi}(r,\vartheta ,\varphi ,x^{5},x^{6},x^{7},x^{9})})]\ ^{3}%
\widetilde{\mathring{w}}_{r}(r,\vartheta ,\varphi ,x^{5},x^{6},x^{7})dr+ \\
&&[1+\varepsilon (\frac{\partial _{r}(\ _{3}^{\Lambda }\chi (r,\vartheta
,\varphi ,x^{5},x^{6},x^{7},x^{9})\ \ ^{3}\mathring{\Phi}(r,\vartheta
,\varphi ,x^{5},x^{6},x^{7},x^{9}))}{\partial _{r}\ \ ^{3}\mathring{\Phi}%
(r,\vartheta ,\varphi ,x^{5},x^{6},x^{7},x^{9})} \\
&&-\frac{\frac{\partial }{\partial x^{9}}(\ _{3}^{\Lambda }\chi (r,\vartheta
,\varphi ,x^{5},x^{6},x^{7},x^{9})\ \ ^{3}\mathring{\Phi}(r,\vartheta
,\varphi ,x^{5},x^{6},x^{7},x^{9}))}{\frac{\partial }{\partial x^{9}}\ ^{3}%
\mathring{\Phi}(r,\vartheta ,\varphi ,x^{5},x^{6},x^{7},x^{9})})]\ ^{3}%
\widetilde{\mathring{w}}_{r}(r,\vartheta ,\varphi ,x^{5},x^{6},x^{7})dr+ \\
&&[1+\varepsilon (\frac{\partial _{\vartheta }(\ _{3}^{\Lambda }\chi
(r,\vartheta ,\varphi ,x^{5},x^{6},x^{7},x^{9})\ \ ^{3}\mathring{\Phi}%
(r,\vartheta ,\varphi ,x^{5},x^{6},x^{7},x^{9}))}{\partial _{\vartheta }\ \
^{3}\mathring{\Phi}(r,\vartheta ,\varphi ,x^{5},x^{6},x^{7},x^{9})} \\
&&-\frac{\frac{\partial }{\partial x^{9}}(\ _{3}^{\Lambda }\chi (r,\vartheta
,\varphi ,x^{5},x^{6},x^{7},x^{9})\ \ ^{3}\mathring{\Phi}(r,\vartheta
,\varphi ,x^{5},x^{6},x^{7},x^{9}))}{\frac{\partial }{\partial x^{9}}\ ^{3}%
\mathring{\Phi}(r,\vartheta ,\varphi ,x^{5},x^{6},x^{7},x^{9})})]\ ^{3}%
\widetilde{\mathring{w}}_{\vartheta }(r,\vartheta ,\varphi
,x^{5},x^{6},x^{7})d\vartheta + \\
&&[1+\varepsilon (\frac{\partial _{\varphi }(\ _{3}^{\Lambda }\chi
(r,\vartheta ,\varphi ,x^{5},x^{6},x^{7},x^{9})\ \ ^{3}\mathring{\Phi}%
(r,\vartheta ,\varphi ,x^{5},x^{6},x^{7},x^{9}))}{\partial _{\varphi }\ \
^{3}\mathring{\Phi}(r,\vartheta ,\varphi ,x^{5},x^{6},x^{7},x^{9})} \\
&&-\frac{\frac{\partial }{\partial x^{9}}(\ _{3}^{\Lambda }\chi (r,\vartheta
,\varphi ,x^{5},x^{6},x^{7},x^{9})\ \ ^{3}\mathring{\Phi}(r,\vartheta
,\varphi ,x^{5},x^{6},x^{7},x^{9}))}{\frac{\partial }{\partial x^{9}}\ ^{3}%
\mathring{\Phi}(r,\vartheta ,\varphi ,x^{5},x^{6},x^{7},x^{9})})]\ ^{3}%
\widetilde{\mathring{w}}_{\varphi }(r,\vartheta ,\varphi
,x^{5},x^{6},x^{7})d\varphi + \\
&&[1+\varepsilon (\frac{\frac{\partial }{\partial x^{5}}(\ _{3}^{\Lambda
}\chi (r,\vartheta ,\varphi ,x^{5},x^{6},x^{7},x^{9})\ \ ^{3}\mathring{\Phi}%
(r,\vartheta ,\varphi ,x^{5},x^{6},x^{7},x^{9}))}{\frac{\partial }{\partial
x^{5}}\ \ ^{3}\mathring{\Phi}(r,\vartheta ,\varphi ,x^{5},x^{6},x^{7},x^{9})}
\\
&&-\frac{\frac{\partial }{\partial x^{9}}(\ _{3}^{\Lambda }\chi (r,\vartheta
,\varphi ,x^{5},x^{6},x^{7},x^{9})\ \ ^{3}\mathring{\Phi}(r,\vartheta
,\varphi ,x^{5},x^{6},x^{7},x^{9}))}{\frac{\partial }{\partial x^{9}}\ ^{3}%
\mathring{\Phi}(r,\vartheta ,\varphi ,x^{5},x^{6},x^{7},x^{9})})]\ ^{3}%
\widetilde{\mathring{w}}_{5}(r,\vartheta ,\varphi ,x^{5},x^{6},x^{7})dx^{5}+
\end{eqnarray*}%
\begin{eqnarray*}
&&[1+\varepsilon (\frac{\frac{\partial }{\partial x^{6}}(\ _{3}^{\Lambda
}\chi (r,\vartheta ,\varphi ,x^{5},x^{6},x^{7},x^{9})\ \ ^{3}\mathring{\Phi}%
(r,\vartheta ,\varphi ,x^{5},x^{6},x^{7},x^{9}))}{\frac{\partial }{\partial
x^{6}}\ \ ^{3}\mathring{\Phi}(r,\vartheta ,\varphi ,x^{5},x^{6},x^{7},x^{9})}
\\
&&-\frac{\frac{\partial }{\partial x^{9}}(\ _{3}^{\Lambda }\chi (r,\vartheta
,\varphi ,x^{5},x^{6},x^{7},x^{9})\ \ ^{3}\mathring{\Phi}(r,\vartheta
,\varphi ,x^{5},x^{6},x^{7},x^{9}))}{\frac{\partial }{\partial x^{9}}\ ^{3}%
\mathring{\Phi}(r,\vartheta ,\varphi ,x^{5},x^{6},x^{7},x^{9})})]\ ^{3}%
\widetilde{\mathring{w}}_{6}(r,\vartheta ,\varphi ,x^{5},x^{6},x^{7})dx^{6}+
\end{eqnarray*}%
\begin{eqnarray*}
&&[1+\varepsilon (\frac{\frac{\partial }{\partial x^{7}}(\ _{3}^{\Lambda
}\chi (r,\vartheta ,\varphi ,x^{5},x^{6},x^{7},x^{9})\ \ ^{3}\mathring{\Phi}%
(r,\vartheta ,\varphi ,x^{5},x^{6},x^{7},x^{9}))}{\frac{\partial }{\partial
x^{7}}\ \ ^{3}\mathring{\Phi}(r,\vartheta ,\varphi ,x^{5},x^{6},x^{7},x^{9})}
\\
&&-\frac{\frac{\partial }{\partial x^{9}}(\ _{3}^{\Lambda }\chi (r,\vartheta
,\varphi ,x^{5},x^{6},x^{7},x^{9})\ \ ^{3}\mathring{\Phi}(r,\vartheta
,\varphi ,x^{5},x^{6},x^{7},x^{9}))}{\frac{\partial }{\partial x^{9}}\ ^{3}%
\mathring{\Phi}(r,\vartheta ,\varphi ,x^{5},x^{6},x^{7},x^{9})})]\ ^{3}%
\widetilde{\mathring{w}}_{7}(r,\vartheta ,\varphi
,x^{5},x^{6},x^{7})dx^{6}]^{2}.
\end{eqnarray*}%
} The above quadratic elements transform into a trivial embedding of the 4-d
Kerr solution into a 10-d spacetime if $\varepsilon \rightarrow 0$ and
primary data (labelled by circles) can be transformed into a diagonal 6-d
internal space.

The solution (\ref{detailed10d}) is equivalent to (\ref{10dsol}) up to
re-definition of coordinates and some integration functions. It is difficult
to see "shell by shell" nonlinear symmetries and construct generic
off-diagonal solutions in the variant with standard variables used in
heterotic string gravity. In former non N-adapted variables, the formulas
are much more cumbersome and less adapted for studying limits to well known
solutions and generalizations with nontrivial backgrounds and extra
dimensional contributions. To elaborate and apply a corresponding "shell by
shell" N-adapted geometric techniques of constructing exact solutions in 4d
and extra dimension theories is important both from a mathematical stand
point and a physical point of view.

\section{ Outlook and Concluding Remarks}

\label{s4}In this work, we have applied the anholonomic frame method, AFDM,
for constructing new classes of stationary solutions of motion equations in
heterotic supergravity. Such solutions have generic off-diagonal metrics for
effective ten dimenisonal, 10-d, spacetimes enabled with generalized
connections and depend on all possible 4-d and extra dimensional space
coordinates. They admit subclasses of solutions with warping on coordinate $%
y^{4}$, nearly almost-K\"{a}hler 6-d internal manifolds in the presence of
nonholonomically deformed gravitational and gauge instantons. The almost-K%
\"{a}hler structure is necessary if we want to generate the Kerr metric with
possible (off-) diagonal and nonholonomic deformations to black ellipsoid
type solutions with locally anisotropic polarized physical constants, small
deformations of horizons, embedding into nontrivial extra dimension vacuum
gravitational fields and/or gauge \ configurations. \ These solutions
preserve two real supercharges corresponding to $N=1/2$ supersymmetry from
the viewpoint of four non-compact dimensions and various nonholonomic
deformations.

Following the AFDM, we can integrate the motion equations in heterotic
supergravity in very general forms with dependence on all 10-d spacetime
coordinates. Such constructions are possible to more general classes of
almost-K\"{a}hler nonholonomic variables and so-called canonical
nonholonomic variables. This allows us to decouple string modified Einstein
equations with effective sources in general form which are similar to
Einstein-Yang-Mills-Higgs, EYMH, systems in higher dimensions and with
generalized gauge like interactions. It is possible to consider associated \
$SU(3)$ structures and solve generalized BPS equations and Bianchi
identities. A crucial difference from former approaches is that our
geometric methods allow us to work with generating and integration functions
for off-diagonal metrics and connections transforming motion equations into
nonlinear systems of partial differential equations, PDEs. In particular, we
can reproduce former results for a diagonalizable ansatz with dependence on
radial and warping coordinates as solutions of ordinary differential
equations, ODEs.

To illustrate the power and importance of the AFDM as the most general
geometric method of constructing analytic solutions of (modified) motion /
gravitational and field equations, we show how this formalism can be applied
for generating N--adapted (i.e. adapted to nonlinear connection structures)
YM and instanton configurations with possible associated $SU(3)$
nonholonomic structures, see the associated work \cite{partner1}. New
classes of exact solutions describing small parametric modifications of Kerr
metrics with effective string sources are provided. We show that in a
certain sense, a large class of physical effects in modified gravity models
like $R^{2}$ can be equivalently modelled/explained by nonholonomic
constraints and effective sources in heterotic string gravity. In explicit \
form, exact/parameteric extra dimension \ deformations of the black hole
metrics in 6-d and 10-d gravity with NS-3 form and 6-d almost-K\"{a}hler
internal spaces are constructed and analysed.

Finally, we emphasize that there are a plethora of future directions which
can be pursued using our methods and results as starting points. This
includes the construction of cosmological solutions in the heterotic string
gravity and/or the study of smooth compact nonholonomic varieties in both
heterotic and geometric flow context. Similar analyses can be performed in
type II string theory in particular, including the Ramond-Ramond sector
and/or considering geometric flows on internal spaces. We worked to the
lowest order of $\alpha ^{\prime }$ but possibilities in the AFDM exist to
extend the constructions to higher orders.

\vskip3pt \textbf{Acknowledgments:} The SV research is for the QGR--Topanga
with a former partial support by IDEI, PN-II-ID-PCE-2011-3-0256 and DAAD. He is grateful  to N. Mavromatos,  D. L\"{u}st, O. Lechtenfeld, S. D. Odintsov and C. Castro Perelman for valuable discussions and support. This work contains also a summary of results of a talk at GR21 in NY.

\end{document}